\documentclass[aps,twocolumn,showpacs,preprintnumbers,floats]{revtex4}\def\@cite#1#2{\textsuperscript{[{#1\if@tempswa , #2\fi}]}}

\usepackage{graphicx}
\usepackage{amsmath}
\usepackage{amsfonts}
\usepackage{amssymb}
\usepackage{color}
\usepackage{subfigure}
\usepackage{epsfig}
\usepackage{morefloats}
\usepackage{multirow}
\usepackage{graphicx,booktabs}
\usepackage{mathrsfs}
\usepackage{txfonts}
\usepackage{indentfirst}
\usepackage{graphicx,booktabs}
\usepackage{booktabs}
\usepackage{longtable,lscape}
\usepackage{multirow}
\usepackage{bm}
\usepackage{graphicx}
\usepackage[colorlinks, citecolor=blue,anchorcolor=red,menucolor=red, linkcolor=red,filecolor=red,urlcolor=blue,frenchlinks=red]{hyperref}

\newcommand{\vsig}{\mbox{\boldmath$\sigma$\unboldmath}}

\begin{document}

\title{Mass spectra and strong decays of charmed and charmed-strange mesons }

\author{Ru-Hui Ni$^{1}$, Qi Li$^{2}$,
Xian-Hui Zhong$^{1,3}$~\footnote {E-mail: zhongxh@hunnu.edu.cn}}

\affiliation{ 1) Department of Physics, Hunan Normal University, and Key Laboratory of Low-Dimensional Quantum Structures and Quantum Control of Ministry of Education, Changsha 410081, China }

\affiliation{ 2)  School of Science, Tianjin Chengjian University, Tianjin 300000, China}
\affiliation{ 3)  Synergetic Innovation Center for Quantum Effects and Applications (SICQEA),
Hunan Normal University, Changsha 410081, China}


\begin{abstract}

 A semi-relativistic potential model is adopted to calculate the mass spectra of charmed and charmed-strange meson states up to the $2D$ excitations.
 The strong decay properties are further analyzed with a chiral quark model by using the numerical wave functions obtained from the potential model.
 By using the strong decay amplitudes extracted from the chiral quark model, we also systematically study the coupled-channel effects on
 the bare masses of the $1P$-wave states, since the masses of $D^*_{s0}(2317)$ and $D_{s1}(2460)$
 cannot be explained with bare $1P$-wave states within the potential model. Based on our good descriptions of the mass and decay properties for the low-lying well-established states, we give a quark model classification for the high mass resonances observed in recent years.
 In the $D$-meson family, $D_0(2550)$ can be classified as the radially excited state $D(2^1S_0)$; $D_3^*(2750)$ and $D_2(2740)$
 can be classified as the second orbital excitations $D(1^3D_3)$ and $D(1D'_2)$, respectively; $D_J^*(3000)$ may be a candidate
 of $D(1^3F_4)$ or $D(2^3P_2)$; while $D_J(3000)$ may favor the high mass mixed state $D(2P'_1)$; however,
 there still exist puzzles for understanding the natures of $D_1^*(2600)$ and $D_1^*(2760)$, whose decay properties
 cannot be well explained with either pure $D(2^3S_1)$ and $D(1^3D_1)$ states or their mixing. In the $D_s$-meson family,
 $D_{s3}^*(2860)$ favors the $D_s(1^3D_3)$ assignment; $D_{s1}^*(2700)$ and $D_{s1}^*(2860)$ may favor the mixed states
 $|(SD)_1\rangle_L$ and $|(SD)_1\rangle_H$ via the $2^3S_1$-$1^3D_1$ mixing, respectively; $D_{sJ}(3040)$ may favor
 $D_s(2P_1)$ or $D_s(2P_1')$, or corresponds to a structure contributed by both $D_s(2P_1)$ and $D_s(2P_1')$; the newly observed resonance $D_{s0}(2590)^+$ as an assignment of $D_s(2^1S_0)$, by including coupled-channel effects the mass of $D_s(2^1S_0)$ is close to
 the observed value, however, the width cannot be well understood in the present study. Many missing excited $D$- and $D_s$-meson states have a relatively narrow width, they are most likely to be observed in their dominant decay channels in future experiments.

\end{abstract}


\maketitle

\section{Introduction}

In the past 15 years, significant progress has been achieved in
the observations of the $D$ and $D_s$ meson spectra. More and more higher
excitations have been found in experiments. In the $D$-meson family,
several new signals $D(2550)^0$, $D^*(2600)^{0,+}$, $D(2750)^0$ and $D^*(2760)^{0,+}$
were observed for the first time by the BaBar collaboration in 2010~\cite{BaBar:2010zpy},
and were confirmed by the LHCb collaboration with slightly different
masses in 2013~\cite{LHCb:2013jjb}. The decay angular distributions
show that both $D(2550)$ and $D(2750)$ should have an unnatural spin parity,
while both $D^*(2600)$ and $D^*(2760)$ favor a natural spin parity.
Furthermore, the LHCb collaboration observed two new higher $D$-meson excitations, $D^*_J(3000)$ and
$D_J(3000)$, with natural and unnatural parities,
respectively~\cite{LHCb:2013jjb}. In 2015, LHCb observed a new state $D^*_1(2760)^0$
with spin-parity numbers $J^P=1^-$ in the $D^+\pi^-$ channel by analyzing the $B^- \to D^{+} K^- \pi^-$ decay~\cite{LHCb:2015eqv}.
In 2016, LHCb also observed a new state $D^*_2(3000)$ with $J^P=2^+$ in the
$D^+\pi^-$ channel by analyzing the $B^- \to D^{+} \pi^- \pi^-$ decay~\cite{Aaij:2016fma}.
The resonance parameters of $D^*_2(3000)$ are inconsistent with the previously
observed resonance $D^*_J(3000)$ in Ref.~\cite{LHCb:2013jjb}. In 2019, a four-body amplitude analysis of
the $B^- \to D^{*+} \pi^- \pi^-$ decay is performed by the LHCb collaboration~\cite{Aaij:2019sqk}.
The spin-parity numbers for $D(2550)$, $D^*(2600)$, $D(2750)$ and $D^*(2760)$ were
systematically determined to be $J^P=0^-$, $1^-$, $2^-$ and $3^-$, respectively.
The determined spin parity numbers for $D^*(2600)$ and $D^*(2760)$ are consistent
with the determinations in the previous experiments~\cite{Aaij:2016fma,Aaij:2015sqa}.
In the Review of Particle Physics (RPP), $D(2550)$, $D^*(2600)$, $D(2750)$ and $D^*(2760)$
are labeled as $D_0(2550)$, $D^*_1(2600)$, $D_2(2740)$ and $D^*_3(2750)$, respectively,
by the particle data group (PDG)~\cite{Zyla:2020zbs}.

In the $D_s$-meson sector, two
new resonances/structures $D_{sJ}(2700)^+$ and $D_{sJ}(2860)^+$ observed in the $DK$ channel by BaBar
in 2006~\cite{BaBar:2006gme}. The $D_{sJ}(2700)^+$ was confirmed by Belle one year
later~\cite{Belle:2007hht}, and its spin-parity numbers were determined
to be $J^P=1^-$. In 2009, BaBar observed the decays $D_{s1}^*(2700)^+\to D^*K$ and $D_{sJ}^*(2860)^+\to D^*K$,
and measured their branching fractions relative to the $DK$ final state~\cite{BaBar:2009rro}.
Meanwhile, a new broad higher $D_s$-meson excitation $D_{sJ}(3040)^+$ was also reported by BaBar.
In 2012, the existence of $D_{s1}^*(2700)^+$ and $D_{sJ}(2860)^+$ was further confirmed by using the $pp$
collision data at LHCb~\cite{LHCb:2012uts}. In 2014, by an analysis of $B_s^0 \rightarrow \bar{D}^0 K^- \pi^+$ decays,
the LHCb collaboration found two resonances $D_{s1}^*(2860)^-$ with $J^P=1^-$
and $D_{s3}^*(2860)^-$ with $J^P=3^-$ in the $\bar{D}^0K^-$ final state~\cite{LHCb:2014ott,LHCb:2014ioa},
which indicates that the $D_{sJ}(2860)$ structure previously observed
by BaBar~\cite{BaBar:2009rro,BaBar:2006gme} and LHCb~\cite{LHCb:2012uts}
consists of at least these two resonances. In 2016, the $D_{s3}^*(2860)^+$ resonance was observed
in the $D^{*+}K^0_s$ channel by LHCb~\cite{Aaij:2016utb}, its resonance parameters and
spin-parity numbers are consistent with the determinations for $D_{s3}^*(2860)^-$ in
Refs.~\cite{LHCb:2014ott,LHCb:2014ioa}. Furthermore, LHCb also found weak evidence of
$D_{sJ}(3040)^+$ consistent with an unnatural parity assignment~\cite{Aaij:2016utb}.
Very recently, the LHCb collaboration observed
a new excited $D_s$ meson $D_{s0}(2590)^+$ with $J^P=0^-$ in $B^0 \rightarrow D^- D^+ K^+ \pi^-$
decays~\cite{LHCb:2020gnv}. More experimental information about the excited charmed and
charmed-strange mesons is collected in Table~\ref{EXP DandDs}.

The experimental progress provides us good opportunities to establish an abundant $D$ and $D_s$-meson spectrum
up to the higher orbital and radial excitations. In theory, to understand the nature of the charmed and charmed-strange mesons,
especially the newly observed states, and to establish the
charmed and charmed-strange meson spectra, in the recent years a lot of studies have been
carried out within various phenomenological models from several
aspects, such as the mass spectrum~\cite{Ebert:2009ua,Liu:2013maa,Liu:2015lka,Liu:2015uya,Liu:2016efm,Badalian:2011tb,Allosh:2021biq,
Patel:2021aas,Chen:2018nnr,Zhou:2014ytp,Chen:2009zt,Zhang:2009nu,Godfrey:2015dva,Song:2015nia,
Song:2015fha,Gandhi:2019lta,Sun:2013qca,Ferretti:2015rsa,Segovia:2015dia,Li:2010vx,Lu:2014zua,Kher:2017wsq,Eshraim:2014eka,Shah:2014caa,Yu:2020khh}, strong decays~\cite{Godfrey:2015dva,Song:2015nia,Song:2015fha,Gandhi:2019lta,Sun:2013qca,Ferretti:2015rsa,Segovia:2015dia,Li:2010vx,
Lu:2014zua,Kher:2017wsq,Eshraim:2014eka,Shah:2014caa, Zhong:2008kd,Xiao:2014ura,Zhong:2009sk,Zhong:2010vq,Sun:2009tg,Sun:2010pg,Wang:2021orp,Song:2014mha,Colangelo:2012xi,Colangelo:2010te,Yu:2016mez,
Wang:2010ydc,Wang:2013tka,Yu:2014dda,Wang:2014jua,Wang:2016hkf,Chen:2015lpa,Chen:2011rr,Zhao:2016mxc,Tian:2017okw,Li:2017sww,
Li:2009qu,Godfrey:2013aaa,Godfrey:2014fga,Gandhi:2020vap,Gupta:2018zlg,
Wang:2016enc,Zhang:2016dom,Li:2017zng,Tan:2018lao,Wang:2018psi,Yu:2020khh}, etc., some previous works can be found in Refs.~\cite{Godfrey:1985xj,Zeng:1994vj,Gupta:1994mw,Lahde:1999ih,DiPierro:2001dwf,Godfrey:2005ww,
Close:2005se,Close:2006gr,Vijande:2006hj,Zhang:2006yj,Li:2007px,Wei:2006wa}.
Furthermore, spectroscopic calculations on the lattice are making steady
progress~\cite{Mohler:2011ke,Moir:2013ub,Kalinowski:2015bwa,Cichy:2016bci}.
In the $D$-meson family, the $D_0(2300)$, $D_1(2420)$, $D_1(2430)$ and $D_2(2460)$
resonances listed in RPP~\cite{Zyla:2020zbs} are usually considered to be the $1P$ states,
apart from a few disputes about $D_0(2300)$ and $D_1(2430)$ in recent works~\cite{Gayer:2021xzv,Albaladejo:2016lbb,Du:2017zvv,Du:2020pui}.
The newly observed resonance $D_0(2550)$ may be classified as the radially excited ($2S$) state $D(2^1S_0)$~\cite{Kher:2017wsq,Badalian:2011tb,Liu:2015uya,Yu:2014dda,Chen:2011rr,Gupta:2018zlg,
Wang:2010ydc,Wang:2013tka,Song:2015fha,Godfrey:2015dva,Lu:2014zua}, although the width is
underestimated in some works~\cite{Li:2010vx,Ferretti:2015rsa,Zhong:2010vq,DiPierro:2001dwf,Sun:2010pg}.
The $D_1^*(2600)$ may favor the the $2S$ state $D(2^3S_1)$~\cite{Ferretti:2015rsa,Kher:2017wsq,Badalian:2011tb,Liu:2016efm,Liu:2015uya,Ebert:2009ua,
Gupta:2018zlg,Yu:2014dda,Chen:2015lpa,Wang:2010ydc,Wang:2013tka,Song:2015fha,Godfrey:2015dva,Lu:2014zua},
or a mixture via the $2^3S_1-1^3D_1$ mixing~\cite{Li:2017sww,Chen:2011rr,Yu:2020khh,Zhong:2010vq,Xiao:2014ura,Sun:2010pg,Li:2010vx,Chen:2015lpa}.
The $D^*_3(2750)$ resonance can be assigned to the $1D$ state $D(1^3D_3)$,
while $D_2(2740)$ may correspond to a $1D$ mixed state with $J^P=2^-$~\cite{Badalian:2011tb,Liu:2016efm,Liu:2015uya,Liu:2015lka,Zhong:2010vq,Xiao:2014ura,Wang:2010ydc,Wang:2013tka,
Song:2015fha,Godfrey:2015dva,Lu:2014zua,Wang:2016enc,Li:2010vx,Gandhi:2019lta,Kher:2017wsq,Yu:2016mez,Yu:2014dda,Gupta:2018zlg}. The $D^*_1(2760)^0$ resonance is a good
candidate for the $1D$ state $D(1^3D_1)$~\cite{Godfrey:2015dva}, while a few components of
$D(2^3S_1)$ may exist via the $2^3S_1-1^3D_1$ mixing. The quark model classification of $D_J(3000)$ and
$D_J^*(3000)$ is still controversial in the literature.
The unnatural parity state $D_J(3000)$ is explained with the $3S$ state
$D(3^1S_0)$~\cite{Chen:2018nnr,Godfrey:2015dva,Lu:2014zua,Song:2015fha}, $2P$ states with $J^P=1^+$~\cite{Yu:2014dda,Xiao:2014ura,Sun:2013qca,Gupta:2018zlg,Li:2017zng}, or $1F$ states
with $J^P=3^+$~\cite{Liu:2016efm,Liu:2015lka,Liu:2015uya};
while the natural parity state $D_J^*(3000)$ is explained with
$D(3^3S_1)$~\cite{Chen:2018nnr,Song:2015fha}, $D(2^3P_0)$~\cite{Gupta:2018zlg,Sun:2013qca},
$D(2^3P_2)$~\cite{Gandhi:2019lta,Liu:2013maa,Liu:2015uya,Liu:2016efm}, $D(1^3F_4)$~\cite{Yu:2014dda,Godfrey:2015dva,Xiao:2014ura},
or $D(1^3F_2)$~\cite{Yu:2014dda}, and so on.

On the other hand, in the $D_s$-meson family, it is not controversial to
classify the $D_{s1}(2536)$ and $D_{s2}(2573)$ resonances as $1P$ states $D_s(1P_1')$ (high mass mixed state)
and $D_s(1^3P_2)$, respectively, however, the other two $1P$ states
$D_s(1^3P_0)$ and $D_s(1P_1)$ (low mass mixed state) classified in the quark model
are not well established. Considering the positive parity resonances $D_{s0}(2317)$
and $D_{s1}(2460)$ first reported by BaBar~\cite{BaBar:2003oey} and CLEO~\cite{CLEO:2003ggt} as the $D_s(1^3P_0)$ and  $D_s(1P_1)$ assignments,
one finds their masses are too low to be comparable with the theoretical expectations.
Some studies suggest that $D_{s0}(2317)$ and $D_{s1}(2460)$ are the mixtures of
bare $1P$ $c\bar{s}$ core and $D^{(*)}K$ component~\cite{Ortega:2016mms,Yang:2021tvc}.
The new resonance $D_{s0}(2590)^+$ with $J^P=0^-$ is suggested to be a strong candidate of the
radial excitation $D_s(2^1S_0)$ by the collaboration~\cite{LHCb:2020gnv}, however,
its measured mass and width are inconsistent with the recent theoretical predictions
in Ref.~\cite{Wang:2021orp}. The $D_{s1}^*(2700)^+$ and $D_{s1}^*(2860)^+$ resonances may be identified
as the $D_s(2^3S_1)$ and $D_s(1^3D_1)$, respectively~\cite{Ferretti:2015rsa,Badalian:2011tb,Liu:2016efm,Liu:2015uya,Ebert:2009ua,
Wang:2014jua,Zhou:2014ytp,Song:2014mha,Segovia:2015dia,Chen:2009zt,Zhang:2009nu,Godfrey:2015dva},
or their mixtures~\cite{Li:2017sww,Li:2010vx,Li:2007px,Close:2006gr,Zhong:2009sk,Chen:2011rr,Song:2015nia}.
The $D_{s3}^*(2860)^+$ resonance can be classified as the $1D$ state $D_s(1^3D_3)$~\cite{Liu:2015uya,Ferretti:2015rsa,Kher:2017wsq,Chen:2009zt,Badalian:2011tb,Liu:2016efm,Liu:2015lka,
Wang:2014jua,Godfrey:2014fga,Zhou:2014ytp,Wang:2016hkf,Wang:2016enc,Song:2014mha,Segovia:2015dia,Godfrey:2015dva,Xiao:2014ura}.
It should be mentioned that there still exist puzzles for the $D_{sJ}$ structures around 2.86 GeV,
it may be contributed by all of the $1D$-wave states with $J^P=1^-,2^-,3^-$
~\cite{Segovia:2015dia,Gandhi:2020vap,Zhong:2009sk}.
The higher resonance $D_{sJ}(3040)^+$ may be a candidate
for the $2P$ states with $J^P=1^+$~\cite{Ferretti:2015rsa,Badalian:2011tb,Kher:2017wsq,Segovia:2015dia,Chen:2009zt,Liu:2016efm,Liu:2015uya,Ebert:2009ua,Li:2010vx,Song:2015nia,
Sun:2009tg,Zhong:2009sk,Li:2017zng,Godfrey:2015dva,Xiao:2014ura}.
More information about the status of the charmed and charmed-strange meson
study can be found in the recent review work~\cite{Chen:2016spr}.

The recent LHCb experiments~\cite{LHCb:2013jjb,LHCb:2015eqv,Aaij:2016fma,Aaij:2019sqk,Aaij:2015sqa,
LHCb:2014ott,LHCb:2014ioa,LHCb:2012uts,Aaij:2016utb,LHCb:2020gnv} have demonstrated the capability of both discovering the $D$ and $D_s$ mesons and
determining their properties. Thus, more and more progress in the observations of the excited $D$- and
$D_s$-meson states will be achieved in forthcoming LHCb experiments. Stimulated by the
recent progress in experiments, we have systematically analyzed the strong
decay properties of the excited $D$- and $D_s$-meson states within a chiral quark model in
Refs.~\cite{Zhong:2008kd,Zhong:2009sk,Zhong:2010vq,Xiao:2014ura}, where the wave functions
for the excited meson states are adopted the simple harmonic oscillator (SHO) forms,
while their masses are referred to the quark model predictions in the literature.
To deepen our study and more reliably understand the $D$ and $D_s$-meson spectrum, in this work we
carry out a combined analysis of both mass spectrum and strong decays.
First, we calculate the mass spectrum within a semi-relativistic potential model,
where the relativistic effects from the light quarks can be reasonably included.
With this model the masses for the observed $D$ and $D_s$-meson states can be described successfully.
Then, by using the available wave functions and masses from the potential model,
we calculate the OZI-allowed two-body strong decays of the excited
$D$ and $D_s$ mesons with the chiral quark model.
This model has been successfully applied to describe the strong decays of the heavy-light mesons and
baryons~\cite{li:2021hss,Xiao:2020oif,Wang:2019uaj,Wang:2018fjm,Xiao:2020gjo,Wang:2020gkn,Xiao:2018pwe,Xiao:2014ura,Zhong:2010vq,Zhong:2008kd,
Zhong:2009sk,Liu:2012sj,Zhong:2007gp,Xiao:2013xi,Nagahiro:2016nsx,Yao:2018jmc,
Wang:2017kfr,Xiao:2017udy,Wang:2017hej,Liu:2019wdr}.
Based on our good descriptions of the mass and decay properties for
the well-established states, we give our quark model classifications of the newly
observed resonances/structures. Finally, according to our assignments for the newly observed resonances,
we attempt to predict the properties of the missing resonances, which may
be useful for future investigations in experiments.

 \begin{table*}
\begin{center}
\caption{The newly observed excited charmed and charmed-strange meson states in recent 15 years.
The ``N" and ``UN" stand for the natural parity and unnatural parity, respectively.}
\label{EXP DandDs}
\scalebox{1.05}{
\begin{tabular}{clccccccc}
\hline\hline
&Resonance &~~~~~~Mass~(MeV)~~~ &~~~~~~Width~(MeV)~~~ &~~~~~~Observed channel~~~ &~~~~~~$J^P$~~~ &~~~~~~Ref.~~~ &~~Time~~~  \tabularnewline
\hline
&~~~$D(2550)^0$ &~~~$2539.4\pm 11.3$ &~~~$130\pm25$&~~~$D^{*+}\pi^-$~~&~~~UN &~~~BaBar~\cite{BaBar:2010zpy} &~~~~~2010~~~  \tabularnewline
&~~~$D_J (2580)^0$ &~~~$2579.5\pm 8.9$ &~~~$177.5\pm63.8$&~~~$D^{*+}\pi^-$~~&~~~UN &~~~LHCb~\cite{LHCb:2013jjb} &~~~~~2013~~~  \tabularnewline
&~~~$D_0 (2550)^0$ &~~~$2518\pm 9$ &~~~$199\pm 22$&~~~$D^{*+}\pi^-$~~&~~~$0^-$ &~~~LHCb~\cite{Aaij:2019sqk} &~~~~~2019~~~  \tabularnewline
\hline
&~~~$D^*(2600)^0$ &~~~$2608.7\pm 4.9$ &~~~$93 \pm 19$&~~~$D^+\pi^-$,$D^{*+}\pi^-$~~&~~~N &~~~BaBar~\cite{BaBar:2010zpy} &~~~~~2010~~~  \tabularnewline
&~~~$D^*(2600)^+$ &~~~$2621.3\pm 7.9$ &~~~$93.0$&~~~$D^0\pi^+$~~&~~~N &~~~BaBar~\cite{BaBar:2010zpy} &~~~~~2010~~~  \tabularnewline
&~~~$D_J^*(2650)^0$ &~~~$2649.2\pm 7.0$ &~~~$140.2\pm35.7$&~~~$D^{*+}\pi^-$~~&~~~N &~~~LHCb~\cite{LHCb:2013jjb} &~~~~~2013~~~  \tabularnewline
&~~~$D_1^*(2680)^0$ &~~~$2681.1\pm 23.6$ &~~~$186.7\pm25.3$&~~~$D^+\pi^-$~~&~~~$1^{-}$ &~~~LHCb~\cite{Aaij:2016fma} &~~~~~2016~~~  \tabularnewline
&~~~$D_1^*(2600)^0$ &~~~$2641.9\pm 6.3$ &~~~$149\pm24$&~~~$D^{*+}\pi^-$~~&~~~$1^{-}$ &~~~LHCb~\cite{Aaij:2019sqk} &~~~~~2019~~~  \tabularnewline
\hline
&~~~$D(2750)^0$ &~~~$2752.4\pm 4.4$ &~~~$71\pm 17$&~~~$D^{*+}\pi^-$~~ &~~~UN &~~~BaBar~\cite{BaBar:2010zpy} &~~~~~2010~~~  \tabularnewline
&~~~$D_J(2740)^0$ &~~~$2737.0\pm 14.7$ &~~~$73.2\pm 38.4$&~~~$D^{*+}\pi^-$~~ &~~~UN &~~~LHCb~\cite{LHCb:2013jjb} &~~~~~2013~~~  \tabularnewline
&~~~$D_2(2740)^0$ &~~~$2751\pm 10$ &~~~$102\pm 32$&~~~$D^{*+}\pi^-$~~ &~~~$2^{-}$ &~~~LHCb~\cite{Aaij:2019sqk} &~~~~~2019~~~  \tabularnewline
\hline
&~~~$D^*(2760)^0$ &~~~$2763.3\pm 4.6$ &~~~$60.9\pm8.7$&~~~$D^{+}\pi^-$~~ &~~~? &~~~BaBar~\cite{BaBar:2010zpy} &~~~~~2010~~~  \tabularnewline
&~~~$D^*(2760)^+$ &~~~$2769.7\pm 5.3$ &~~~$60.9      $ &~~~$D^{0}\pi^+$~~&~~~? &~~~BaBar~\cite{BaBar:2010zpy} &~~~~~2010~~~  \tabularnewline
&~~~$D_J^*(2760)^0$ &~~~$2760.1\pm 4.8$ &~~~$74.4 \pm22.5$&~~~$D^{+}\pi^-$,$D^{*+}\pi^-$~~ &~~~N &~~~LHCb\cite{LHCb:2013jjb} &~~~~~2013~~~  \tabularnewline
&~~~$D_J^*(2760)^+$ &~~~$2771.7\pm 5.5$ &~~~$66.7 \pm17.1$&~~~$D^{0}\pi^+$~~ &~~~N &~~~LHCb~\cite{LHCb:2013jjb} &~~~~~2013~~~  \tabularnewline
&~~~$D_3^*(2760)^-$ &~~~$2798\pm 15$ &~~~$105 \pm 47$&~~~$\bar{D}^{0}\pi^-$~~ &~~~$3^{-}$ &~~~LHCb~\cite{Aaij:2015sqa} &~~~~~2015~~~  \tabularnewline
&~~~$D_3^*(2760)^0$ &~~~$2775.5\pm 13.7$ &~~~$95.3 \pm50.6$&~~~$D^{+}\pi^-$~~ &~~~$3^{-}$ &~~~LHCb~\cite{Aaij:2016fma} &~~~~~2016~~~  \tabularnewline
&~~~$D_3^*(2750)^0$ &~~~$2753\pm 10$ &~~~$66\pm24$&~~~$D^{*+}\pi^-$~~ &~~~$3^{-}$ &~~~LHCb~\cite{Aaij:2019sqk}&~~~~~2019~~~  \tabularnewline
&~~~$D_1^*(2760)^0$ &~~~$2781\pm 35$ &~~~$177\pm 59$&~~~$D^{+}\pi^-$~~ &~~~$1^{-}$ &~~~LHCb~\cite{LHCb:2015eqv}&~~~~~2015~~~  \tabularnewline
\hline
&~~~$D_J(3000)^0$ &~~~$2971.8\pm 8.7$ &~~~$188.1\pm44.8$&~~~$D^{*+}\pi^-$~~ &~~~UN &~~~LHCb~\cite{LHCb:2013jjb} &~~~~~2013~~~  \tabularnewline
\hline
&~~~$D_J^*(3000)^0$ &~~~$3008.0\pm4.0$ &~~~$110.5\pm11.5$&~~~$D^{+}\pi^-$~~ &~~~? &~~~LHCb~\cite{LHCb:2013jjb} &~~~~~2013~~~  \tabularnewline
&~~~$D_J^*(3000)^+$ &~~~$3008.1(fixed)$ &~~~$110.5(fixed)$&~~~$D^{0}\pi^+$~~ &~~~? &~~~LHCb~\cite{LHCb:2013jjb} &~~~~~2013~~~  \tabularnewline
&~~~$D_2^*(3000)^0$ &~~~$3214\pm 98$ &~~~$186 \pm135$&~~~$D^{+}\pi^-$~~  &~~~$2^+$ &~~~LHCb~\cite{Aaij:2016fma} &~~~~~2016~~~  \tabularnewline
\hline\hline
&~~~$D_{s0}(2590)^+$ &~~~$2591\pm13$ &~~~$89\pm28$ &~~~$D^-K^+$~~&~~~$0^-$ &~~~LHCb~\cite{LHCb:2020gnv} &~~~~~2020~~~  \tabularnewline
&~~~$D_{sJ}(2700)^+$ &~~~$2688\pm7$ &~~~$112\pm 43$&~~~$DK$~~&~~~? &~~~BaBar~\cite{BaBar:2006gme} &~~~~~2006~~~  \tabularnewline
&~~~$D_{s1}^*(2700)^+$ &~~~$2708_{-19}^{+20}$ &~~~$108_{-54}^{+59}$&~~~$D^0K^+$~~&~~~$1^-$ &~~~Belle~\cite{Belle:2007hht} &~~~~~2007~~~  \tabularnewline
&~~~$D_{s1}^*(2700)^+$ &~~~$2710_{-9}^{+14}$ &~~~$149_{-59}^{+46}$ &~~~$DK,D^{*}K$~~&~~~N &~~~BaBar~\cite{BaBar:2009rro} &~~~~~2009~~~  \tabularnewline
&~~~$D_{s1}^*(2700)^+$ &~~~$2709.2\pm6.4$ &~~~$115.8\pm19.4$&~~~$DK$~~&~~~? &~~~LHCb~\cite{LHCb:2012uts} &~~~~~2012~~~  \tabularnewline
&~~~$D_{s1}^*(2700)^+$ &~~~$2699^{+14}_{-7}$ &~~~$127^{+24}_{-19}$&~~~$D^0K^+$~~ &~~~$1^-$ &~~~BaBar~\cite{Lees:2014abp} &~~~~~2014~~~  \tabularnewline
&~~~$D_{s1}^*(2700)^+$ &~~~$2732.3\pm 10.1$ &~~~$136\pm 43$&~~~$D^{*0}K^+$~~ &~~~$1^-$ &~~~LHCb~\cite{Aaij:2016utb} &~~~~~2016~~~  \tabularnewline
\hline
&~~~$D_{sJ}(2860)^+$ &~~~$2856.6\pm6.5$ &~~~$47\pm 17$&~~~$DK$~~&~~~? &~~~BaBar~\cite{BaBar:2006gme} &~~~~~2006~~~  \tabularnewline
&~~~$D_{sJ}^*(2860)^+$ &~~~$2862_{-4}^{+7}$ &~~~$48\pm9$&~~~$DK,D^{*}K$~~ &~~~~N &~~~BaBar~\cite{BaBar:2009rro} &~~~~~2009~~~  \tabularnewline
&~~~$D_{sJ}^*(2860)^+$ &~~~$2866.1\pm7.3$ &~~~$69.9\pm9.8$&~~~$DK$~~ &~~~~? &~~~LHCb~\cite{LHCb:2012uts} &~~~~~2012~~~  \tabularnewline
&~~~$D_{s1}^*(2860)^+$ &~~~$2859.0\pm27.0$ &~~~$159\pm80$&~~~$...$~~ &~~~~$1^-$ &~~~LHCb~\cite{LHCb:2014ott,LHCb:2014ioa} &~~~~~2014~~~  \tabularnewline
&~~~$D_{s3}^*(2860)^+$ &~~~$2860.5\pm7.0$ &~~~$53.0\pm10.0$&~~~$...$~~ &~~~~$3^-$ &~~~LHCb~\cite{LHCb:2014ott,LHCb:2014ioa} &~~~~~2014~~~  \tabularnewline
&~~~$D_{s3}^*(2860)^+$ &~~~$2867.1\pm6.2$ &~~~$50\pm24$&~~~$D^{*0}K^+$~~ &~~~~$3^-$ &~~~LHCb~\cite{Aaij:2016utb} &~~~~~2016~~~  \tabularnewline
\hline
&~~~$D_{sJ}(3040)^+$ &~~~$3044^{+31}_{-9}$ &~~~$239.0\pm60.0$&~~~$D^{*0}K^+$~~ &~~~~? &~~~BaBar~\cite{BaBar:2009rro} &~~~~~2009~~~  \tabularnewline
\hline\hline
\end{tabular}}
\end{center}
\end{table*}

This work is organized as follows. In Sec.~\ref{spectrum}, the mass spectra for the charmed and charmed-strange mesons are calculated within a semi-relativistic quark model. In Sec.~\ref{STRONG DECAY}, the strong decays are estimated within the chiral quark model.
In Sec.~\ref{discussion}, some discussions based on our numerical results
are carried out. Finally, a summary is given in Sec.~\ref{Summary}.

\section{mass spectrum}\label{spectrum}

\subsection{model}

In Ref.~\cite{li:2021hss}, we adopt a nonrelativistic linear potential model to calculate the $B$- and $B_s$-meson
mass spectrum. It is found that the masses for the $B$- and $B_s$-meson states can be
reasonably described within the nonrelativistic quark
model~\cite{li:2021hss}, however, the effective harmonic oscillator parameters, $\beta_{eff}$, which are obtained
by equating the root-mean-square radius of the harmonic oscillator wavefunction for the
specified $(n,l)$ quantum numbers to the root-mean-square radius of
the wavefunctions, are notably smaller than those from the relativized quark model~\cite{Godfrey:2015dva}.
To consistently include the relativistic effects on the wavefunctions, the nonrelativistic
Hamiltonian $H=\mathbf{p}^2/(2\mu)+m_1+m_2+V(r)$ is replaced with the relativistic one
\begin{equation}
 H=\sqrt{\mathbf{p}^2+m_1^2}+\sqrt{\mathbf{p}^2+m_2^2}+V(r),
\end{equation}
where $\mathbf{p}=\mathbf{p}_1=-\mathbf{p}_2$ is the quark momentum in the center-of-mass system,
$r$ is the distance between two quarks; $m_1$ and $m_2$ are the masses of light and
heavy quarks, respectively; and the reduced mass $\mu=m_1m_2/(m_1+m_2)$.

The effective potential
$V(r)$ includes the spin-independent part $V_0(r)$ and spin-dependent part $V_{sd}(r)$. The
spin-independent part $V_0(r)$ is adopted the standard Cornell form~\cite{Eichten:1978tg}
\begin{eqnarray}\label{H0}
V_0(r)=-\frac{4}{3}\frac{\alpha_s(r)}{r}+br+C_{0},
\end{eqnarray}
which includes the color Coulomb interaction and linear confinement, and zero point energy $C_0$.
The spin-dependent part $V_{sd}(r)$ is adopted the widely used form~\cite{Godfrey:1985xj,Swanson:2005,Godfrey:2004ya}
\begin{eqnarray}\label{H0}
V_{sd}(r)=H_{SS}+H_{T}+H_{LS},
\end{eqnarray}
where
\begin{eqnarray}\label{ss}
H_{SS}= \frac{32\pi \alpha_{s}(r) \cdot \sigma^3 e^{-\sigma^2 r^2}}{9 \sqrt{\pi}  \tilde{m}_1m_2} \mathbf{S_1} \cdot \mathbf{S_2}
\end{eqnarray}
is the spin-spin contact hyperfine potential. The tensor potential $H_T$ is adopted as
\begin{eqnarray}\label{t}
H_{T}= \frac{4}{3}\frac{\alpha_s(r)}{\tilde{m}_1 m_2}\frac{1}{r^3}\left(\frac{3\mathbf{S}_{1}\cdot \mathbf{r}\mathbf{S}_{2}\cdot \mathbf{r}}{r^2}-\mathbf{S}_{1}\cdot\mathbf{S}_{2}\right).
\end{eqnarray}
The spin-orbit interaction $H_{LS}$ can be decomposed
into symmetric part $H_{sym}$ and antisymmetric part
$H_{anti}$:
\begin{eqnarray}\label{vs}
H_{LS}=H_{sym}+H_{anti},
\end{eqnarray}
with
\begin{eqnarray}\label{vs}
H_{sym}=\frac{\mathbf{S_{+}\cdot L}}{2}\left[ \left(\frac{1}{2  \tilde{m}_1^2}+\frac{1}{2m_2^2} \right) \left( \frac{4 \alpha_s(r)}{3r^3}-\frac{b}{r}   \right)+\frac{8 \alpha_s(r)}{3  \tilde{m}_1 m_2r^3} \right],\\
H_{anti}=\frac{\mathbf{S_{-}\cdot L}}{2}\left(\frac{1}{2  \tilde{m}_1^2}-\frac{1}{2m_2^2} \right) \left( \frac{4 \alpha_s(r)}{3r^3}-\frac{b}{r}   \right) .\ \ \ \ \ \ \ \ \ \ \ \ \ \ \ \ \ \ \ \ \ \ \
\end{eqnarray}
In these equations, $\mathbf{L}$ is the relative orbital angular momentum of the $q\bar{q}$
system; $\mathbf{S}_1$ and $\mathbf{S}_{2}$ are the spins of the light and heavy quarks, respectively, and $\mathbf{S}_{\pm}\equiv\mathbf{S}_1\pm \mathbf{S}_{2}$. The running coupling constant $\alpha_s(r)$ in
the coordinate space is adopted a parameterized form as suggested in
Ref.~\cite{Godfrey:1985xj}
\begin{equation}
\alpha_s(r) =\sum_{i} \alpha_i \frac{2}{\sqrt{\pi}}\int_0^{\gamma_i r}e^{-x^2}dx.
\end{equation}
The parameters $\alpha_i$ and $\gamma_i$ are free parameters which can be fitted
to make the behavior of the running coupling constant at short distance be consistent with the coupling constant
in momentum space predicted by QCD. In this work we take $\alpha_1=0.30$, $\alpha_2=0.15$, $\alpha_3=0.20$, $\gamma_1=\frac{1}{2}$, $\gamma_2=\frac{\sqrt{10}}{2}$, $\gamma_3=\frac{\sqrt{1000}}{2}$, which are the same as those adopted in
Refs.~\cite{Godfrey:1985xj}, except that the parameter $\alpha_1$ is slightly
adjusted to better describe the mass spectrum. It should be mentioned that in the spin-dependent potentials we have
replaced the light quark mass $m_1$ with $\tilde{m}_1$ to include some relativistic corrections
to the potentials as suggested in Ref.~\cite{Liu:2013maa}.
The parameter set \{ $b$, $\sigma$, $m_1$, $\tilde{m}_1$, $m_2$, $C_0$\}
in the above potentials is determined by fitting the mass spectrum.

For the heavy-light meson system, the antisymmetric
part of the spin-orbit potential, $H_{anti}$, can cause a configuration mixing between
spin triplet $n^{3}L_{J}$ and spin singlet $n^{1}L_{J}$ defined in the $L$-$S$ coupling scheme.
Thus, the physical states $nL_J$ and $nL'_J$ are expressed as
\begin{equation}\label{mixsta}
\left(
  \begin{array}{c}
   nL_J\\
   nL'_J\\
  \end{array}\right)=
  \left(
  \begin{array}{cc}
   \cos\theta_{nL} &\sin\theta_{nL}\\
  -\sin\theta_{nL} &\cos\theta_{nL}\\
  \end{array}
\right)
\left(
  \begin{array}{c}
  n^{1}L_{J}\\
  n^{3}L_{J}\\
  \end{array}\right).
\end{equation}
where $J=L=1,2,3\cdots$, and the $\theta_{nL}$ is the mixing angle.
In this work $nL'_J$ corresponds to the higher mass mixed state
as often adopted in the literature. The mixing angle $\theta_{nL}$ is perturbatively
determined with the non-diagonal matrix
element $\langle n^{1}L_{J} |H_{anti}| n^{3}L_{J}\rangle$.
It should be mentioned that the coupled-channel effects can cause
a configuration mixing as well, we neglect these effects on the the
mixing angle in our calculations.

\subsection{numerical method}

In this work we use the Gaussian expansion method~\cite{Hiyama:2003cu} to solve the radial Schrodinger equation for
a meson system with quantum numbers $LM$ of the orbital angular momentum and
its $z$ component,
\begin{equation}
\begin{split}
(H-E) \psi_{LM}(r)=0  \label{eqH}.
\end{split}
\end{equation}
The spatial wave function $\psi_{LM}(r)$ is expanded with a set of Gaussian basis functions,
\begin{equation}
\psi_{LM}(r)=\sum_{m=1}^{m_{max}}C_m \phi_{mL}(r) Y_{LM}(\hat{\mathbf{r}})\label{EUQ}.
\end{equation}
The Gaussian function $\phi_{mL}(r)$ with given range parameters is writhen as
\begin{equation}
\phi_{mL}(r)= \left(\frac{2^{L+2} (2v_m)^{L+\frac{3}{2}}}{\sqrt{\pi} (2L+1)!!}  \right)^{\frac{1}{2}}   r^L e^{-v_m r^2}.
\end{equation}
Transforming $\phi_{mL}(r)$ to the momentum space, one has
\begin{equation}
\phi_{mL}(p)=(-i)^L \left(  \frac{\sqrt{2} v_m^{-\frac{3}{2}}}{\sqrt{\pi}(2L+1)!!  } \right)^{\frac{1}{2}}\left( \frac{p}{\sqrt{v_m}}  \right)^L e^{-\frac{p^2}{4v_m}}.
\end{equation}
The size parameters $v_m$ are set to be a geometric progression form~\cite{Hiyama:2003cu}
\begin{equation}
\begin{split}
 v_m=&\frac{1}{r_m^2}, \\
 r_m=&r_1 a^{m-1} \ \ (m=1-m_{max}).
\end{split}
\end{equation}
There are three parameters $ \{r_1,a,m_{max}\}$. These parameters, the eigenenergy $E$, and the expansion
coefficients $\{C_m\}$ can be determined with the Rayleigh-Ritz
variational principle by solving the generalized eigenvalue problem
\begin{equation}
\begin{split}
  \sum_{m'} (H_{m m'}-E N_{m m'}) C_{m'}=0 \ \ (m=1-m_{max}),
\end{split}
\end{equation}
where $H_{mm'}=\langle \phi_{mL}|H |\phi_{m'L} \rangle$ and $N_{nn'}=\langle \phi_{mL}|1 |\phi_{m'L} \rangle$.

\subsection{parameters}

The model parameters adopted in this work are listed in Table~\ref{parameters}.
The parameter set \{$b$, $\sigma$, $m_1$, $\tilde{m}_1$, $m_2$, $C_0$, $\alpha_1$\}
for the $D$-meson spectrum is determined by fitting the masses of the well established
states $D(1865)^0$, $D^*(2007)^0$, $D_1(2420)^0$, $D_1(2430)^0$, $D_2(2460)^0$ and $D_3(2750)^0$.
While for the $D_s$-meson sector, the parameter set is determined by fitting the masses of the well established
states $D_s(1969)$, $D^*_s(2112)$, $D_{s1}(2536)$ and $D_{s2}(2573)$ together with the newly observed
state $D^*_{s3}(2860)$.
In the present work, the slope parameter $b$ and the running coupling constant
$\alpha_1$ for the $D$-meson spectrum are set to be the same as those for the $D_s$-meson spectrum,
considering that they may be independent on a specific quark flavor.
It should be pointed out that the zero-point-energy parameter $C_0$ is taken to be zero
for the $c\bar{c}$, $b\bar{b}$, $b\bar{c}$ heavy quarkonium systems in the literature
~\cite{Deng:2016ktl,Deng:2016stx,Li:2019tbn,Li:2019qsg}.
For these heavy quarkonium systems, the zero point energy can be absorbed into
the constituent quark masses because it only affects the heavy quark masses slightly.
However, if the zero point energy is absorbed into the meson systems containing
light quarks, it can significantly change the light constituent quark masses,
which play an important role in the spin-dependent potentials.
Thus, to obtain a good description of both the masses and the hyperfine/fine splittings
for the meson systems containing light quarks, a zero-point-energy parameter
$C_0$ is usually adopted in the calculations. The slope parameter of the linear
potential $b=0.18$ GeV$^2$ determined in the present work is consistent with that of the relativized
quark model~\cite{Godfrey:1985xj}, while is slightly larger than $b\simeq 0.12-0.14$ GeV$^2$ adopted in
the non-relativistic quark model~\cite{Deng:2016ktl,Deng:2016stx,Li:2019tbn,Li:2019qsg}.

It should be mentioned that we cannot obtained stable solutions for some states due to
the singular behavior of $1/r^3$ in the spin-dependent potentials. To overcome the singular
behavior, following the method of our previous works~\cite{Deng:2016ktl,Deng:2016stx,Li:2019tbn,
Li:2019qsg,Li:2020xzs}, we introduce a cutoff distance
$r_c$ in the calculation. Within a small range $r\in (0,r_c)$, we let $1/r^3=1/r_c^3$.
By introducing the cutoff distance $r_c$, we can nonperturbatively include the corrections from these
spin-dependent potentials containing $1/r^3$ to both
the mass and wave function of a meson state, which are crucial
for our predicting the decay properties. It is found that the mass of the $1^3P_0$ state
is more sensitive to the cutoff distance $r_c$ due to its relatively larger factor
$\langle\mathbf{S_{+}\cdot L}\rangle$ than the other excited meson states. Thus,
the cutoff parameters $r_c$ for the $D$- and $D_s$-meson spectra are determined
by fitting the masses of the $D(1^3P_0)$ and $D_s(1^3P_0)$. Note that when the other parameters
are well determined, the masses of these $1^3P_0$ states can be
reliably worked out with the perturbation method~\cite{li:2021hss} without introducing the cutoff
distance $r_c$, although the wave functions obtain no corrections from the
spin-dependent potentials containing $1/r^3$. We obtain the masses $M=2313$ and
$2409$ MeV for $D(1^3P_0)$ and $D_s(1^3P_0)$, respectively. These masses calculated with the perturbation
method are comparable with the predictions in Refs.~\cite{Zeng:1994vj,Lahde:1999ih}.
By fitting the masses $2313$ and $2409$ MeV of the $D(1^3P_0)$ and $D_s(1^3P_0)$ states obtained with the perturbation
method, we determine the cutoff distance parameters to be $r_c=0.327$ and $0.311$ fm for the
$D$- and $D_s$-meson spectra, respectively.

\begin{table}[htbp]
\begin{center}
\caption{Potential model parameters.}
\label{parameters}
\scalebox{1.0}{
\begin{tabular}{cccccccccc}
\hline
\hline
&\qquad&$m_c$~(GeV)&$m_{u,d}$~(GeV)&$\tilde{m}_{u,d}$~(GeV)&$m_s$~(GeV)&$\tilde{m}_s$~(GeV) \tabularnewline
\hline
&$D$      &$1.70$   &$0.40$       &$0.62$       &$...$              &$...$             \tabularnewline

&$D_s$    &$1.70$   &$... $       &$... $       &$0.50$              &$0.70$   \tabularnewline
\hline\hline
&\qquad&$\alpha_1$&$b~(\textrm{GeV}^2)$&$\sigma$~(GeV)&$C_0$~(MeV)&$r_c$~(fm) \tabularnewline
\hline
&$D$      &$0.30$   &$0.18$     &$1.02$     &$-493.0$  &$0.327$  \tabularnewline
&$D_s$    &$0.30$    &$0.18$    &$1.11$    &$-452.1$   &$0.311$ \tabularnewline
\hline
\hline
\end{tabular}}
\end{center}
\end{table}

\begin{table*}
\caption{Our predicted charmed meson masses (MeV) compared with the data and some other quark model predictions.
The mixing angles of the $D_L -D'_L$ states defined in Eq.(\ref{mixsta}) in this work are determined to be $\theta_{1P}=-34.0^{\circ}$, $\theta_{2P}=-23.5^{\circ}$, $\theta_{1D}=-40.2^{\circ}$, $\theta_{2D}=-40.2^{\circ}$, $\theta_{1F}=-41.0^{\circ}$.}
\label{D}
\scalebox{1.05}{
\begin{tabular}{clccccccc}
\hline\hline
~~~~State~~~~&$J^P~~~~$&~~~~Ours~~~~ &~~~~Exp~\cite{Zyla:2020zbs}~~~~&~~~~GM~\cite{Godfrey:2015dva}~~~~&~~~~EFG~\cite{Ebert:2009ua}~~~~&~~~~ZVR~\cite{Zeng:1994vj}~~~~&~~~~ LJM~\cite{Li:2010vx}~~~~&~~~~LNR~\cite{Lahde:1999ih}~~~~\tabularnewline
\hline
$~1^1S_0 $~ &$~0^-$~ &$~1865$~ &$~1865$~       &$~1877$~ &$~1871$~ &$~1850$~ &$~1867$~ &$~1874$~ \tabularnewline
$~1^3S_1 $~ &$~1^-$~ &$~2008$~ &$~2008$~       &$~2041$~ &$~2010$~ &$~2020$~ &$~2010$~ &$~2006$~ \tabularnewline
$~2^1S_0 $~ &$~0^-$~ &$~2547$~ &$~2564\pm20$~ &$~2581$~ &$~2581$~ &$~2500$~ &$~2555$~ &$~2540$~ \tabularnewline
$~2^3S_1 $~ &$~1^-$~ &$~2636$~ &$~2627\pm10$~ &$~2643$~ &$~2632$~ &$~2620$~ &$~2636$~ &$~2601$~ \tabularnewline
$~3^1S_0 $~ &$~0^-$~ &$~3029$~ &$~...$~        &$~3068$~ &$~3062$~ &$~2980$~ &$~...$~  &$~2904$~ \tabularnewline
$~3^3S_1 $~ &$~1^-$~ &$~3093$~ &$~...$~        &$~3110$~ &$~3096$~ &$~3070$~ &$~...$~  &$~2947$~ \tabularnewline
\hline
$~1^3P_0 $~ &$~0^+$~ &$~2313$~ &$~2349/2300$~  &$~2399$~ &$~2406$~ &$~2270$~ &$~2252$~  &$~2341$~ \tabularnewline
$~1P     $~ &$~1^+$~ &$~2424$~ &$~2412\pm 9$~ &$~2456$~ &$~2426$~ &$~2400$~ &$~2402$~  &$~2389$~ \tabularnewline
$~1P'    $~ &$~1^+$~ &$~2453$~ &$~2422$~ &$~2467$~ &$~2469$~ &$~2410$~ &$~2417$~  &$~2407$~ \tabularnewline
$~1^3P_2 $~ &$~2^+$~ &$~2475$~ &$~2461$~ &$~2502$~ &$~2460$~ &$~2460$~ &$~2466$~  &$~2477$~ \tabularnewline
$~2^3P_0 $~ &$~0^+$~ &$~2849$~ &$~...$~        &$~2931$~ &$~2919$~ &$~2780$~ &$~2752$~  &$~2758$~ \tabularnewline
$~2P     $~ &$~1^+$~ &$~2900$~ &$~...$~        &$~2924$~ &$~2932$~ &$~2890$~ &$~2886$~  &$~2792$~   \tabularnewline
$~2P'    $~ &$~1^+$~ &$~2936$~ &$~...$~        &$~2961$~ &$~3021$~ &$~2890$~ &$~2926$~  &$~2802$~ \tabularnewline
$~2^3P_2 $~ &$~2^+$~ &$~2955$~ &$~...$~        &$~2957$~ &$~3012$~ &$~2940$~ &$~2971$~  &$~2860$~ \tabularnewline
\hline
$~1^3D_1 $~ &$~1^-$~ &$~2754$~ &$~...$~        &$~2817$~ &$~2788$~ &$~2710$~ &$~2740$~  &$~2750$~ \tabularnewline
$~1D     $~ &$~2^-$~ &$~2755$~ &$~...$~        &$~2816$~ &$~2806$~ &$~2740$~ &$~2693$~  &$~2689$~   \tabularnewline
$~1D'    $~ &$~2^-$~ &$~2827$~ &$~2747\pm 6$~        &$~2845$~ &$~2850$~ &$~2760$~ &$~2789$~  &$~2727$~ \tabularnewline
$~1^3D_3 $~ &$~3^-$~ &$~2782$~ &$~2763.1\pm3.2$~ &$~2833$~ &$~2863$~ &$~2780$~ &$~2719$~  &$~2688$~ \tabularnewline
$~2^3D_1 $~ &$~1^-$~ &$~3143$~ &$~...$~        &$~3231$~ &$~3228$~ &$~3130$~ &$~3168$~  &$~3052$~ \tabularnewline
$~2D     $~ &$~2^-$~ &$~3168$~ &$~...$~        &$~3212$~ &$~3307$~ &$~3160$~ &$~3145$~  &$~2997$~ \tabularnewline
$~2D'    $~ &$~2^-$~ &$~3221$~ &$~...$~        &$~3248$~ &$~3359$~ &$~3170$~ &$~3215$~  &$~3029$~ \tabularnewline
$~2^3D_3 $~ &$~3^-$~ &$~3202$~ &$~...$~        &$~3226$~ &$~3335$~ &$~3190$~ &$~3170$~  &$~2999$~ \tabularnewline
\hline
$~1^3F_2 $~ &$~2^+$~ &$~3096$~ &$~...$~        &$~3132$~ &$~3090$~ &$~3000$~ &$~...$~   &$~...$~  \tabularnewline
$~1F     $~ &$~3^+$~ &$~3022$~ &$~...$~        &$~3108$~ &$~3129$~ &$~3010$~ &$~...$~   &$~...$~  \tabularnewline
$~1F'    $~ &$~3^+$~ &$~3129$~ &$~...$~        &$~3143$~ &$~3145$~ &$~3030$~ &$~...$~   &$~...$~  \tabularnewline
$~1^3F_4 $~ &$~4^+$~ &$~3034$~ &$~...$~        &$~3113$~ &$~3187$~ &$~3030$~ &$~...$~   &$~...$~  \tabularnewline
\hline\hline
\end{tabular}}
\end{table*}

\begin{table*}
\begin{center}
\caption{Our predicted charmed-strange meson masses (MeV) compared with the data and some other quark model predictions.
The mixing angles of the $D_{sL} -D'_{sL}$ states defined in Eq.(\ref{mixsta}) in this work are determined to be $\theta_{1P}=-36.8^{\circ}$, $\theta_{2P}=-21.0^{\circ}$, $\theta_{1D}=-40.7^{\circ}$, $\theta_{2D}=-41.3^{\circ}$, $\theta_{1F}=-40.7^{\circ}$.}
\label{Ds}
\scalebox{1.05}{
\begin{tabular}{clccccccc}
\hline\hline
~~~~State~~~~&$J^P~~~~$&~~~~Ours~~~~ &~~~~Exp~\cite{Zyla:2020zbs}~~~~&~~~~GM~\cite{Godfrey:2015dva}~~~~&~~~~EFG~\cite{Ebert:2009ua}~~~~&~~~~ZVR~\cite{Zeng:1994vj}~~~~&~~~~ LJM~\cite{Li:2010vx}~~~~&~~~~LNR~\cite{Lahde:1999ih}~~~~\tabularnewline
\hline
$~1^1S_0$~ &$~0^-$~ &$~1969$~ &$~1969$~        &$~1979$~ &$~1969$~ &$~1940$~ &$~1969$~ &$~1975$~ \tabularnewline
$~1^3S_1$~ &$~1^-$~ &$~2112$~ &$~2112$~        &$~2129$~ &$~2111$~ &$~2130$~ &$~2107$~ &$~2108$~ \tabularnewline
$~2^1S_0$~ &$~0^-$~ &$~2649$~ &$~...$~         &$~2673$~ &$~2688$~ &$~2610$~ &$~2640$~ &$~2659$~ \tabularnewline
$~2^3S_1$~ &$~1^-$~ &$~2737$~ &$~2714\pm5$~ &$~2732$~ &$~2731$~ &$~2730$~ &$~2714$~ &$~2722$~ \tabularnewline
$~3^1S_0$~ &$~0^-$~ &$~3126$~ &$~...$~         &$~3154$~ &$~3219$~ &$~3090$~ &$~...$~  &$~3044$~ \tabularnewline
$~3^3S_1$~ &$~1^-$~ &$~3191$~ &$~...$~         &$~3193$~ &$~3242$~ &$~3190$~ &$~...$~  &$~3087$~ \tabularnewline
\hline
$~1^3P_0$~ &$~0^+$~ &$~2409$~ &$~2317$~       &$~2484$~ &$~2509$~ &$~2380$~ &$~2344$~ &$~2455$~ \tabularnewline
$~1P    $~ &$~1^+$~ &$~2528$~ &$~2459$~       &$~2549$~ &$~2536$~ &$~2510$~ &$~2488$~ &$~2502$~ \tabularnewline
$~1P'   $~ &$~1^+$~ &$~2545$~ &$~2535$~        &$~2556$~ &$~2574$~ &$~2520$~ &$~2510$~ &$~2522$~ \tabularnewline
$~1^3P_2$~ &$~2^+$~ &$~2575$~ &$~2569$~        &$~2592$~ &$~2571$~ &$~2580$~ &$~2559$~ &$~2586$~ \tabularnewline
$~2^3P_0$~ &$~0^+$~ &$~2940$~ &$~...$~         &$~3005$~ &$~3054$~ &$~2900$~ &$~2830$~ &$~2901$~ \tabularnewline
$~2P    $~ &$~1^+$~ &$~3002$~ &$~...$~         &$~3018$~ &$~3067$~ &$~3000$~ &$~2958$~ &$~2928$~ \tabularnewline
$~2P'   $~ &$~1^+$~ &$~3026$~ &$~...$~         &$~3038$~ &$~3154$~ &$~3010$~ &$~2995$~ &$~2942$~ \tabularnewline
$~2^3P_2$~ &$~2^+$~ &$~3053$~ &$~ ...$~        &$~3048$~ &$~3142$~ &$~3060$~ &$~3040$~ &$~2988$~ \tabularnewline
\hline
$~1^3D_1$~ &$~1^-$~ &$~2843$~ &$~2859\pm27$~  &$~2899$~ &$~2913$~ &$~2820$~ &$~2804$~  &$~2845$~ \tabularnewline
$~1D    $~ &$~2^-$~ &$~2857$~ &$~...$~         &$~2900$~ &$~2931$~ &$~2860$~ &$~2788$~  &$~2838$~ \tabularnewline
$~1D'   $~ &$~2^-$~ &$~2911$~ &$~...$~         &$~2926$~ &$~2961$~ &$~2880$~ &$~2849$~  &$~2856$~ \tabularnewline
$~1^3D_3$~ &$~3^-$~ &$~2882$~ &$~2860\pm7$~   &$~2917$~ &$~2971$~ &$~2900$~ &$~2811$~  &$~2857$~ \tabularnewline
$~2^3D_1$~ &$~1^-$~ &$~3233$~ &$~...$~         &$~3306$~ &$~3383$~ &$~3250$~ &$~3217$~  &$~3172$~ \tabularnewline
$~2D    $~ &$~2^-$~ &$~3267$~ &$~...$~         &$~3298$~ &$~3403$~ &$~3280$~ &$~3217$~  &$~3144$~ \tabularnewline
$~2D'   $~ &$~2^-$~ &$~3306$~ &$~...$~         &$~3323$~ &$~3456$~ &$~3290$~ &$~3260$~  &$~3167$~ \tabularnewline
$~2^3D_3$~ &$~3^-$~ &$~3299$~ &$~...$~         &$~3311$~ &$~3469$~ &$~3310$~ &$~3240$~  &$~3157$~ \tabularnewline
\hline
$~1^3F_2$~ &$~2^+$~ &$~3176$~ &$~...$~         &$~3208$~ &$~3230$~ &$~3120$~ &$~...$~   &$~...$~  \tabularnewline
$~1F    $~ &$~3^+$~ &$~3123$~ &$~...$~         &$~3186$~ &$~3254$~ &$~3130$~ &$~...$~   &$~...$~  \tabularnewline
$~1F'   $~ &$~3^+$~ &$~3205$~ &$~...$~         &$~3218$~ &$~3266$~ &$~3150$~ &$~...$~   &$~...$~  \tabularnewline
$~1^3F_4$~ &$~4^+$~ &$~3134$~ &$~...$~         &$~3190$~ &$~3300$~ &$~3160$~ &$~...$~   &$~...$~  \tabularnewline
\hline\hline
\end{tabular}}
\end{center}
\end{table*}

\begin{figure*}
\centering \epsfxsize=15.8 cm \epsfbox{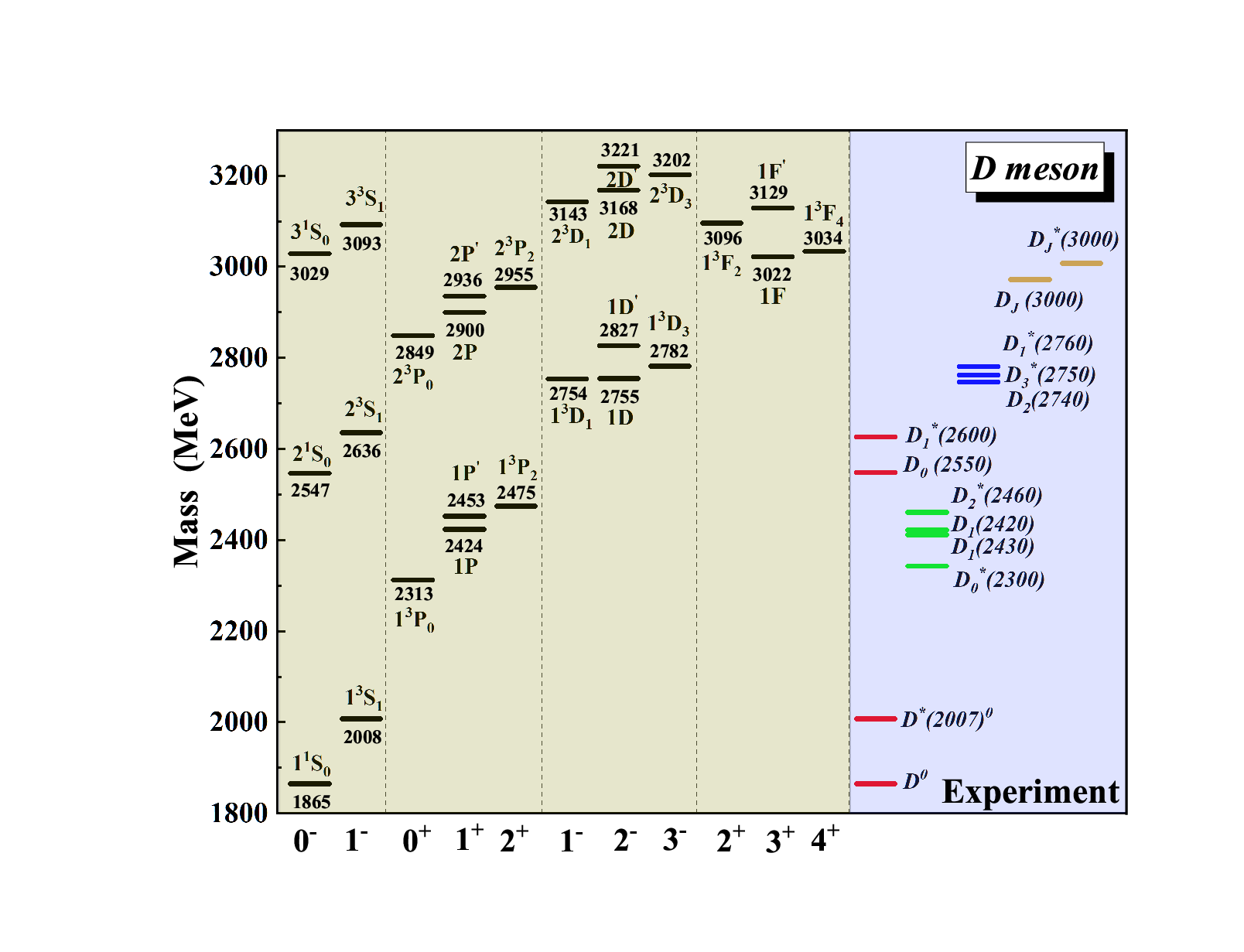} \vspace{-1.2 cm} \caption{Predicted charmed meson mass spectrum compared with the observations.}\label{Dmassspectrum}
\end{figure*}

\begin{figure*}
\centering \epsfxsize=15.8 cm \epsfbox{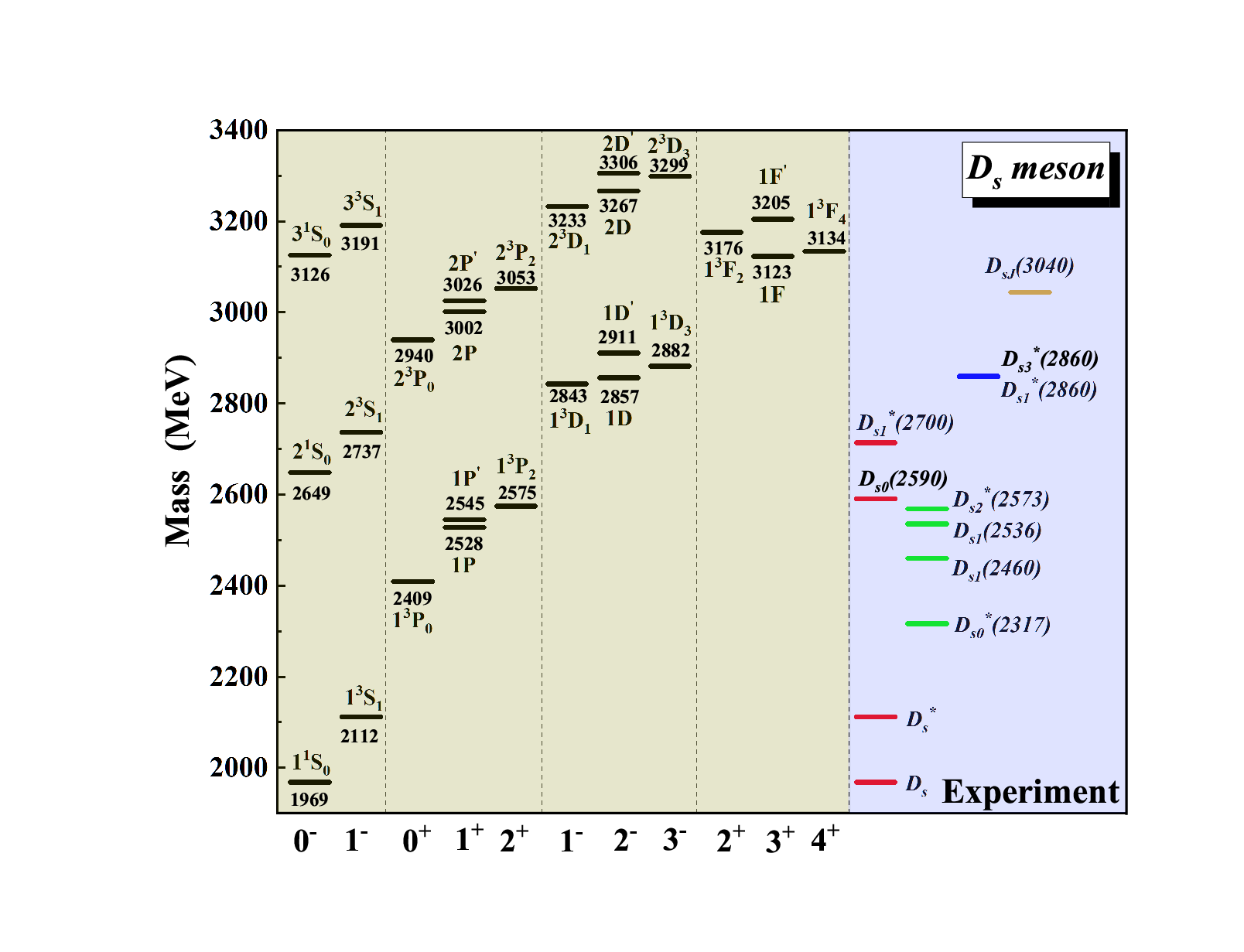} \vspace{-1.2 cm} \caption{Predicted charmed-strange meson mass spectrum compared with the observations.}\label{Dsmassspectrum}
\end{figure*}

\subsection{results}

With the determined model parameters listed in Table~\ref{parameters}, by solving the radial Schr\"{o}dinger equation
with the Gaussian expansion method~\cite{Hiyama:2003cu} we obtain the masses of the $D$ and $D_s$ meson states,
which are listed in Table~\ref{D} and
Table~\ref{Ds}, respectively. For comparison, some other model predictions in Refs.~\cite{Zyla:2020zbs,Godfrey:2015dva,Ebert:2009ua,Zeng:1994vj,Li:2010vx,Lahde:1999ih}
and the data from RPP~\cite{Zyla:2020zbs} are listed in the same table as well.
Furthermore, for clarity, the spectra are also shown in Figs.~\ref{Dmassspectrum} and \ref{Dsmassspectrum}.
It is shown that the masses for the well-established states together with the
newly observed states can be reasonably described within the semi-relativistic
quark model. Our results are also in good agreement with other
quark model predictions, although there are some model dependencies in the
predicted masses for the higher $2D$- and $1F$-wave states.

To compare the meson wave functions obtained in the present work with
those obtained with the relativized quark model~\cite{Godfrey:2015dva}, we also extract the
effective harmonic oscillator parameters $\beta_{eff}$ of the harmonic oscillator
wave functions by equating the rms radius of the harmonic oscillator wave function for the
specified $(n,l)$ quantum numbers to the rms radius of the wave functions calculated
from our potential model. Our obtained $\beta_{eff}$ parameters together those from the relativized
quark model~\cite{Godfrey:2015dva} are given in Table~\ref{DDsbetaeff}.
It is found that the $\beta_{eff}$ parameters of the harmonic oscillator
wave functions estimated in this work are consistent with those determined with the relativized
quark model~\cite{Godfrey:2015dva}.

\begin{table}[]
\caption{Predicted effective harmonic oscillator parameters $\beta_{eff}$~(GeV) of the
harmonic oscillator wave functions for the charmed and charmed-strange meson states. For comparison,
the values predicted from the relativized quark model are also listed.}
\label{DDsbetaeff}
\scalebox{1.00}{
\begin{tabular}{cccccccccccccccccccccccccccc}
\hline\hline
&
&\multicolumn{2}{c}{$\underline{~~~~~~~~~~~~~~~~~D~~~~~~~~~~~~~~~~~}$}
&\multicolumn{2}{c}{$\underline{~~~~~~~~~~~~~~~~~D_s~~~~~~~~~~~~~~~~~}$}
\\
&~ State &Ours    & GM~\cite{Godfrey:2015dva}     &Ours & GM~\cite{Godfrey:2015dva} \tabularnewline
\hline
&~$1^1S_0~$ &~0.597  &~0.601                       ~  &~0.616      &~0.651~~    \tabularnewline
&~$1^3S_1~$ &~0.499  &~0.516                       ~  &~0.514      &~0.562~~    \\
&~$2^1S_0~$ &~0.451  &~0.450                       ~  &~0.461      &~0.475~~    \\
&~$2^3S_1~$ &~0.424  &~0.434                       ~  &~0.432      &~0.458~~    \\
&~$3^1S_0~$ &~0.403  &~0.407                       ~  &~0.409      &~0.424~~    \\
&~$3^3S_1~$ &~0.390  &~0.399                       ~  &~0.395      &~0.415~~    \\
\hline
&~$1^3P_0~$ &~0.538  &~0.516                       ~  &~0.549      &~0.542~~    \\
&~$1P ~$ &~0.459,0.460  &~0.475,~0.482~            ~  &~0.468,0.469     &~0.498,~0.505~~   \\
&~$1P'~$ &~0.459,0.460  &~0.475,~0.482~            ~  &~0.468,0.469     &~0.498,~0.505~~    \\
&~$1^3P_2~$ &~0.421  &~0.437                       ~  &~0.431      &~0.464~~    \\
&~$2^3P_0~$ &~0.427  &~0.431                       ~  &~0.436      &~0.444~~    \\
&~$2P ~$ &~0.405,0.414  &~0.417,~0.419~            ~  &~0.413,0.420     &~0.433,~0.434~~    \\
&~$2P'~$ &~0.405,0.414  &~0.417,~0.419~            ~  &~0.413,0.420     &~0.433,~0.434~~    \\
&~$2^3P_2~$ &~0.391  &~0.402                       ~  &~0.398      &~0.420~~    \\
\hline
&~$1^3D_1~$ &~0.473  &~0.456                       ~  &~0.478      &~0.469~~   \\
&~$1D ~$ &~0.416,0.420  &~0.428,~0.433~            ~  &~0.424,0.428     &~0.444,~0.448~~    \\
&~$1D'~$ &~0.416,0.420  &~0.428,~0.433~            ~  &~0.424,0.428     &~0.444,~0.448~~    \\
&~$1^3D_3~$ &~0.397  &~0.407                       ~  &~0.405      &~0.426~~    \\
&~$2^3D_1~$ &~0.419  &~0.410                       ~  &~0.425      &~0.419~~    \\
&~$2D ~$ &~0.390,0.391  &~0.396,~0.399~            ~  &~0.396,0.398     &~0.408,~0.410~~    \\
&~$2D'~$ &~0.390,0.391  &~0.396,~0.399~            ~  &~0.396,0.398     &~0.408,~0.410~~    \\
&~$2^3D_3~$ &~0.374  &~0.385                       ~  &~0.381      &~0.400~~    \\
\hline
&~$1^3F_2~$ &~0.422  &~0.423                       ~  &~0.426      &~0.432~~    \\
&~$1F ~$ &~0.396,0.398  &~0.404,~0.407~            ~  &~0.402,0.404     &~0.417,~0.419~~    \\
&~$1F'~$ &~0.396,0.398  &~0.404,~0.407~            ~  &~0.402,0.404     &~0.417,~0.419~~    \\
&~$1^3F_4~$ &~0.388  &~0.390                       ~  &~0.394      &~0.405~~    \\
\hline\hline
\end{tabular}}
\end{table}

\section{STRONG DECAY}\label{STRONG DECAY}

\subsection{ model }

In this work, the Okubo-Zweig-Iizuka (OZI)-allowed two-body strong
decays of the excited $D$ and $D_s$ meson states are calculated within
a chiral quark model. The details of this model can be found in Refs.~\cite{Zhong:2008kd,Zhong:2007gp,
Zhong:2009sk,Xiao:2014ura}. In the chiral quark model~\cite{Manohar:1983md}, the low-energy quark-pseudo-scalar-meson interactions
in the SU(3) flavor basis are described by the effective Lagrangian~\cite{Li:1994cy,Li:1997gda,Zhao:2002id}
\begin{eqnarray}
{\cal L}_{Pqq}&=&\sum_j
\frac{1}{f_m}\bar{\psi}_j\gamma^{j}_{\mu}\gamma^{j}_{5}\psi_j\partial^{\mu}\phi_m,\label{coup}.
\end{eqnarray}
While the quark-vector-meson interactions in the SU(3) flavor basis are
described by the effective Lagrangian~\cite{Zhao:1998fn,Zhao:2000tb,Zhao:2001jw}
\begin{eqnarray}
{\cal
L}_{Vqq}&=&\sum_j\bar{\psi}_j(a\gamma^{j}_{\mu}+\frac{ib}{2m_j}\sigma_{\mu\nu}q^\nu)V^\mu\psi_j
\label{coup2}.
\end{eqnarray}
In the above effective Lagrangians, $\psi_j$ represents the $j$th quark field in the
hadron, $\phi_m$ is the pseudoscalar meson field, $f_m$ is the
pseudoscalar meson decay constant, and $V^\mu$ represents the vector
meson field. Parameters $a$ and $b$ denote the vector and tensor
coupling strength, respectively.

To match the nonrelativistic wave functions of the heavy-light mesons, we should adopt the
nonrelativistic form of the Lagrangians in the calculations.
The nonrelativistic form of Eq.~(\ref{coup}) is given by~\cite{Li:1994cy,Li:1997gda,Zhao:2002id}
\begin{eqnarray}\label{ccpk}
H_{m}=\sum_j\left(A\vsig_j \cdot \textbf{q}
+\frac{\omega_m}{2\mu_q}\vsig_j\cdot \textbf{p}_j\right)I_j
\varphi_m,
\end{eqnarray}
in the center-of-mass system of the initial hadron, where we have
defined $A\equiv -(1+\frac{\omega_m}{E_f+M_f})$. On the other hand,
from Eq.~(\ref{coup2}), the nonrelativistic transition operators for the emission of a transversely and
longitudinally polarized vector meson are derived by~\cite{Zhao:1998fn,Zhao:2000tb,Zhao:2001jw}
\begin{eqnarray}\label{vc}
H_m^T=\sum_j \left[i\frac{b'}{2m_q}\vsig_j\cdot
(\mathbf{q}\times\mathbf{\epsilon})+\frac{a}{2\mu_q}\mathbf{p}_j\cdot
\mathbf{\epsilon}\right]I_j\varphi_m,
\end{eqnarray}
and
\begin{eqnarray}
H_m^L=\sum_j \frac{a M_v}{|\mathbf{q}|}I_j\varphi_m \ .
\end{eqnarray}
In the above equations,  $\textbf{q}$ is the
three-vector momentum of the final state pseudoscalar/vector meson;
$\omega_m$ is the energy of final state pseudoscalar meson;
$\textbf{p}_j$ is the internal momentum operator of
the $j$th quark in the heavy-light meson rest frame; $\vsig_j$ is
the spin operator for the $j$th quark of the
heavy-light system; and $\mu_q$ is a reduced mass given by
$1/\mu_q=1/m_j+1/m'_j$ with $m_j$ and $m'_j$ for the masses of the
$j$th quark in the initial and final mesons, respectively. $E_f$ and $M_f$ represent the energy and
mass of the final state heavy hadron, $M_v$ is the mass of the emitted vector meson. The plane wave part of the
emitted light meson is $\varphi_m=e^{-i\textbf{q}\cdot
\textbf{r}_j}$, and $I_j$ is the flavor operator defined for the
transitions in the SU(3) flavor space
\cite{Li:1997gda,Zhao:2002id,Zhao:1998fn,Zhao:2000tb,Zhao:2001jw}. The
parameter $b'$ in Eq.~(\ref{vc}) is defined as $b'\equiv b-a$.
The chiral quark model has been successfully
applied to describe the strong decays of the heavy-light mesons and baryons~\cite{li:2021hss,Xiao:2020oif,Wang:2019uaj,Wang:2018fjm,Xiao:2020gjo,Wang:2020gkn,Xiao:2018pwe,Xiao:2014ura,Zhong:2010vq,Zhong:2008kd,
Zhong:2009sk,Liu:2012sj,Zhong:2007gp,Xiao:2013xi,Nagahiro:2016nsx,Yao:2018jmc,
Wang:2017kfr,Xiao:2017udy,Wang:2017hej,Liu:2019wdr}.
It should be mentioned that the nonrelativistic form of quark-pseudoscalar-meson interactions
expressed in Eq.~(\ref{ccpk}) is similar to that of the pseudoscalar
emission model~\cite{DiPierro:2001dwf,Godfrey:1985xj,Koniuk:1979vy,Capstick:2000qj,Goity:1998jr}, except that the factors $A\equiv -(1+\frac{\omega_m}{E_f+M_f})$ and $h\equiv \frac{\omega_m}{2\mu_q}$ in this work have an explicit dependence on the energies of final hadrons.

For a light pseudoscalar meson emission in heavy-light
meson strong decays, the partial decay width can be
calculated with
\begin{equation}
\Gamma_P = \left( \frac{\delta}{f_m} \right)^2 \frac{(E_f+M_f)\mid \bm{q} \mid }{4 \pi M_i(2J_i+1)}\sum_{J_{iz},J_{fz}}\mid \mathcal{M}_{ J_{iz},J_{fz} } \mid ^2,
\end{equation}
where $\mathcal{M}_{J_{iz},J_{fz}}$ is the transition amplitude, and
$J_{iz}$ and $J_{fz}$ stand for the third components of the total
angular momenta of the initial and final heavy-light mesons,
respectively. $\delta$ as a global parameter accounts for the strength of the quark-meson couplings.
Here, we take the same value as that determined in Refs.\cite{Xiao:2014ura,Zhong:2007gp,Zhong:2008kd}, i.e., $\delta = 0.557 $.
While, for a light vector meson emission in heavy-light
meson strong decays, the partial decay width can be
calculated with
\begin{equation}
\Gamma_V = \frac{(E_f+M_f)\mid \bm{q} \mid }{4 \pi M_i(2J_i+1)}\sum_{J_{iz},J_{fz}}\mid \mathcal{M}_{ J_{iz},J_{fz} } \mid ^2.
\end{equation}

To be consistent with the parameters of the mass calculations within the potential model, the masses of the component quarks are adopted as
$m_c=1.7$~GeV, $m_{u/d }=0.4$~GeV and $m_s=0.50$~GeV. The decay constants for $\pi$, $K$ and $\eta$ mesons are taken as
$f_{\pi}= 132 $~MeV, $f_K= f_{\eta}= 160 $~MeV, respectively. For the quark-vector-meson
coupling strength which still suffers relatively large
uncertainties, we adopt the values extracted from vector meson
photoproduction, i.e. $a\simeq -3$ and $b'\simeq
5$~\cite{Zhao:1998fn,Zhao:2000tb,Zhao:2001jw}. The masses of the
mesons used in the calculations are adopted from RPP~\cite{Zyla:2020zbs} if there are
observations, otherwise, the meson masses are adopted our predictions.

\section{Discussion}\label{discussion}

\begin{table}[htp]
\begin{center}
\caption{Partial decay widths (MeV) and their branching fractions for the $S$-wave charmed mesons.}
\label{DDecaySwave}
\scalebox{0.82}{
\begin{tabular}{clccccccc}
\hline\hline
~~&~~State~~        &~~Channel~~     &~~$\Gamma_i$~~              &~~Br (\%)~~ &~~$\Gamma_{exp}$~\cite{Zyla:2020zbs}~~          \tabularnewline
\hline
&$D(1^3S_1)$      &$D^0 \pi^0$     &$48.0$~~keV           &$100$          &$<2.1$~~MeV                  \tabularnewline
&as~$D^*(2007)^0$ &\textbf{Total}  &$\mathbf{48.0}$~~keV  &$\mathbf{100}$ &$\mathbf{<2.1}$~~MeV         \tabularnewline
\hline
&$D(1^3S_1)$      &$ D^0 \pi^+$     &$68.7$~~keV           &$69.2$         &$56.5\pm 1.2$~~keV           \tabularnewline
&as~$D^*(2010)^+$ &$ D^+ \pi^0$     &$30.6$~~keV           &$30.8$         &$25.6\pm 0.6$~~keV           \tabularnewline
&$            $   &\textbf{Total}   &$\mathbf{99.3}$~~keV  &$\mathbf{100}$ &$\mathbf{83.4 \pm 1.8}$~~keV \tabularnewline
\hline
&$D(2^1S_0)$      &$D^*\pi$        &$ 39.2$               &$43.9$          &$ ...$                  \tabularnewline
&as~$D_0(2550)$   &$D_0^*(2300)\pi$&$ 50.0$               &$56.1$          &$ ...$                  \tabularnewline
&$            $   &\textbf{Total}  &$\mathbf{89.2}$      &$\mathbf{100}$  &$\mathbf{135 \pm 17}$    \tabularnewline
\hline
&$D(2^3S_1)$       &$D \pi$          &$ 19.3$              &$47.2$     &$...$    \tabularnewline
&as~$D_1^*(2600)$  &$D_s K$          &$ 0.5$              &$1.2$      &$...$    \tabularnewline
&$  $              &$D \eta$         &$ 0.2$             &$0.5$      &$...$    \tabularnewline
&$  $              &$D^* \pi$        &$ 0.6$             &$1.5$     &$...$    \tabularnewline
&$  $              &$D_s^* K$        &$ 0.1$              &$0.2$      &$...$    \tabularnewline
&$  $              &$D^* \eta$       &$ 0.5$              &$1.2$      &$...$    \tabularnewline
&$  $              &$D_2^*(2460)\pi$ &$ 0.03$             &$0.07$     &$...$    \tabularnewline
&$  $              &$D_1(2430) \pi$  &$ 16.8$             &$41.1$     &$...$    \tabularnewline
&$  $              &$D_1(2420) \pi$  &$ 2.9$              &$7.1$      &$...$    \tabularnewline
&$  $              &\textbf{Total}   &$\mathbf{40.9}$     &$\mathbf{100}$    &$\mathbf{139 \pm 31}$    \tabularnewline
\hline
&$D(3^1S_0)$ &$ D^* \pi$              &$ 84.2$             &$20.5$      &$...$    \tabularnewline
&$3029$      &$ D_s^*  K$             &$ 20.0$             &$4.9$      &$...$    \tabularnewline
&$  $        &$ D^* \eta$             &$ 9.5$              &$2.3$       &$...$    \tabularnewline
&$  $        &$ D^* \eta'$            &$ 0.4$             &$0.1$     &$..$    \tabularnewline
&$  $        &$ D(2^3S_1)(2627)  \pi$     &$ 22.5$             &$5.5$       &$...$    \tabularnewline
&$  $        &$ D_0^*(2300)  \pi$     &$ 120.1$             &$29.2$      &$...$    \tabularnewline
&$  $        &$ D_{s}(1^3P_0)(2409)K$ &$ 19.5$    &$4.7$    &$...$ \tabularnewline
&$  $        &$ D_0^*(2300) \eta$                  &$ 12.3$    &$3.0$    &$...$    \tabularnewline
&$  $        &$ D_2^*(2460)  \pi$                  &$ 97.5$   &$23.7$   &$...$    \tabularnewline
&$  $        &$ D_2^*(2460)  \eta$                 &$ 0.03$   &$7\times10^{-3}$  &$...$    \tabularnewline
&$  $        &$ D  \rho$                           &$ 1.4$    &$0.3$    &$...$    \tabularnewline
&$  $        &$ D  \omega$                         &$ 0.4$    &$0.1$    &$...$    \tabularnewline
&$  $        &$ D_s  K^*$                          &$ 0.3$    &$0.07$   &$...$    \tabularnewline
&$  $        &$ D^*  \rho$                         &$ 17.6$   &$4.3$    &$...$    \tabularnewline
&$  $        &$ D^*  \omega$                       &$ 5.6$   &$1.4$    &$...$    \tabularnewline
&$  $        &$ D_s^*  K^*$                        &$ 0.4$    &$0.1$    &$...$    \tabularnewline
&$  $        &\textbf{Total}                       &$\mathbf{411.7}$  &$\mathbf{100}$ &$ ...$    \tabularnewline
\hline
&$D(3^3S_1)$ &$ D \pi$              &$ 12.7$             &$6.1$     &$...$    \tabularnewline
&$3093$      &$ D_s K$              &$ 1.8$              &$0.9$     &$...$    \tabularnewline
&$  $        &$ D \eta$             &$ 0.4$              &$0.2$   &$...$    \tabularnewline
&$  $        &$ D \eta'$            &$ 5\times10^{-3}$   &$2\times10^{-3}$ &$...$    \tabularnewline
&$  $        &$ D_0(2550)\pi$       &$ 3.3$              &$1.6$     &$...$    \tabularnewline
&$  $        &$ D^* \pi$            &$ 6.7$             &$3.2$    &$...$    \tabularnewline
&$  $        &$ D_s^* K$            &$ 3.6$              &$1.7$     &$...$    \tabularnewline
&$  $        &$ D^* \eta$           &$ 1.9$              &$0.9$     &$...$    \tabularnewline
&$  $        &$ D^* \eta'$          &$ 0.5$              &$0.2$     &$...$    \tabularnewline
&$  $        &$ D(2^3S_1)(2627)\pi$     &$ 1.7$              &$0.8$     &$...$  \tabularnewline
&$  $        &$ D_1(2430) \pi$      &$ 64.2$             &$30.7$    &$...$    \tabularnewline
&$  $        &$ D_{s}(1P_1)(2528)K$  &$ 6.5$    &$3.1$  &$...$    \tabularnewline
&$  $        &$ D_1(2430) \eta$                  &$ 4.0$    &$1.9$  &$...$    \tabularnewline
&$  $        &$ D(2P_1)(2900)\pi$   &$ 23.2$   &$11.1$ &$...$    \tabularnewline
&$  $        &$ D_1(2420) \pi$                   &$ 9.4$    &$4.5$  &$...$    \tabularnewline
&$  $        &$ D_{s1}(2535) K$                  &$ 0.4$    &$0.2$  &$...$    \tabularnewline
&$  $        &$ D_1(2420) \eta$                  &$ 0.6$    &$0.3$  &$...$    \tabularnewline
&$  $        &$ D(2P'_1)(2936) \pi$ &$ 3.4$    &$1.6$  &$...$    \tabularnewline
&$  $        &$ D_2^*(2460)\pi$                  &$ 0.07$   &$0.03$ &$...$    \tabularnewline
&$  $        &$ D_{s2}^*(2573)K$                 &$ 0.01$   &$5\times10^{-3}$ &$...$    \tabularnewline
&$  $        &$ D_2^*(2460)\eta$                 &$ 0.1$    &$0.05$   &$...$    \tabularnewline
&$  $        &$ D  \rho$                         &$ 41.3$   &$19.7$    &$...$    \tabularnewline
&$  $        &$ D  \omega$                       &$ 12.9$    &$6.2$    &$...$    \tabularnewline
&$  $        &$ D_s  K^*$                        &$ 8.1$    &$3.9$   &$...$    \tabularnewline
&$  $        &$ D^*  \rho$                       &$ 2.1$    &$1.0$    &$...$    \tabularnewline
&$  $        &$ D^*  \omega$                     &$ 0.6$    &$0.3$    &$...$    \tabularnewline
&$  $        &$ D_s^*  K^*$                      &$ 0.01$   &$5\times10^{-3}$ &$...$    \tabularnewline
&$  $        &\textbf{Total}                     &$\mathbf{209.4}$ &$\mathbf{100}$  &$...$    \tabularnewline
\hline \hline
\end{tabular}}
\end{center}
\end{table}

\begin{table}[htp]
\begin{center}
\caption{Partial decay widths  and their branching fractions for the $S$-wave charmed-strange mesons.}
\label{DsDecaySwave}
\scalebox{0.85}{
\begin{tabular}{clcccccccc}
\hline\hline
~~&~~State~~     &~~Channel~~ &~~$\Gamma_i$ (MeV)~~  &~~Br~(\%)~~   &~~$\Gamma_{exp}$ (MeV)~\cite{Zyla:2020zbs}~~   \tabularnewline
\hline
&$D_s(2^1S_0)$        &$D^* K$         &$37.1$             &$100$            &$...$   \tabularnewline
&$2649$               &\textbf{Total}  &$\mathbf{37.1}$    &$\mathbf{100}$   &$...$   \tabularnewline
\hline
&$D_s(2^3S_1)$        &$D K$           &$4.8$          &$41.0$   &$...$   \tabularnewline
&as~$D_{s_1}^*(2700)$ &$D_s \eta$      &$0.1$           &$0.9$   &$...$   \tabularnewline
&$~$                  &$D^* K$         &$6.4$          &$54.7$  &$...$   \tabularnewline
&$~$                  &$D_s^* \eta$    &$0.4$           &$3.4$   &$...$   \tabularnewline
&$~$                  &\textbf{Total}  &$\mathbf{11.7}$ &$\mathbf{100}$   &$\mathbf{120\pm11}$   \tabularnewline
\hline
&$D_s(3^1S_0)$        &$D^*K$          &$1.5$    &$0.9$  &$...$   \tabularnewline
&$3126$               &$D_s^* \eta$    &$12.2$   &$7.2$  &$...$   \tabularnewline
&$~$                  &$D_s^* \eta'$   &$0.4$    &$0.2$  &$...$   \tabularnewline
&$~$                  &$D(2^3S_1)(2627)K$  &$1.5$    &$0.9$  &$...$ \tabularnewline
&$~$                  &$D_0^*(2300)K$  &$93.5$   &$55.5$  &$...$ \tabularnewline
&$~$                  &$D_{s}(1^3P_0)(2409)\eta$  &$15.2$     &$9.0$     &$...$ \tabularnewline
&$~$                  &$D_2^*(2460) K$                         &$27.1$     &$16.1$     &$...$   \tabularnewline
&$~$                  &$D K^*$                                 &$0.6$      &$0.4$     &$...$   \tabularnewline
&$~$                  &$D_s  \phi$                             &$0.01$     &$0.006$     &$...$   \tabularnewline
&$~$                  &$D^* K^*$                               &$17.9$     &$10.6$     &$...$   \tabularnewline
&$~$                  &\textbf{Total}                          &$\mathbf{169.9}$   &$\mathbf{100}$     & $...$   \tabularnewline
\hline
&$D_s(3^3S_1)$        &$D K$            &$6.3$                &$5.1$   &$...$   \tabularnewline
&$3191$               &$D_s \eta$       &$0.5$                &$0.4$   &$...$   \tabularnewline
&$~$                  &$D_s \eta'$      &$9 \times 10^{-5}$   &$4 \times 10^{-5}$   &$...$ \tabularnewline
&$~$                  &$D^* K$          &$10.0$               &$8.1$   &$...$   \tabularnewline
&$~$                  &$D_s^* \eta$     &$2.6$                &$2.1$   &$...$   \tabularnewline
&$~$                  &$D_s^* \eta'$    &$0.5$                &$0.4$   &$...$   \tabularnewline
&$~$                  &$D_0(2550)K$     &$0.2$                &$0.2$   &$...$   \tabularnewline
&$~$                  &$D(2^3S_1)(2627)K$   &$7.0$                &$5.7$   &$...$   \tabularnewline
&$~$                  &$D_1(2430)K$     &$47.9$               &$39.0$   &$...$   \tabularnewline
&$~$                  &$D_{s}(1P_1)(2528)\eta$   &$5.2$    &$4.2$  &$...$ \tabularnewline
&$~$                  &$D_1(2420)K$                         &$3.6$    &$2.9$   &$...$   \tabularnewline
&$~$                  &$D_{s1}(2535) \eta$                  &$0.4$    &$0.3$   &$...$   \tabularnewline
&$~$                  &$D_2^*(2460)K$                       &$2.3$    &$1.9$   &$...$   \tabularnewline
&$~$                  &$D_{s2}^*(2573)\eta$                 &$0.05$   &$0.04$   &$...$   \tabularnewline
&$~$                  &$D K^*$                              &$31.5$   &$25.6$   &$...$   \tabularnewline
&$~$                  &$D_s  \phi$                          &$4.1$    &$3.3$   &$...$   \tabularnewline
&$~$                  &$D^*  K^*$                           &$0.7$    &$0.6$  &$...$   \tabularnewline
&$~$                  &$D_s^*  \phi$                        &$9\times 10^{-3}$   &$7\times 10^{-3}$    &$...$\tabularnewline
&$~$                  &\textbf{Total}                       &$\mathbf{122.9}$    &$\mathbf{100}$  &$...$  \tabularnewline
\hline \hline
\end{tabular}}
\end{center}
\end{table}

\subsection{$1S$-wave vector states}

In the $D$ and $D_s$ families, the ground vector ($1^3S_1$) charmed and charmed-strange
states, $D^*$ and $D_s^*$, are well established. The strong decay transition $D_s^*\to DK$ is kinematic forbidden.
The charged state $D^*(2010)^+$ can decay into both the $D^+ \pi^0$
and $D^0 \pi^+$ final states. While the decays of the neutral $D^*(2007)^0$ are governed by
the $D^0 \pi^0$ channel, however, the $D^*(2007)^0\to D^+\pi^-$ is kinematic forbidden. With the
numerical wavefunctions determined from the potential model, the strong decays of $D^*(2007)^0$ and $D^*(2010)^+$ are
calculated within the chiral quark model. As shown in Table~\ref{DDecaySwave},
our predicted decay partial width of $\Gamma[D^*(2007)^0\to D^0\pi^0]\simeq 48$ keV
is consistent with the observation. While for $D^*(2010)^+$, the predicted width of $\Gamma\simeq 99$ keV
and the partial width ratio
\begin{equation}
R=\frac{\Gamma(D^0 \pi^+)}{\Gamma(D^+\pi^0)} \simeq 2.25
\end{equation}
are in remarkable agreement with the experimental data $\Gamma_{exp}=83.4\pm 1.8$ keV and $R_{exp}=2.21$,
respectively~\cite{Zyla:2020zbs}.

\subsection{$2S$-wave states}

\subsubsection{$2^1S_0$}

\begin{figure}
\centering \epsfxsize=8.9 cm \epsfbox{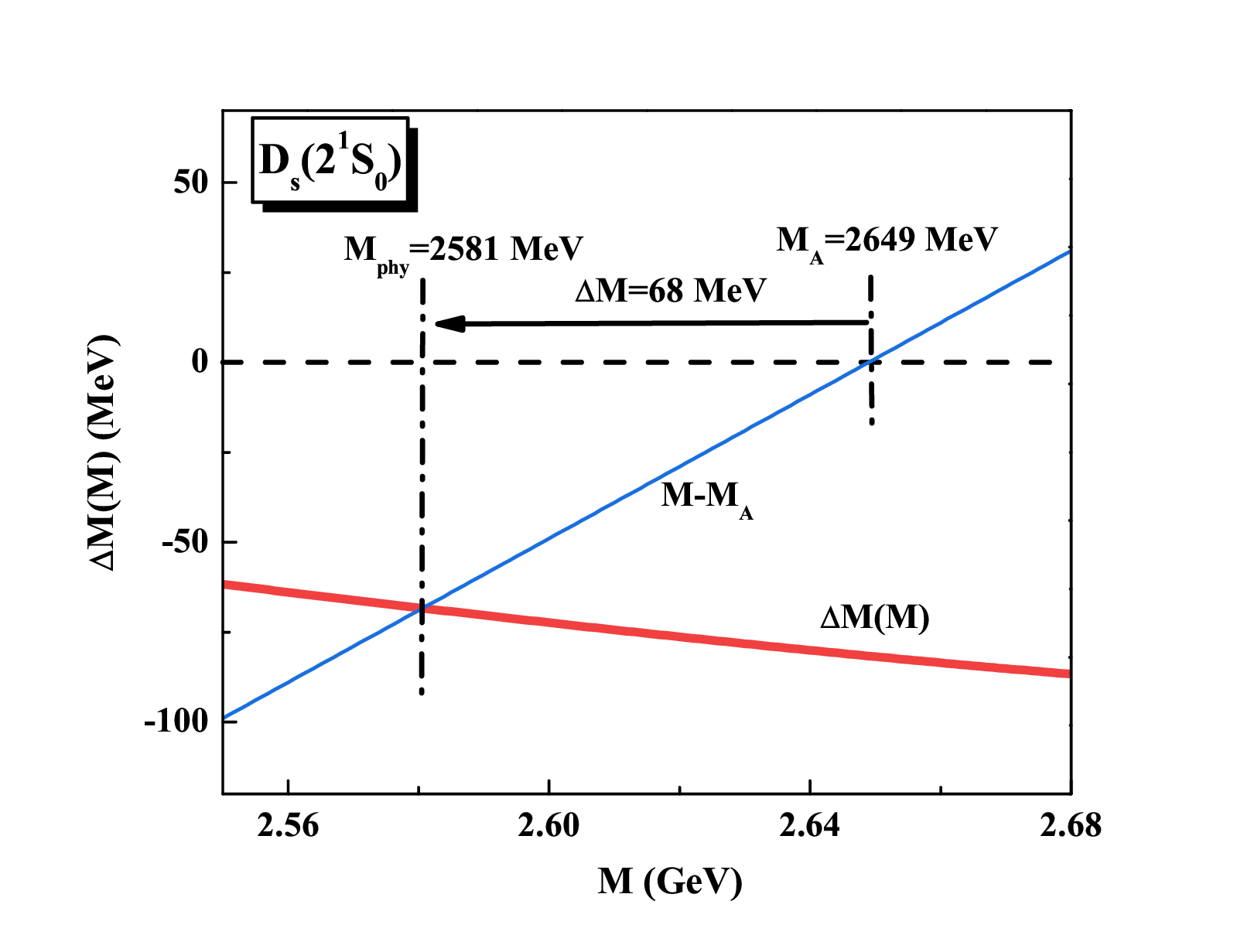} \vspace{-0.6 cm} \caption{
The determination of the physical mass for the $D_s(2^1S_0)$ state.
The mass shift function $\Delta M(M)$ and liner function $M-M_A$ are shown by
the thick and thin lines, respectively. $M_A$ stands for the bare mass of $D_s(2^1S_0)$,
and the physical mass $M_{phy}$ of the dressed $D_s(2^1S_0)$ state as a solution of the coupled-channel
equation Eq.(\ref{M=MA+Delta M}) is located at the
intersection point of two solid lines. }\label{Ds21S0massshift}
\end{figure}

In the $D$-meson family, our predicted mass for the $D(2^1S_0)$ state is $M=2547$ MeV,
which is comparable with the predictions in the literature~\cite{Zeng:1994vj,Lahde:1999ih,Ebert:2009ua,Liu:2013maa,Liu:2015lka,Liu:2015uya,Liu:2016efm,Badalian:2011tb,Allosh:2021biq,
Patel:2021aas,Chen:2018nnr,Godfrey:2015dva,Song:2015fha,Gandhi:2019lta,Sun:2013qca,Ferretti:2015rsa,Li:2010vx,Lu:2014zua,
Kher:2017wsq}. The $D(2^1S_0)$ may dominantly
decay into the $D^{*}\pi$ and $D_0(2300)\pi$ channels with a width of
$\Gamma\simeq 89$ MeV. The partial width ratio between $D^{*}\pi$ and
$D_0(2300)\pi$ is predicted to be
\begin{equation}
R=\frac{\Gamma[D^* \pi]}{\Gamma[D_0(2300) \pi]} \simeq 0.78.
\end{equation}
Our predicted width of $D(2^1S_0)$ with the chiral quark model is close
to the predictions within the $^3P_0$ models~\cite{Chen:2011rr,Yu:2014dda,Song:2015fha,Godfrey:2015dva,Lu:2014zua},
however, in these works the predicted decay rate into the
$D_0(2300)\pi$ channel is tiny.

The $D_0(2550)$ listed in RPP~\cite{Zyla:2020zbs} may be classified as the radially excited state $2^1S_0$
in the $D$-meson family. Its average measured mass and width are $M_{exp}=2549\pm 19$ MeV and $\Gamma_{exp}=165\pm 24$ MeV,
respectively~\cite{Zyla:2020zbs}. This state was first observed by the BaBar collaboration in the $D^{*+}\pi^-$ channel
in 2010~\cite{BaBar:2010zpy}, and was confirmed by the LHCb collaboration with significance by using $pp$
collision data~\cite{LHCb:2013jjb,Aaij:2019sqk}. As the assignment of $D(2^1S_0)$,
the mass of $D_0(2550)$ is consistent with various quark model predictions~\cite{Kher:2017wsq,Badalian:2011tb,Liu:2015uya,Song:2015fha,Godfrey:2015dva,Lu:2014zua}.
In Refs.~\cite{Zhong:2008kd,Zhong:2010vq,
Xiao:2014ura} we have studied the strong decays of the $D_0(2550)$ as the $2^1S_0$
state by using the SHO wave function, the obtained width, $\Gamma\simeq 20-70$ MeV, is too narrow to
be comparable with the data. In the present work, with the genuine wave function determined by the potential model
our predicted decay width of $D(2^1S_0)$,
\begin{equation}
\Gamma\simeq 89  \ \ \ \mathrm{MeV },
\end{equation}
is close to the data $\Gamma_{exp}=130\pm25$ MeV measured by BaBar~\cite{BaBar:2010zpy}.
The $D_0(2550)$ is also explained as the $D(2^1S_0)$ state based on the strong decay analyses in the literature~\cite{Yu:2014dda,Chen:2011rr,Gupta:2018zlg,
Wang:2010ydc,Wang:2013tka,Song:2015fha,Godfrey:2015dva,Lu:2014zua}.
According to our chiral quark model predictions, the $D_0(2550)$ has large decay rate ($\sim56\%$) into
the $D_0(2300)\pi$ channel. Thus, to better understand the nature of the $D_0(2550)$ state and to test various model predictions,
further observations of the missing $D_0(2300)\pi$ channel are needed in future experiments.

In the charmed-strange sector, our predicted mass for the $D_s(2^1S_0)$ state is $M=2649$ MeV,
which is comparable with the predictions in the literature~\cite{Zeng:1994vj,Lahde:1999ih,Ebert:2009ua,Liu:2013maa,Liu:2015lka,Liu:2015uya,Liu:2016efm,Badalian:2011tb,Allosh:2021biq,
Patel:2021aas,Chen:2018nnr,Godfrey:2015dva,Song:2015nia,Ferretti:2015rsa,Segovia:2015dia,Li:2010vx,
Kher:2017wsq}. The $D^* K$ decay channel is
the only OZI-allowed two body strong channel for $D_s(2^1S_0)$. With the
$D_s(2^1S_0)$ wave function obtained from our potential model calculations,
its width is predicted to be
\begin{equation}
\Gamma\simeq 37  \ \ \ \mathrm{MeV }.
\end{equation}
The $D_s(2^1S_0)$ state is also predicted to be a narrow state with a width of 10s MeV in the literature ~\cite{Godfrey:2015dva,Song:2015nia,Colangelo:2012xi,Wang:2012wk,Zhong:2008kd,
Xiao:2014ura}.

Recently, the LHCb collaboration observed a new excited $D_s^+$ state, $D_{s0}(2590)^+$, in the $D^+K^+\pi^-$ invariant mass spectrum of the $B^0 \to D^+D^+K^+\pi^-$ decay~\cite{LHCb:2020gnv}. Its mass, width and the spin parity numbers are measured to be $M_{exp}=2591 \pm 6 \pm 7$ MeV, $ \Gamma_{exp} = 89 \pm 16 \pm 12$ MeV and $J^P=0^-$, respectively. The $D_{s0}(2590)^+$ is suggested to be a candidate of the missing $D_s(2^1S_0)$ state~\cite{LHCb:2020gnv}. However, considering $D_{s0}(2590)^+$ as the $D_s(2^1S_0)$ state, it is found the observed mass is
about $60$ MeV lower than most potential model predictions.

The coupling of the $D_s(2^1S_0)$ $c\bar{s}$ core to
the two hadron final states was considered within the $^3P_0$ model in
the literature~\cite{Xie:2021dwe,Ortega:2021fem}. It is found that when taking into
account the $D^{*}K$ loop correction to the bare $c\bar{s}$ state, the
physical mass will be close to that of $D_{s0}(2590)^+$. In this work,
we also estimate the mass shift of $D_s(2^1S_0)$ by including the
$D^*K$ coupled-channel interaction within our chiral quark model.
The details about the coupled-channel quark model are given in Appendix.~\ref{Coupled-channel quark model}.
The mass shift of $D_s(2^1S_0)$ is shown in Fig.~\ref{Ds21S0massshift}.
Our result shows that the coupled-channel interaction induces a
mass shift of $\sim68$ MeV. The bare mass $M=2649$ MeV of $D_s(2^1S_0)$ will be shifted
to the physical mass $M_{phy}=2581$ MeV, which is consistent with the measured mass of $D_{s0}(2590)^+$.
Our coupled-channel calculation within the chiral quark model are
consistent with that in Refs.~\cite{Xie:2021dwe,Ortega:2021fem}.

Assigning the newly observed resonance $D_{s0}(2590)^+$ to
the $D_s(2^1S_0)$ state, the higher mass problem can be overcome by taking into
account the $D^{*}K$ loop correction, however, the width of $D_{s0}(2590)^+$
cannot be well understood within our chiral quark model. Adopting the observed mass $M=2591$ MeV,
our predicted width,
\begin{equation}
\Gamma\simeq 19  \ \ \ \mathrm{MeV },
\end{equation}
is about a factor of $5$ smaller than the center value of the data $\Gamma_{exp} = 89$ MeV.
Our predicted width is consistent with the recent predictions with the relativistic wave functions
obtained by solving the full Salpeter equation~\cite{Wang:2021orp} and the $^3P_0$ model~\cite{Xie:2021dwe}.
To establish the $D_s(2^1S_0)$ state and uncover the nature of the $D_{s0}(2590)^+$,
more observations are needed in future experiments.

\subsubsection{$2^3S_1$}\label{ab}


In the $D$-meson family, our predicted mass for the $D(2^3S_1)$ state is $M=2636$ MeV,
which is comparable with the predictions in the literature~\cite{Zeng:1994vj,Lahde:1999ih,Ebert:2009ua,Liu:2013maa,Liu:2015lka,Liu:2015uya,Liu:2016efm,Badalian:2011tb,Allosh:2021biq,
Patel:2021aas,Chen:2018nnr,Godfrey:2015dva,Song:2015fha,Gandhi:2019lta,Sun:2013qca,Ferretti:2015rsa,Li:2010vx,Lu:2014zua,
Kher:2017wsq}. According to our chiral quark
model predictions, the $D(2^3S_1)$ may be a narrow state with a width of
\begin{equation}
\Gamma\simeq 41  \ \ \ \mathrm{MeV },
\end{equation}
and dominantly decays into $D\pi$ and $D_1(2430)\pi$ channels with branching fractions about $47\%$ and $41\%$, respectively.
However, the decay rate into the $D^*\pi$ channel is tiny ($\sim 2\%$).
In Refs.~\cite{Song:2015fha,Godfrey:2015dva,Lu:2014zua,Li:2017sww,Chen:2015lpa}, the $D(2^3S_1)$ is predicted
to be a broader state with a width of $\Gamma\simeq 60-200$ MeV. Combined with our previous study~\cite{Zhong:2010vq},
we find that the strong decay properties of $D(2^3S_1)$ are very sensitive to the details of the wave function
due to the nodal effects.

From the point of view of mass, the $D_{1}^*(2600)$ resonance listed in RPP~\cite{Zyla:2020zbs}
may be a candidate of the $D(2^3S_1)$ state. This resoancne was first observed by BaBar in the
$D\pi$ and $D^*\pi$ decay channels in 2010~\cite{BaBar:2010zpy}. The measured mass and width are
$M_{exp}=2609\pm 4$ MeV and $\Gamma_{exp}=96\pm6\pm13$ MeV, respectively, and the measured partial width ratio between
$D\pi$ and $D^*\pi$ is $R=\Gamma(D\pi)/\Gamma(D^*\pi ) = 0.32\pm0.11$.
In 2013, in the $D^{*+}\pi^-$ final state the LHCb collaboration observed a
similar resonance $D_J^*(2650)^0$ with a mass of $M_{exp}=2649 \pm 7$ MeV and a width of
$\Gamma_{exp}=140.2\pm35.7$ MeV~\cite{LHCb:2013jjb}. In 2016, LHCb collaboration carried out an amplitude analysis of the
$B^{-} \to D^{+} \pi^{-} \pi^{-}$ decays, they extracted a $J^P=1^-$ resonance $D_1^*(2680)$ with mass and
width of  $M_{exp}=2681\pm 23.6$ MeV and $\Gamma_{exp}=186.7 \pm25.3$ MeV~\cite{Aaij:2016fma}.
Recently, from the $B^{-} \to D^{*+} \pi^{-} \pi^{-}$ decays, the LHCb collaboration
also extracted a $J^P=1^-$ resonance $D_1^*(2600)^0$ with mass and
width of $M_{exp}=2641.9\pm 6.3$ MeV and $\Gamma_{exp}=149 \pm24$ MeV~\cite{Aaij:2019sqk}. The resonances
observed in different experiments might be the same state, which is
denoted by $D_1^*(2600)$ in RPP~\cite{Zyla:2020zbs}, although there are some differences
in the observations of different experiments.

Considering $D_1^*(2600)$ as the $D(2^3S_1)$ assignment, the strong decay properties
have been analyzed in the literature. The strong decay analyses in
Refs.~\cite{Ferretti:2015rsa,Kher:2017wsq,Gupta:2018zlg,Yu:2014dda,Chen:2015lpa,Wang:2010ydc,Wang:2013tka,Song:2015fha,Godfrey:2015dva,Lu:2014zua}
support this assignment. However, with the $D(2^3S_1)$ assignment our
predicted width $\Gamma\simeq 41$ MeV is too small to be comparable
with the average measured value $\Gamma_{exp}=141\pm 23$ MeV.
To well explain the decay properties, the $D_1^*(2600)$ is also
suggested to be a mixed state via the $2^3S_1-1^3D_1$ mixing in the literature
~\cite{Li:2017sww,Chen:2011rr,Yu:2020khh,Zhong:2010vq,Xiao:2014ura,Sun:2010pg,Li:2010vx,Chen:2015lpa}.
In Ref.~\cite{Colangelo:2012xi}, the study within effective Lagrangian method
indicates that it is impossible to explain the ratio $R=\Gamma(D\pi)/\Gamma(D^*\pi ) = 0.32\pm0.11$~\cite{BaBar:2010zpy}
measured by BaBar with a pure $D(2^3S_1)$ state.

In the charmed-strange sector, our predicted mass for the $D_s(2^3S_1)$ state is $M=2737$ MeV,
which is comparable with the predictions in the literature~\cite{Zeng:1994vj,Lahde:1999ih,Ebert:2009ua,Liu:2013maa,Liu:2015lka,Liu:2015uya,Liu:2016efm,Badalian:2011tb,Allosh:2021biq,
Patel:2021aas,Chen:2018nnr,Godfrey:2015dva,Song:2015nia,Ferretti:2015rsa,Segovia:2015dia,Li:2010vx,
Kher:2017wsq}. According to our chiral quark
model predictions, the $D_s(2^3S_1)$ may be a narrow state with a width of
\begin{equation}
\Gamma\simeq 12  \ \ \ \mathrm{MeV },
\end{equation}
and mainly decays into the $DK$ and $D^*K$ final states. Our predictions are consistent with those predicted with a $^3P_0$ model~\cite{Zhang:2006yj}.
However, in other works~\cite{Godfrey:2015dva,Close:2005se,Chen:2015lpa,Song:2015nia}
the $D_s(2^3S_1)$ is predicted to be a relatively broad state with a width of $\Gamma\simeq 100 \sim 200$ MeV.

From the point of view of mass, the $D_{s1}^*(2700)$ resonance listed in RPP~\cite{Zyla:2020zbs}
can be assigned to the $D_s(2^3S_1)$ state. The $D_{s1}^*(2700)$ was first observed in the $DK$ final state by the
BaBar collaboration in 2006~\cite{BaBar:2006gme}, and one year later its
quantum numbers $J^P=1^-$ were determined by the Belle collaboration~\cite{Belle:2007hht}.
The average measured mass and width are $M_{exp}=2714\pm 5$ MeV and $\Gamma_{exp}=122\pm 10$ MeV, respectively~\cite{Zyla:2020zbs}.
More experimental information about $D_{s1}^*(2700)$ is collected in Table~\ref{EXP DandDs}.
Some phenomenological analyses in the literature~\cite{Ferretti:2015rsa,Badalian:2011tb,Liu:2016efm,Liu:2015uya,Ebert:2009ua,
Wang:2014jua,Zhou:2014ytp,Song:2014mha,Segovia:2015dia,Chen:2009zt,Zhang:2009nu,Godfrey:2015dva} support $D_{s1}^*(2700)$ as the $D_s(2^3S_1)$ assignment. However, considering $D_{s1}^*(2700)$ as $D_s(2^3S_1)$, the measured width $\Gamma_{exp}=122\pm 10$ MeV and ratio
$\Gamma(D^*K)/\Gamma(DK ) = 0.91\pm0.25$ cannot be explained within our chiral quark model.
To well explain the observations, the $D_{s1}^*(2700)$ is also suggested to be a mixed state via the $2^3S_1-1^3D_1$ mixing in the literature~\cite{Li:2017sww,Li:2010vx,Li:2007px,Close:2006gr,Zhong:2009sk,Chen:2011rr,Song:2015nia}.
The $2^3S_1-1^3D_1$ mixing in the $D$- and $D_s$-meson families will be further discussed later.

\subsection{$3S$-wave states}

\subsubsection{$3^1S_0$}

In the $D$-meson family, the mass for the second radial excitation $D(3^1S_0)$ is predicted to
be $M=3029$ MeV within our potential model calculations. The mass gap between $D(3^1S_0)$
and $D(2^1S_0)$ is estimated to be $\Delta M\simeq 480$ MeV. Our predictions are consistent with
those in Refs.~\cite{Godfrey:2015dva,Ebert:2009ua,Zeng:1994vj}. The $D(3^1S_0)$ may be a
broad state with a width of
\begin{equation}
\Gamma\simeq 410  \ \ \ \mathrm{MeV },
\end{equation}
and dominantly decays into the
$D_0^*(2300)\pi$ (29\%), $D_2^*(2460)\pi$ (24\%) and $D^*\pi$ (21\%) final states.
More details can be seen in Table \ref{DDecaySwave}.
It should be mentioned that there exist large model dependencies in the decay properties
predicted in the literature.

In Refs.~\cite{Lu:2014zua,Song:2015fha}, the authors suggested that the unnatural parity state $D_J(3000)$ observed in the
$D^*\pi$ final state by the LHCb collaboration~\cite{LHCb:2013jjb} might be explained with $D(3^1S_0)$ according to
their strong decay analysis within the $^3P_0$ model. However, our predicted width
of $D(3^1S_0)$ is too broad to be comparable with the measured width $\Gamma_{exp}=188.1\pm44.8$ MeV,
although the predicted mass is consistent with the data.
To establish the $D(3^1S_0)$, more observations of the other main decay channels,
such as $D_0^*(2300)\pi$ and $D_2^*(2460)\pi$, are needed in future experiments.

In the $D_s$-meson family, the mass for the second radial excitation $D_s(3^1S_0)$ is predicted to
be $M=3126$ MeV within our potential model calculations, which is about 100 MeV larger than
that of the charmed partner $D(3^1S_0)$. Our prediction is consistent with
that of the relativized quark model~\cite{Godfrey:2015dva}. From Table~\ref{DsDecaySwave}, one sees that
the $D_s(3^1S_0)$ state may have a width of
\begin{equation}
\Gamma\simeq 170  \ \ \ \mathrm{MeV },
\end{equation}
and dominantly decays into the
$D_0^*(2300)K$ and $D_2^*(2460)K$ final states. The main decay mode
$D_2^*(2460)K$ predicted within our chiral quark model are consistent with that predicted within
the $^3P_0$ model~\cite{Godfrey:2015dva}, however, our predicted width is about
a factor of 2.2 larger than that within the $^3P_0$ model~\cite{Godfrey:2015dva}. To look for the
missing $D_s(3^1S_0)$ state, the $D_2^*(2460)K$ final state is worth to observing in future experiments.

\subsubsection{$3^3S_1$}

\begin{figure}
\centering \epsfxsize=7.4 cm \epsfbox{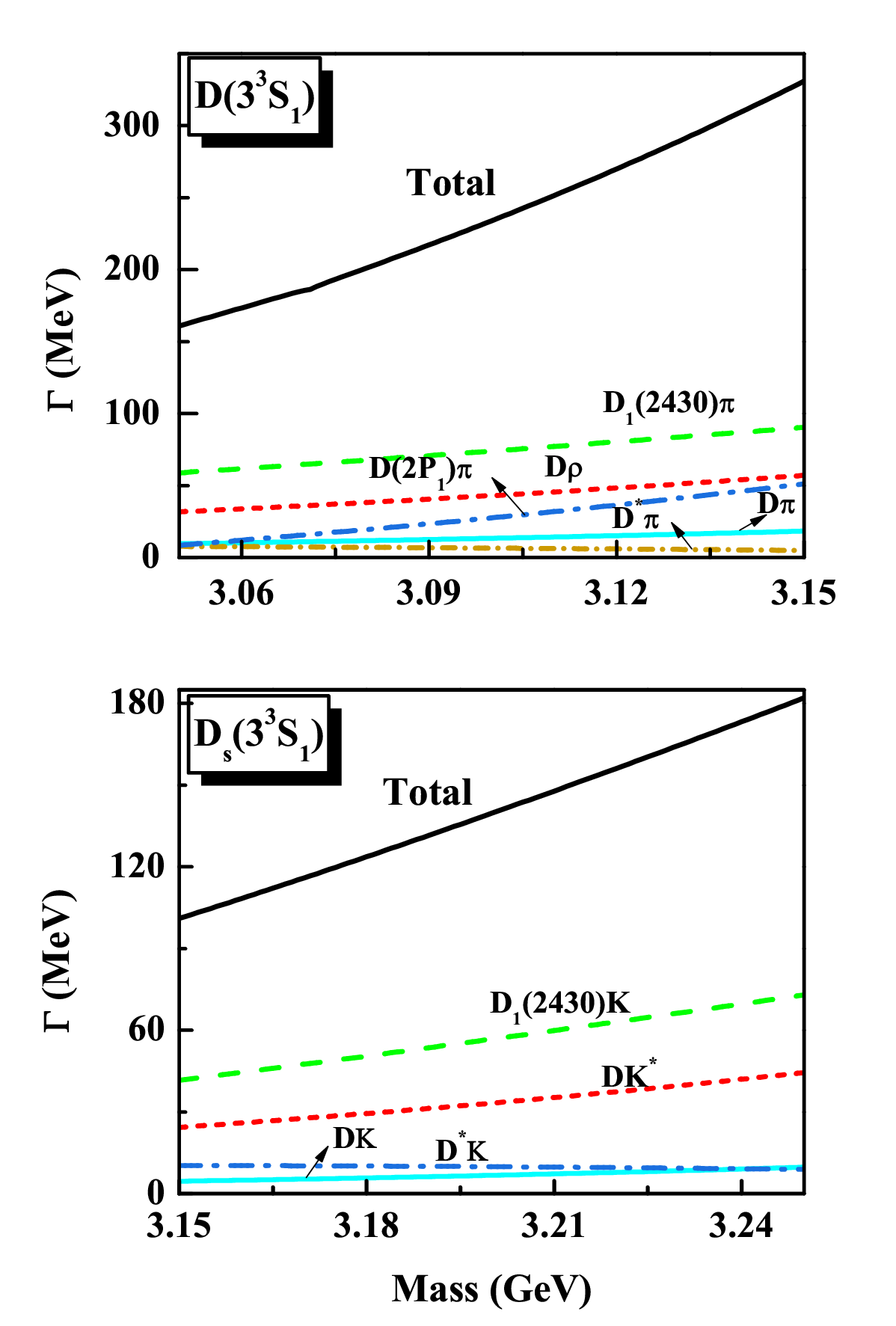} \vspace{-0.2 cm} \caption{Total width and partial widths of the main decay channels for $D(3^3S_1)$ and $D_s(3^3S_1)$ as the functions of their masses.}\label{D33S1Mirunning}
\end{figure}

In the $D$-meson family, the mass for the second radial excitation $D(3^3S_1)$ is predicted to
be $M=3093$ MeV within our potential model calculations, which is consistent with
the predictions in Refs.~\cite{Godfrey:2015dva,Ebert:2009ua,Zeng:1994vj}.
The mass splitting between $D(3^3S_1)$ and $D(3^1S_0)$ is estimated to
be $\Delta M\simeq 64$ MeV. From Table~\ref{DDecaySwave}, it is seen that the $D(3^3S_1)$ may be a
broad state with a width of
\begin{equation}
\Gamma\simeq 210  \ \ \ \mathrm{MeV },
\end{equation}
and has large decay rates into
$D\pi$ and $D^*\pi$ channels. To see the dependence of the decay properties of $D(3^3S_1)$ on its mass,
the partial widths of the main decay channels together with the total width as functions of the mass
are also plotted in Fig.~\ref{D33S1Mirunning}. It is found that the partial and total decay widths increase smoothly with the
mass. With a mass uncertainty of $50$ MeV, the total width of $D(3^3S_1)$ varies in the range $\sim 160-320$ MeV.

In Ref.~\cite{Song:2015fha}, the authors suggested that the natural parity state $D_J^*(3000)$ observed in the
$D\pi$ final state by the LHCb collaboration~\cite{LHCb:2013jjb} may be explained with $D(3^3S_1)$ according to
their mass and strong decay analysis. However, our predicted mass and width of $D(3^3S_1)$, $M= 3093$ MeV
and $\Gamma \simeq 210$ MeV, are notably larger than the data $M_{exp}=3008.1\pm4.0$ MeV
and $\Gamma_{exp} =110.5 \pm11.5$ MeV measured by LHCb~\cite{LHCb:2013jjb}. Furthermore, the predicted decay rates into the
$D\pi$ channel is tiny, which is inconsistent with the fact that $D_J^*(3000)$ was first observed
in the $D\pi$ final state. To establish the $D(3^3S_1)$, more observations of the other decay channels,
such as $D_1(2430)\pi$ and $D \rho$, are needed in future experiments.

In the $D_s$-meson sector, the mass for the second radial excitation $D_s(3^3S_1)$ is predicted to
be $M=3191$ MeV within our potential model calculations, which is consistent with
the predictions in Refs.~\cite{Godfrey:2015dva,Zeng:1994vj}. The mass splitting between
$D_s(3^3S_1)$ and $D_s(3^1S_0)$ is estimated to be $\Delta M\simeq 65$ MeV, which
is nearly equal to that for the charmed sector. The strong decay properties of $D_s(3^3S_1)$ have been
shown in Table~\ref{DsDecaySwave}, it is seen that $D_s(3^3S_1)$ may be a relatively narrow state with a width of
\begin{equation}
\Gamma\simeq 120  \ \ \ \mathrm{MeV },
\end{equation}
and mainly decays into $D_1(2430)K$ and $DK^*$ channels. The dependence of the decay
properties of $D_s(3^3S_1)$ on its mass is also shown in Fig.~\ref{D33S1Mirunning}.
It is found that the partial and total decay widths increase smoothly with the
mass. With a mass uncertainty of $50$ MeV, the total width of $D_s(3^3S_1)$ varies in the range $\sim 100-180$ MeV.
Our predicted decay properties are comparable with those of $^3P_0$ model in Refs.~\cite{Godfrey:2015dva,Song:2015nia}.
It should be mentioned that for the higher excited states, the predicted decay properties have
large model dependencies.

\subsection{$1P$-wave states }

\begin{table}[htp]
\begin{center}
\caption{Partial decay widths and their branching fractions for the $1P$-wave charmed mesons.
The decay widths in square brackets are the results for the mixed states $D(1P_1)$ and $D(1P'_1)$ predicted with the mixing angle $\theta_{1P}=-(55\pm5)^{\circ}$ in the heavy quark symmetry limit.}
\label{Dmeson1PDecay}
\scalebox{0.95}{
\begin{tabular}{clccccccc}
\hline\hline
&State~~~~ &Channel&~~~~$\Gamma_i$ (MeV)~~~~&~~Br~(\% )~~& $\Gamma_{exp}$ (MeV)~\cite{Zyla:2020zbs} \tabularnewline
\hline
&$ D(1^{3}P_0) $   &$ D \pi$        &$538.0$          &$100$          & $ ...$  \tabularnewline
&as~$D_0^*(2300)$  &\textbf{Total}  &$\mathbf{538.0}$ &$\mathbf{100}$ & $\mathbf{229 \pm 16}$  \tabularnewline
\hline
&$ D(1^{3}P_2) $  &$D \pi$         &$26.0$           &$63.0$    & $ ...$  \tabularnewline
&as~$D_2^*(2460)$ &$D \eta $       &$0.05$           &$0.1$    & $ ...$  \tabularnewline
&$ $              &$D^* \pi$       &$15.3$           &$36.9$    & $ ...$  \tabularnewline
&$ $              &\textbf{Total}  &$\mathbf{41.3}$  &$\mathbf{100}$   & $\mathbf{47.3 \pm0.8}$  \tabularnewline
\hline
&$ D(1P_1) $        &$D^* \pi$       &$214.5~[241.0\pm 1.0]$            &$100$            & $ ...$  \tabularnewline
&as~$ D_1(2430)$   &\textbf{Total}  &$\mathbf{214.5~[241.0\pm 1.0]}$   &$\mathbf{100}$    & $\mathbf{314\pm29}$  \tabularnewline
\hline
&$ D(1P'_1) $        &$D^* \pi$       &$42.5~[16.8\pm 1.0]$              &$100$        & $ ...$  \tabularnewline
&as~$ D_1(2420)$   &\textbf{Total}     &$\mathbf{42.5~[16.8\pm 1.0]}$  &$\mathbf{100}$     & $\mathbf{31.3 \pm 1.9}$  \tabularnewline
\hline \hline
\end{tabular}}
\end{center}
\end{table}

\begin{table}[htp]
\begin{center}
\caption{Partial decay widths  and their branching fractions for the $1P$-wave charmed-strange mesons.
The decay widths in square brackets are the results for the mixed states $D_s(1P_1)$ and $D_s(1P'_1)$ predicted with the mixing angle $\theta_{1P}=-(55\pm5)^{\circ}$ in the heavy quark symmetry limit.}
\label{Dsmeson1PDecay}
\scalebox{0.95}{
\begin{tabular}{clcccccc}
\hline\hline
&State~~~~ &Channel&~~~~$\Gamma_i$ (MeV)~~~&~Br~(\%)~~& ~~$\Gamma_{exp}$ (MeV)~\cite{Zyla:2020zbs}  \tabularnewline
\hline
&$D_s(1^{3}P_0) $   &$ D K$        &$437.5$          &$100$          & $ ...$  \tabularnewline
&$2409$             &\textbf{Total}  &$\mathbf{437.5}$ &$\mathbf{100}$ & $\mathbf{...}$  \tabularnewline
\hline
&$D_s(1^{3}P_2) $     &$D K$           &$11.7$          &$88.0$    & $ ...$  \tabularnewline
&as~$D_{s2}^*(2573)$  &$D_s \eta$      &$0.1$           &$0.8$     & $ ...$  \tabularnewline
&$ $                  &$D^* K$         &$1.5$           &$11.2$    & $ ...$  \tabularnewline
&$ $                  &\textbf{Total}  &$\mathbf{13.3}$  &$\mathbf{100}$   & $\mathbf{16.9 \pm0.7}$  \tabularnewline
\hline
&$D_s(1P_1) $        &$D^*K$          &$192.2~[210.7\pm 1.0]$            &$100$             & $ ...$  \tabularnewline
&$2528$            &\textbf{Total}  &$\mathbf{192.2~[210.7\pm 1.0]}$   &$\mathbf{100}$    & $\mathbf{...}$  \tabularnewline
\hline
&$D_s(1P'_1) $        &$D^* K$          &$22.2~[1.4\pm 1.0]$              &$100$             & $ ...$  \tabularnewline
&as~$D_{s1}(2536)$   &\textbf{Total}  &$\mathbf{22.2~[1.4\pm 1.0]}$    &$\mathbf{100}$     & $\mathbf{0.92\pm0.05}$  \tabularnewline
\hline \hline
\end{tabular}}
\end{center}
\end{table}

\begin{figure*}
\centering \epsfxsize=18.4 cm \epsfbox{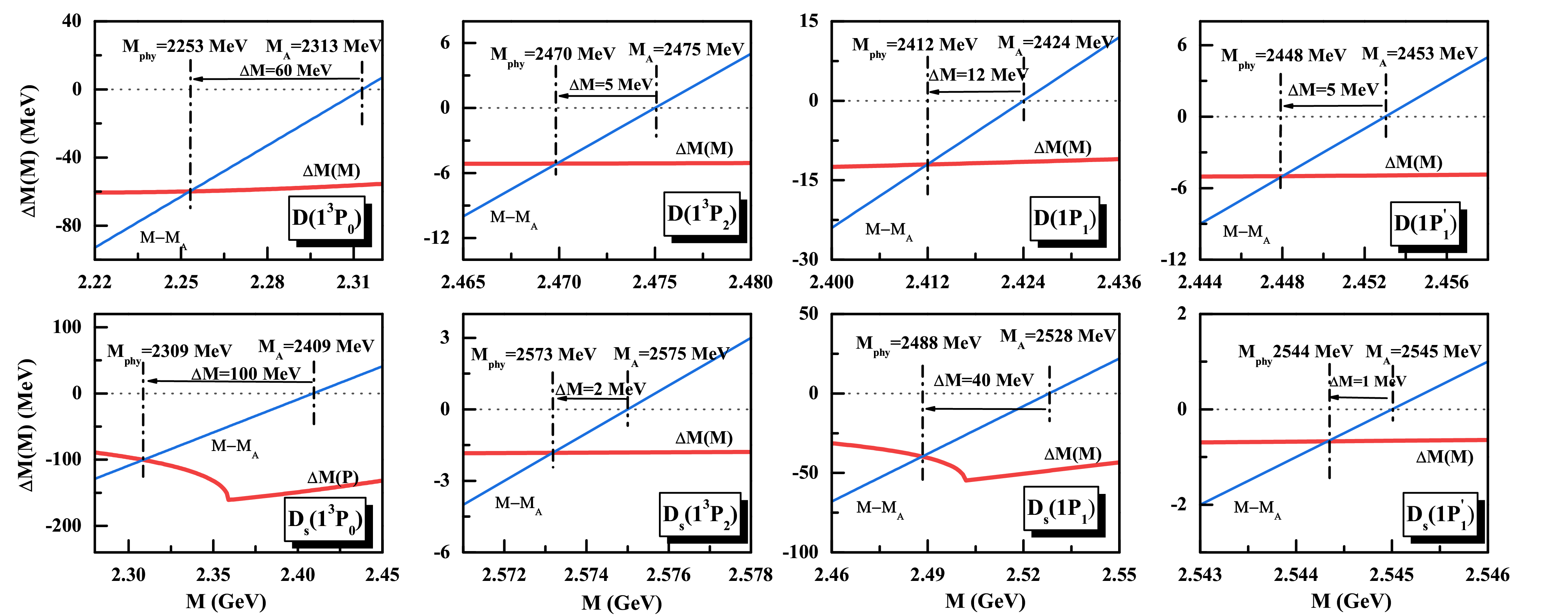} \vspace{-0.2 cm} \caption{
The determinations of the physical masses for the $1P$-wave charmed and charmed-strange meson states.
The mass shift function $\Delta M(M)$ and liner function $M-M_A$ are shown by
the thick and thin lines, respectively. $M_A$ stands for the bare mass of the $1P$-wave states,
and the physical mass $M_{phy}$ of a dressed $1P$-wave state as a solution of
the coupled-channel equation Eq.(\ref{M=MA+Delta M}) is located at the
intersection point of two solid lines.}\label{DDs1Ppurestatesmassshift}
\end{figure*}

\begin{figure}
\centering \epsfxsize=8.0 cm \epsfbox{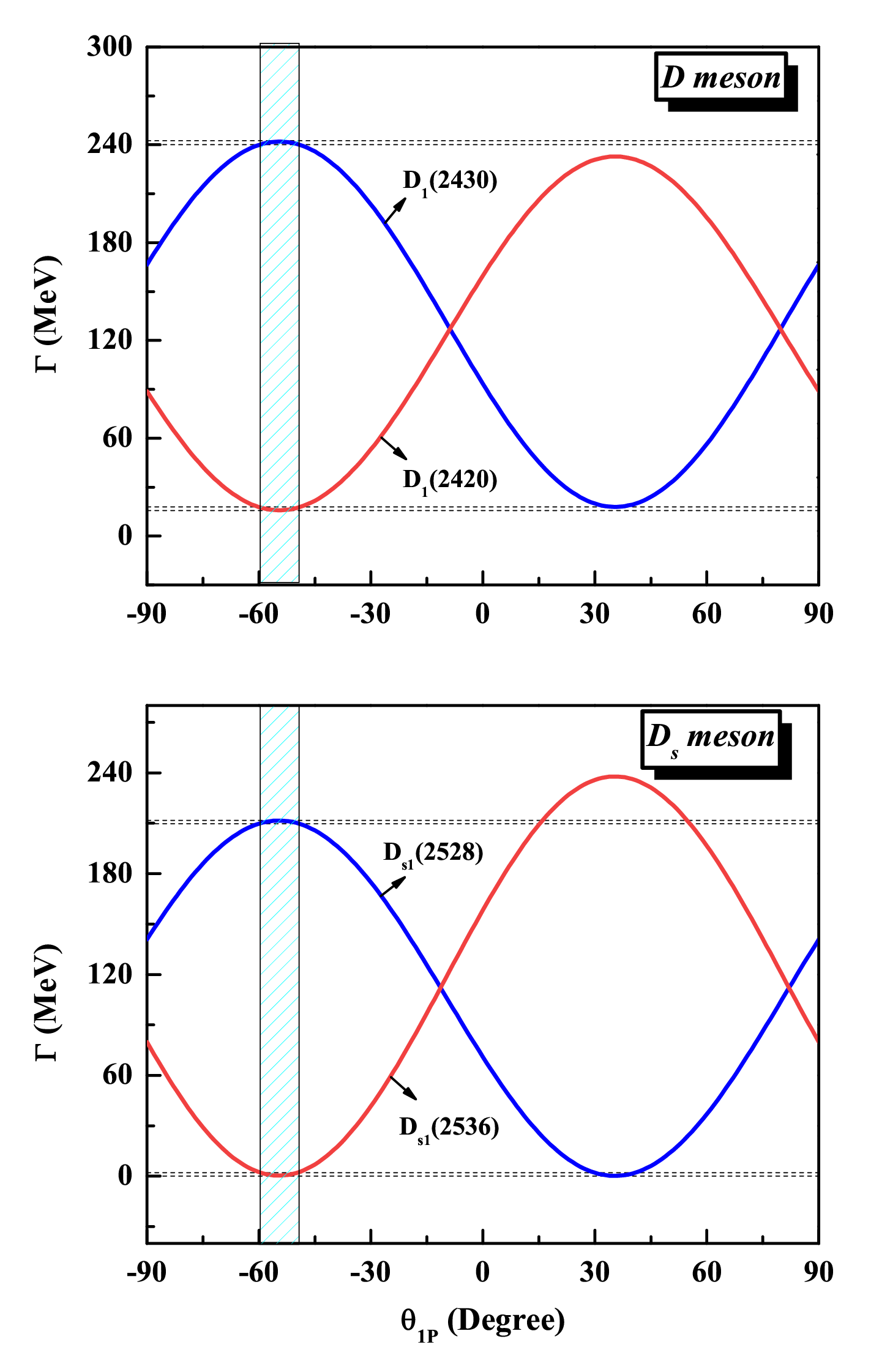} \vspace{-0.6 cm} \caption{ Decay widths of $D_{(s)}(1P_1)$ and $D_{(s)}(1P'_1)$ as functions of the mixing angle $\theta_{1P}$. In the vertical direction, the shaded region represents the possible range of the mixing angle $\theta_{1P}\simeq -(55\pm 5)^\circ$ extracted in the heavy quark symmetry limit.}\label{D1P1PpiWidthandphi}
\end{figure}

\subsubsection{$1^3P_0$  }

The broad $D_0^*(2300)$ resonance listed in RPP~\cite{Zyla:2020zbs} is generally considered to be the
$D(1^3P_0)$ state in $D$-meson family. The neutral state $D_0^*(2300)^0$ with $J^P=0^+$ was first observed
in the $D^+\pi^-$ channel by the Belle collaboration in 2003~\cite{Belle:2003nsh}, and was confirmed by the
BaBar collaboration in 2009~\cite{BaBar:2009pnd}. The charged state $D_0^*(2300)^+$
was also established in the $\bar{D}^0\pi^+$ channel by the FOCUS collaboration in 2003~\cite{FOCUS:2003gru},
and was confirmed by the LHCb collaboration in 2015~\cite{Aaij:2015sqa}.
In experiments, only the $D\pi$ channel is observed since the other OZI-allowed
two-body strong channels are forbidden. The average measured mass and width of $D_0^*(2300)$
from RPP are $M_{exp}=2343\pm 10$ MeV and $\Gamma_{exp}=229\pm 16$ MeV, respectively~\cite{Zyla:2020zbs},
which are consistent with the quark model expectations~\cite{Close:2005se,Tan:2018lao,Zhong:2008kd,Godfrey:2005ww,
DiPierro:2001dwf,Goity:1998jr,Song:2015nia,Ebert:2009ua,Lu:2014zua}.

Our predicted mass of $D(1^3P_0)$ is $M=2313$ MeV. Taking into account the $D\pi$ loop correction to
bare mass of the $D(1^3P_0)$ $c\bar{s}$ core, from Fig.~\ref{DDs1Ppurestatesmassshift}
it is seen that the physical mass is reduced to $M_{phy}=2253$~MeV, there is a mass shift of
$\Delta M\simeq 60$ MeV compared with the bare mass. The physical mass of the dressed
$D(1^3P_0)$ state is about 90 MeV lower than the PDG average mass $M_{exp}=2343\pm 10$ MeV~\cite{Zyla:2020zbs},
however, is close to measured value $\sim 2300$ MeV from Belle and BaBar experiments~\cite{Belle:2003nsh,BaBar:2009pnd}.

Taking the measured mass
$M_{exp}=2343$ MeV and the wave function extracted from the potential model, we predict that
the $D_0^*(2300)$ is a broad state with a width of
$\Gamma\simeq 540$ MeV, which is about a factor of 2.3 larger
than the average data $\Gamma_{exp}=229\pm 16$ MeV from RPP~\cite{Zyla:2020zbs}.
It should be mentioned that the mass of
$D_0^*(2300)$ measured from different collaborations is quite different,
while the measured width also bears large uncertainties.
In some recent works, the $D_0(2300)$ resonance was suggested to be a two-pole structure
in chiral dynamics~\cite{Albaladejo:2016lbb,Du:2017zvv,Du:2020pui}.
The recent Lattice calculations of the $D\pi$ scattering amplitudes obtain a complex $D_0^*$ state resonance pole
with a mass $M\simeq 2200$ MeV and a width $\Gamma\simeq 400$ MeV~\cite{Gayer:2021xzv}.
The mass and width are in contrast to the currently reported experimental results.
To better understand the nature of $D_0^*(2300)$, more accurate measurements are needed
to be carried out in future experiments.

In the $D_s$-meson sector, our predicted mass of $D_s(1^3P_0)$ is $M=2409$ MeV,
which is comparable with the predictions in the Refs.~\cite{Godfrey:2015dva,Ebert:2009ua,Lahde:1999ih}.
The mass of $D_s(1^3P_0)$ state is about 100 MeV overestimated by
the potential model if considering $D_{s0}^*(2317)$ as the $D_s(1^3P_0)$ state. The mass calculated with
lattice QCD also is significantly higher than than of $D_{s0}^*(2317)$~\cite{Moir:2013ub}.
In Ref.~\cite{Segovia:2015dia}, the study indicates that including the one-loop corrections of
the OGE potential the mass of $D_s(1^3P_0)$ will be reduced by about 130 MeV. Then, the mass
of $D_s(1^3P_0)$, $M\simeq2279$ MeV, is close to the that of $D_{s0}^*(2317)$.
The study of $DK$ scattering in full lattice QCD supports the interpretation of
the $D_{s0}^*(2317)$ as a $DK$ molecule~\cite{Liu:2012zya}.
Recently, Zhi Yang \emph{et al.} studied the positive parity $D_s$ resonant states
within the Hamiltonian effective field theory by combining it with the quark model,
they found that $D_{s0}^*(2317)$ may consist of the $D_s(1^3P_0)$ state,
but at the same time couple to the $DK$ channel~\cite{Yang:2021tvc}.

In this work, we also estimate the mass shift of $D_s(1^3P_0)$ by including the
$DK$ coupled-channel interaction within our chiral quark model.
The mass shift is determined in Fig.~\ref{DDs1Ppurestatesmassshift}.
From the figure, we can see a cusp singularity in the mass shift curve of $D_s(1^3P_0)$,
this is a typical $S$-wave mass shift, the formation mechanism was discussed in Ref.~\cite{Isgur:1998kr}.
Our result shows that the large $S$-wave coupling to $DK$ channel of $D_s(1^3P_0)$ induces a
mass shift of about $\Delta M\simeq 100$ MeV. The the physical mass of the dressed $D_s(1^3P_0)$ state
is estimated to be $M_{phy}=2309$ MeV, which is very close to the measured mass of $D_{s0}^*(2317)$.
Our coupled-channel analysis within the chiral quark model is consistent with that
in Refs.~\cite{Segovia:2015dia,Ortega:2016mms,Yang:2021tvc}. Since the mass of $D_{s0}^*(2317)$ is below the mass threshold of the
$DK$ channel, its extremely narrow width can be understood in theory.

\subsubsection{$1^3P_2$  }

The $D_2^*(2460)$ resonance is assigned as the $1^3P_2$ state of the $D$-meson family.
Our theoretical mass $M=2475$ MeV and width $\Gamma\simeq 41.3$ MeV are
in good agreement with the average measured values $M_{exp}=2461.1\pm 0.7$ MeV and
$\Gamma \simeq 47.3 \pm0.8 $ MeV from RPP~\cite{Zyla:2020zbs}. The $D_2^*(2460)$ dominantly decays into the $D\pi$ and $D^*\pi$ channels. The partial width ratio between $D\pi$ and $D^*\pi$ is predicted to be
\begin{equation}
R=\frac{\Gamma(D \pi)}{\Gamma(D^* \pi)} \simeq 1.70,
\end{equation}
which is also in good agreement with the data $R=1.52 \pm 0.14$~\cite{Zyla:2020zbs}.
The decay properties predicted in this work are consistent with our
previous predictions with the SHO wave functions~\cite{Zhong:2008kd} and other predictions in the
$^3P_0$ models~\cite{Close:2005se,Godfrey:2005ww}, and the PCAC and low energy theorem~\cite{Zhang:2016dom}.

The coupled-channel effects on the mass shift of $D(1^3P_2)$
are also studied. The results are shown in From Fig.~\ref{DDs1Ppurestatesmassshift}.
One can seen that the mass shift (i.e., $\Delta M\simeq5$ MeV) is tiny when including the $D\pi$ and $D^*\pi$ loop
corrections. There are two main reasons for the negligibly small coupled-channel contribution: (i)
the mass of $D(1^3P_2)$ is far from the $D\pi$ and $D^*\pi$ thresholds; (ii) $D(1^3P_2)$
couples to $D\pi$ and $D^*\pi$ channels via a weak $D$-wave coupling.

In the $D_s$-meson family, the $1^3P_2$ state has been well established in experiments.
The narrow resonance $D_{s2}^*(2573)$ listed in RPP~\cite{Zyla:2020zbs} should belong
to the $D_s(1^3P_2)$ state. Our theoretical mass $M=2575$ MeV and width $\Gamma\simeq 13.3$ MeV can well reproduce
the average measured values $M_{exp}=2569$ MeV and
$\Gamma \simeq 16.9 \pm0.7 $ MeV from RPP~\cite{Zyla:2020zbs}.
The $D_{s2}^*(2573)$ mainly decays into
the $DK$ channel, while the decay rate into $D^*K$ channel is sizeable.
Our predicted partial width ratio between $D^*K$ and $DK$,
\begin{eqnarray}
R= \frac{\Gamma(D^*K)}{\Gamma(DK)} \simeq 0.13,
\end{eqnarray}
is also consistent with the data $R_{exp}< 0.33$~\cite{Zyla:2020zbs}.
The decay properties predicted with the genuine wave function
from our potential model calculations in this work are consistent with our
previous predictions with the SHO wave functions~\cite{Zhong:2008kd}, and other predictions by using the
$^3P_0$ models~\cite{Close:2005se,Godfrey:2005ww} and the PCAC and low energy theorem~\cite{Zhang:2016dom}.

Finally, we also study the coupled-channel
effects on the mass of the bare $D_s(1^3P_2)$ state. From the strong decay analysis
we know that $D_s(1^3P_2)$ mainly decays into $DK$ and $D^*K$ channels.
Considering the $D_s(1^3P_2)$ $c\bar{s}$ core coupling to these channels, a tiny mass shift $\Delta M=2$~MeV can be seen in Fig.~\ref{DDs1Ppurestatesmassshift}. The tiny couple-channel effects on $D_s(1^3P_2)$
are mainly due to a weak $D$-wave coupling to $DK$ and $D^*K$ channels.
Our conclusion is consistent with that of the recent study~\cite{Yang:2021tvc}.

\subsubsection{ $1P_1$ and $1P'_1$  }

In the $1P$-wave states, the two $J^P=1^+$ states $1^1P_1 $ and $1^3P_1 $ should be
mixed with each other by the antisymmetric part of the spin-orbit potential. The physical states
$1P_1$ and $1P'_1$ states are expressed as
\begin{equation}
\left(
  \begin{array}{c}
   1P_1\\
   1P'_1\\
  \end{array}\right)=
  \left(
  \begin{array}{cc}
   \cos\theta_{1P} &\sin\theta_{1P}\\
  -\sin\theta_{1P} &\cos\theta_{1P}\\
  \end{array}
\right)
\left(
  \begin{array}{c}
  1^{1}P_{1}\\
  1^{3}P_{1}\\
  \end{array}\right).
\end{equation}
In this work, the $1P_1$ and $1P'_1$ correspond to the low-mass and high-mass mixed states, respectively.

In the $D$-meson family, the masses for the two mixed states
$D(1P_1)$ and $D(1P'_1)$ are determined to be $M=2424$ MeV and $2453$ MeV, respectively.
The mass splitting between $D(1P_1)$ and $D(1P'_1)$ is estimated to be $\Delta M\simeq 30$ MeV,
which is close to the prediction in Ref.~\cite{Ebert:2009ua}. The mixing angle
$\theta_{1P}=-34.0^{\circ}$ determined in this work is similar to the determinations
in Refs.~\cite{Lu:2014zua,Godfrey:2015dva}, however, is about a factor of $\sim 1.6$
smaller than the value $\theta_{1P}=-54.7^{\circ}$ extracted in the heavy quark limit.
The low mass state $D(1P_1)$ should  be a broad state with a width of about several
hundred MeV, while the high mass state $D(1P_1')$ is a narrow state with a width of
about several tens MeV. The $D^*\pi$ channel is the only OZI-allowed two-body
strong decay channel of $D(1P_1)$ and $D(1P'_1)$.

The $D_1(2430)$ and $D_1(2420)$ resonances listed in RPP~\cite{Zyla:2020zbs} can be assigned to the
$1P$-wave mixed state $D(1P_1)$ and $D(1P'_1)$, respectively.
For the broad resonance $D_1(2430)$, its average measured mass and width
are $M_{exp}=2412\pm 9$ MeV and $\Gamma_{exp}=314\pm 29$ MeV, respectively~\cite{Zyla:2020zbs}.
While for the narrow resonance $D_1(2420)$, its average measured mass and width
are $M_{exp}=2422.1\pm 0.8$ MeV and $\Gamma_{exp}=31.3\pm 1.9$ MeV, respectively~\cite{Zyla:2020zbs}. The average mass splitting from the measurements, $\Delta M_{exp}\simeq 10$ MeV,
is smaller than our potential model prediction $\Delta M\simeq 30$ MeV.

Taking the physical masses and predicted mixing angle $\theta_{1P}=-34^{\circ}$ for $D(1P_1)$ and $D(1P'_1)$,
we calculate the strong decay properties, our results are listed in Table~\ref{Dmeson1PDecay}.
It is seen that our predicted decay width of $D_1(2420)$ are in good agreement with
the data, however, the predicted width of $D_1(2430)$ is slightly smaller than the
lower limit of the measured width $\Gamma_{exp}=314\pm 29$ MeV. With the mixing angle
$\theta_{1P}=-54.7^{\circ}$ extracted in the heavy quark limit,
the theoretical width of $D_1(2430)$, $\Gamma\simeq 240$ MeV,
is more close to the data. The predicted width of $D_1(2420)$,
$\Gamma\simeq 17$ MeV, is slightly smaller than the PDG average value
$\Gamma_{exp}=31.3\pm 1.9$ MeV, however, is in good agreement
with the measurements, $\sim 20$ MeV, from Belle, BESIII, CDF, CLEO listed by PDG~\cite{Zyla:2020zbs}.
In Fig.~\ref{D1P1PpiWidthandphi}, we show the dependence of the decay widths of the $D_1(2430)$
and $D_1(2420)$ resonances on the mixing angle. It is found that the decay widths are
sensitive to the mixing angle. Taking the mixing angle around value extracted in the heavy quark limit,
i.e. $\theta_{1P}=-(55\pm5)^{\circ}$, the predicted decay properties of the $D_1(2430)$ and
$D_1(2420)$ resonances are consistent with the measurements.

The coupled-channel effects on the masses of the two $1^+$
states $D(1P_1)$ and $D(1P'_1)$ are further studied. Our results are shown in
Fig~\ref{DDs1Ppurestatesmassshift}. It is seen that including the $D^*\pi$ loop correction, the mass shifts of
$D(1P_1)$ and $D(1P'_1)$ are predicted to be $\Delta M\simeq 12$ and $5$~MeV, respectively.
There are only small corrections of the coupled-channel effects to
the masses of $D(1P_1)$ and $D(1P'_1)$, because their masses are far from the $D^*\pi$ threshold.

In the $D_s$-meson family, the masses for the two mixed states
$D_s(1P_1)$ and $D_s(1P'_1)$ are estimated to be $M=2528$ MeV and $M=2545$ MeV, respectively.
Their masses is very close to the mass threshold of $D^*K$.
The mass splitting between $D_s(1P_1)$ and $D_s(1P'_1)$ is estimated to be
$\Delta M\simeq 17$ MeV, which is close to the predictions in Refs.~\cite{Zeng:1994vj,Lahde:1999ih}. The mixing angle
$\theta_{1P}=-36.8^{\circ}$ is almost the same as that of the charmed sector, and is
consistent with the determinations in Ref.~\cite{Godfrey:2015dva}. Adopting this mixing angle, masses, and
wave functions determined from our potential model calculations, we study the two-body
OZI-allowed strong decays, our results are listed in Table~\ref{Dsmeson1PDecay}. It is found that the low mass state
$D_s(1P_1)$ may be a broad state with a width of $\Gamma\simeq 192$ MeV, while the high mass state
$D_s(1P'_1)$ may be a narrow state with a width of $\Gamma\simeq 22$ MeV.
The $D^*K$ channel is the only OZI-allowed two-body strong decay channel of $D_s(1P_1)$ and $D_s(1P'_1)$.

The $D_{s1}(2536)$ resonance can be assigned to the high mass state $D_s(1P'_1)$. When the mixing angle $\theta_{1P}=-36.8^{\circ}$ predicted by our potential model is taken into account, the width of $D_{s1}(2536)$ is $\Gamma \simeq 22$ MeV, which is significantly larger than $\Gamma_{exp} \simeq 0.92 \pm 0.05$ MeV. From Fig.~\ref{D1P1PpiWidthandphi}, one can find that the decay width of $D_{s1}(2536)$ is very sensitive to the mixing angle. Taking the mixing angle around value extracted in the heavy quark limit,
i.e. $\theta_{1P}=-(55\pm5)^{\circ}$, we find that the theoretical width
$\Gamma =1.4\pm1.0$ MeV is consistent with the measured width of
$\Gamma_{exp}=0.92\pm 0.05$ MeV. Our recent analysis of
$B_1(5721)$ and $B_{s1}(5830)$ also shows that as the $1P$-wave
mixed states their mixing angle more favors $\theta_{1P}=-(55\pm5)^{\circ}$~\cite{li:2021hss}.
Thus, for the heavy-light mesons, the mixing angle of the $1P$-wave states seems to be close to the
value $\theta_{1P}=-54.7^{\circ}$ extracted in the heavy quark limit.
It should be mentioned that it is still a puzzle for the mixing between
the $^1P_1$ and $^3P_1$ states, which cannot be well understood within the various potential models.

The simple potential model overestimates the mass of the low-mass $D_s(1P_1)$ state
if considering $D_{s1}(2460)$ as a candidate of it. The mass calculated with
lattice QCD also is significantly higher than that of $D_{s1}(2460)$~\cite{Moir:2013ub}.
The coupled-channel effects may play an important role because of the closeness
behavior for the bare $c\bar{s}$ and $D^*K$ threshold.
Recently, Zhi Yang \emph{et al.} studied these effects within the Hamiltonian effective field theory
by combining it with the quark model~\cite{Yang:2021tvc}. They found that the $D_{s1}(2460)$ resonance may
consist of the $D_s(1P_1)$ state, but at the same time couples to the $D^*K$ channel~\cite{Yang:2021tvc}.
In the present work, including the $D^{*}K$ loop we also study the coupled-channel effects on the mass
shifts of the two $1^+$ $D_s$ states. Our results are shown in Fig.~\ref{DDs1Ppurestatesmassshift}.
It is found that there is a $\sim 40$ MeV correction to the bare mass of
$D_s(1P_1)$ due to a strong $S$-wave $D^{*}K$ interaction. The physical mass of the dressed
$D_s(1P_1)$ is reduced to $M_{phy}=2488$~MeV, which is close to the measured mass of $D_{s1}(2460)$.
It should be mentioned that coupled-channel effects on $D_{s1}(2536)$ are negligibly small due to its
weak coupling with the $D^*K$ channel. Our coupled-channel analysis within the chiral quark model is consistent with that
in Ref.~\cite{Yang:2021tvc}. Since the mass of $D_{s1}(2460)$ is below the $D^{*}K$ threshold,
it becomes an extremely narrow state.

\subsection{$2P$-wave states}

\begin{table*}[htp]
\begin{center}
\caption{Partial decay widths and their branching fractions for the $2P$-wave charmed mesons. }
\label{DDecay2Pwave}
\scalebox{1.00}{
\begin{tabular}{cccccccccccccccccccccccccc}
\hline\hline
&
~~~~~&\multicolumn{2}{c}{$\underline{~~~~~~D(2^3P_0)[2849]~~~~~~}$}      ~~~~~&\multicolumn{2}{c}{$\underline{~~~~~~D(2^3P_2)[2955]~~~~~~}$}
~~~~~&\multicolumn{2}{c}{$\underline{~~~~~~D(2P_1)[2900]~~~~~~}$}
~~~~~&\multicolumn{2}{c}{$\underline{~~~~~~D(2P'_1)[2936]~~~~~~}$}
~~~~~\\
&Channel &$\Gamma_i$~(MeV) &Br~(\%) &$\Gamma_i$~(MeV) &Br~(\%) &$\Gamma_i$~(MeV) &Br~(\%) &$\Gamma_i$~(MeV)    &Br~(\%)\\
\hline
& $D\pi$              &609.8    &56.6       &52.0    &26.9     &--      &--      &--     &--   \tabularnewline
& $D_sK$              &190.8    &17.7       &6.0     &3.1      &--      &--      &--     &--   \tabularnewline
& $D\eta$             &82.8     &7.7        &2.9     &1.5      &--      &--      &--     &--   \tabularnewline
& $D\eta'$            &17.8     &1.7        &0.03    &0.02     &--      &--      &--     &--   \tabularnewline
& $D^*\pi$            &--      &--        &23.7    &12.3     &199.5    &31.4     &77.2    &32.0   \tabularnewline
& $D_s^*K$            &--      &--        &1.0     &0.5      &53.6     &8.4      &21.3    &8.8   \tabularnewline
& $D^*\eta$           &--      &--        &0.5     &0.3      &27.6     &4.3      &10.6     &4.4  \tabularnewline
& $D_0(2550)\pi$      &151.0    &14.0      &7.5     &3.9      &--      &--       &--     &--   \tabularnewline
& $D(2^3S_1)(2627)\pi$    &--      &--        &5.9     &3.1      &88.9     &14.0      &45.6    &18.9   \tabularnewline
& $D_0^*(2300)\pi$    &--      &--        &--     &--      &22.8     &3.6       &2.0     &0.8   \tabularnewline
& $D_{s}(1^3P_0)(2409)K$               &--       &--    &--     &--   &--    &--    &$6\times10^{-4}$    &$2\times10^{-4}$   \tabularnewline
& $D_0^*(2300)\eta$   &--      &--        &--     &--      &0.7      &0.1     &$1\times10^{-5}$   &$4\times10^{-6}$   \tabularnewline
& $D_2^*(2460)\pi$    &--      &--        &9.4     &4.9      &21.5     &3.4     &23.6    &9.8   \tabularnewline
& $D_1(2430)\pi$      &2.5      &0.2        &4.1     &2.1      &3.2      &0.5       &11.3    &4.7   \tabularnewline
& $D_1(2420)\pi$      &14.7     &1.4        &6.4     &3.3      &2.2      &0.3       &1.8     &0.7    \tabularnewline
& $D\rho$             &--      &--        &14.6    &7.5      &138.4    &21.8     &4.7     &1.9   \tabularnewline
& $D\omega$           &--      &--        &4.3     &2.2      &44.4     &7.0      &1.7     &0.7   \tabularnewline
& $D_s K^*$           &--      &--        &0.4     &0.2      &25.6      &4.0      &5.1     &2.1  \tabularnewline
& $D^*\rho$           &6.5      &0.6        &41.4    &21.4     &5.0      &0.8      &28.7    &11.9   \tabularnewline
& $D^*\omega$         &2.3      &0.2        &13.3    &6.9      &1.4      &0.2      &7.7     &3.2   \tabularnewline
\hline
& \textbf{Total}      &$\mathbf{1078.2}$    &$\mathbf{100}$    &$\mathbf{193.4}$    &$\mathbf{100}$   &$\mathbf{634.8}$    &$\mathbf{100}$   &$\mathbf{241.3}$    &$\mathbf{100}$   \tabularnewline
\hline\hline
\end{tabular}}
\end{center}
\end{table*}

\begin{table*}[htp]
\begin{center}
\caption{Partial decay widths and their branching fractions for the $2P$-wave charmed-strange mesons. }
\label{DsDecay2Pwave}
\scalebox{1.00}{
\begin{tabular}{ccccccccccccccccccccccccccc}
\hline\hline
&
~~~~~&\multicolumn{2}{c}{$\underline{~~~~~~D_s(2^3P_0)[2940]~~~~~~}$}      ~~~~~&\multicolumn{2}{c}{$\underline{~~~~~~D_s(2^3P_2)[3053]~~~~~~}$}
~~~~~&\multicolumn{2}{c}{$\underline{~~~~~~D_s(2P_1)[3002]~~~~~~}$}
~~~~~&\multicolumn{2}{c}{$\underline{~~~~~~D_s(2P'_1)[3026]~~~~~~}$}
~~~~~\\
&Channel &$\Gamma_i$~(MeV) &Br~(\%) &$\Gamma_i$~(MeV) &Br~(\%) &$\Gamma_i$~(MeV) &Br~(\%) &$\Gamma_i$~(MeV)    &Br~(\%)\\
\hline
&~~ $DK$                   &549.8    &78.3     &30.9    &29.8     &--      &--       &--     &--   \tabularnewline
&~~ $D_s\eta$              &117.4    &16.7     &3.0     &2.9      &--      &--       &--     &--   \tabularnewline
&~~ $D_s\eta'$             &18.0     &2.6      &0.03    &0.03     &--      &--       &--     &--   \tabularnewline
&~~ $D^*K$                 &--      &--        &9.1     &8.8      &176.5   &44.4     &77.4    &56.7   \tabularnewline
&~~ $D_s^*\eta$            &--      &--        &0.4     &0.4      &36.7    &9.2      &16.6    &12.2   \tabularnewline
&~~ $D_0^*(2300)K$         &--      &--        &--     &--      &18.7      &4.7      &0.2     &0.1   \tabularnewline
&~~ $D_{s}(1^3P_0)(2409)\eta$      &--      &--        &--  &--      &0.6     &0.2       &$4\times10^{-3}$     &$3\times10^{-3}$  \tabularnewline
&~~ $D_2^*(2460)K$        &--      &--       &2.0     &1.9     &11.6     &2.9      &4.0     &2.9   \tabularnewline
&~~ $D_1(2430)K$          &0.5     &0.07    &6.7     &6.5      &0.6      &0.2      &4.4     &3.2   \tabularnewline
&~~ $D_1(2420)K$          &3.9     &0.6     &2.1     &2.0      &1.1      &0.3      &7.8     &5.7 \tabularnewline
&~~ $DK^*$                &--      &--       &10.2    &9.8      &137.9    &34.7      &5.2     &3.8   \tabularnewline
&~~ $D_s\phi$             &--      &--       &0.09    &0.09     &11.2     &2.8       &2.5     &1.8   \tabularnewline
&~~ $D^*K^*$              &13.0    &1.9      &39.4    &38.0     &2.5      &0.6       &18.5    &13.5   \tabularnewline
\hline
&~~ \textbf{Total}      &$\mathbf{702.6}$    &$\mathbf{100}$    &$\mathbf{103.8}$    &$\mathbf{100}$   &$\mathbf{397.4}$    &$\mathbf{100}$   &$\mathbf{136.6}$    &$\mathbf{100}$   \tabularnewline
\hline\hline
\end{tabular}}
\end{center}
\end{table*}

\begin{table}[htp]
\begin{center}
\caption{ Partial and total decay widths~(MeV) of $D_J^*(3000)$ as the $2^3P_2$ state.}
\label{DJStars(3000)as23P2}
\scalebox{1.0}{
\begin{tabular}{llcccccccccccccccccccccc}
\hline\hline
& $D  \pi$  &  $D_s K$    &$ D \eta$  &$ D \eta'$ &$ D_0(2550)\pi$  \tabularnewline
& $ 67.7$   &  $ 8.9$     &$ 4.2$     &$ 0.1$     &$ 13.1$           \tabularnewline
\cline{1-6}
$   $  &  $ D^* \pi $  & $ D_s^* K$  & $ D^* \eta$   & $ D(2^3S_1)(2627) \pi $  & $ D_2^*(2460) \pi $\tabularnewline
$   $  &  $ 34.9$      & $ 2.1$       & $ 1.0$       & $ 12.6$           & $ 17.5$   \tabularnewline
\cline{1-6}
$   $   & $ D_1(2430) \pi $   & $ D_1(2420)\pi $  & $ D \rho $  & $ D \omega $  & $ D_s K^* $ \tabularnewline
$   $   & $ 5.1$              & $ 10.2 $          & $ 23.2$     & $ 7.1$        & $ 1.4 $      \tabularnewline
\cline{1-6}
$   $    & $ D^* \rho $  & $ D^* \omega $    &  \textbf{Total}   &   & \tabularnewline
$   $    & $ 47.9$       & $ 15.5$           & $\mathbf{272.5}$     &   & \tabularnewline
\hline\hline
\end{tabular}}
\end{center}
\end{table}

\begin{figure}
\centering \epsfxsize=8.0 cm \epsfbox{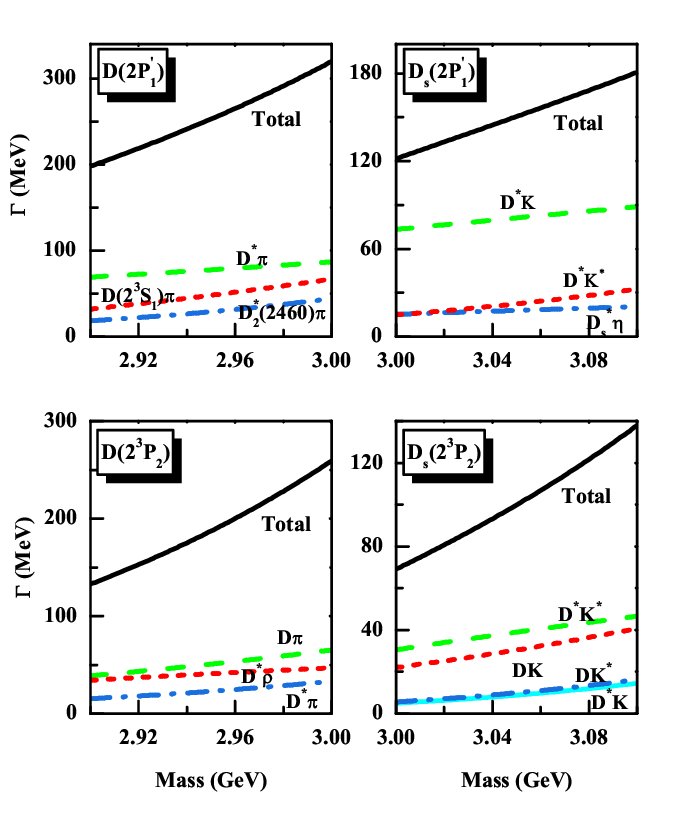} \vspace{-0 cm} \caption{Total decay widths and the main partial decay widths for $D(2P'_1)$, $D_s(2P'_1)$, $D(2^3P_2)$ and $D_s(2^3P_2)$ as functions of their masses.}\label{DDs2PwaveMirunning}
\end{figure}

\subsubsection{$2^3P_0$  }

In the $D$-meson family, our predicted mass for the $D(2^3P_0)$ state, $M=2849$ MeV,
is comparable with the predictions in Refs.~\cite{Liu:2015uya,Liu:2016efm,
Ebert:2009ua,Song:2015fha}, however, is about 100 MeV larger than the predictions
in Refs.~\cite{Zeng:1994vj,Li:2010vx,Lahde:1999ih}. The mass gap between $D(2^3P_0)$
and $D(1^3P_0)$ is estimated to be $\Delta M\simeq 540$ MeV, which is consistent with
those predicted in Refs.~\cite{Godfrey:2015dva,Ebert:2009ua,Zeng:1994vj}.
The $D(2^3P_0)$ may be a very broad state with a width of
\begin{equation}
\Gamma\simeq 1080  \ \ \ \mathrm{MeV },
\end{equation}
and dominantly decays into the
$D\pi$, $D_s K$ and $D_0(2550)\pi$ final states. More details of the decay properties
can be seen in Table~\ref{DDecay2Pwave}. In Refs.~\cite{Xiao:2014ura,Lu:2014zua}, the
$D(2^3P_0)$ is also predicted to be a very broad state with a width of $\Gamma\simeq600-800$ MeV, although
the predicted partial widths show some model dependencies.

In Ref.~\cite{Sun:2013qca}, the authors suggested that the natural parity state $D_J^*(3000)$ observed in the
$D\pi$ final state by the LHCb collaboration~\cite{LHCb:2013jjb} may be explained with $D(2^3P_0)$ according to
their mass and strong decay analysis. However, both our predicted mass and width of $D(2^3P_0)$, $M=2849$ MeV
and $\Gamma\simeq 1080$ MeV, are inconsistent with the data $M_{exp}=3008.1\pm4.0$ MeV
and $\Gamma_{exp} =110.5 \pm11.5$ MeV~\cite{LHCb:2013jjb}. According to our prediction, the $D(2^3P_0)$ may be difficult
to be established in experiments due to its rather broad width.

In the $D_s$-meson sector, our predicted mass for the $D_s(2^3P_0)$,
$M=2940$ MeV, is comparable with the predictions in Refs.~\cite{Liu:2016efm,Zeng:1994vj,Lahde:1999ih}.
Our predicted strong decay properties have been shown in Table~\ref{DsDecay2Pwave}. It is found that
the $D_s(2^3P_0)$ may be a broad state with a width of
\begin{equation}
\Gamma\simeq 700  \ \ \ \mathrm{MeV },
\end{equation}
and dominantly decays into
the $DK$ and $D_s \eta$ final states. The width predicted in the present work is
notably than our previous prediction $\Gamma \simeq 150-200$ MeV with a SHO wave
function~\cite{Xiao:2014ura}, which indicates that the decay properties of $D_s(2^3P_0)$ are very
sensitive to the details of its wave function adopted in the calculations.

It should be mentioned that there are strong model dependencies
in the strong decay predictions. For example, within the $^3P_0$ model
the $D(2^3P_0)$ and $D_s(2^3P_0)$ states may be relatively narrow
states with a width of $\sim 100-200$ MeV~\cite{Godfrey:2015dva}.

\subsubsection{$2^3P_2$ }

In the $D$-meson family, our predicted mass of $D(2^3P_2)$ is $M=2955$ MeV,
which is consistent with the predictions in Refs.~\cite{Li:2010vx,Liu:2015uya,Liu:2016efm,Godfrey:2015dva,Zeng:1994vj}.
Our predicted strong decay properties have been shown in Table~\ref{DDecay2Pwave}.
We find that the $D(2^3P_2)$ state is a relatively narrow state with a width of
\begin{equation}
\Gamma\simeq 190 \ \ \ \mathrm{MeV },
\end{equation}
and dominantly decays into the $D\pi $ and $D^*\rho $ final states
with branching fractions $26.9\%$ and $21.4\%$, respectively. There are some differences between the
predictions in this work and those obtained with the SHO wave function
in our previous work~\cite{Xiao:2014ura}, where the decay rate into $D\pi $ is tiny.
It indicates that the decay properties are sensitive to the details of the wave function of $D(2^3P_2)$.
Furthermore, to see the dependence of the decay properties of $D(2^3P_2)$ on its mass,
we also plot the main partial widths and the total width as functions of the mass in
Fig.~\ref{DDs2PwaveMirunning}. It is found that the partial and total decay widths increase smoothly with the
mass. With a mass uncertainty of $50$ MeV, the total width of $D(2^3P_2)$ varies in the range of $\sim 140-260$ MeV.
The decay properties predicted within various models show large model dependencies.
Within the $^3P_0$ model the width of $D(2^3P_2)$ is predicted to be in the range of $\Gamma \simeq15 - 120$ MeV~\cite{Godfrey:2015dva,Sun:2013qca,Song:2015fha}.

From the point of view of mass, the $D_J^*(3000)$ resonance with a natural parity observed by the LHCb collaboration in
2013~\cite{LHCb:2013jjb} might be a candidate of the $D(2^3P_2)$ state. Our predicted mass $M=2955$ MeV of $D(2^3P_2)$ is close to the
measured value $M_{exp}=3008.1\pm4.0$ MeV of $D_J^*(3000)$. Taking $D_J^*(3000)$ as the $D(2^3P_2)$ state,
we further study its strong decay properties, our results are listed in Table~\ref{DJStars(3000)as23P2}.
It is found that $D_J^*(3000)$ should dominantly decay into the $D\pi$ channel with a branching fraction
of $25\%$, which is consistent with the observations. However, our predicted width,
\begin{equation}
\Gamma\simeq 270 \ \ \ \mathrm{MeV },
\end{equation}
is notably larger than the data $\Gamma_{exp} =110.5 \pm11.5$ MeV measured by LHCb~\cite{LHCb:2013jjb}.
The $D_J^*(3000)$ as the $D(2^3P_2)$ assignment is suggested in Refs.~\cite{Gandhi:2019lta,Liu:2013maa,
Liu:2015uya,Liu:2016efm}. If $D_J^*(3000)$ corresponds to $D(2^3P_2)$ indeed, it may have a large decay rate
into the $D\pi $ channel as well, the partial width ratio between $D^*\pi $ and $D\pi $ is predicted to be
\begin{equation}
R=\frac{\Gamma(D^* \pi)}{\Gamma(D \pi)} \simeq 0.52,
\end{equation}
which may be useful to test the nature of $D_J^*(3000)$.

In the $D_s$-meson sector, the mass of $D_s(2^3P_2)$ is predicted to be
$M=3053$ MeV, which is comparable with the predictions in
Refs.~\cite{Godfrey:2015dva,Ebert:2009ua,Li:2010vx}. Our predicted strong decay properties
have been shown in Table~\ref{DsDecay2Pwave}. It is found that the $D_s(2^3P_2)$ may be a
relatively narrow state with a width of
\begin{equation}
\Gamma\simeq 100 \ \ \ \mathrm{MeV },
\end{equation}
and dominantly decays into the
$D^*K^*$ and $DK$ final states with branching fractions $\sim38\%$ and $\sim30\%$
respectively. There may be a sizeable decay rate ($\sim10\%$) into the $DK^*$ channel
as well. The $DK$, $DK^*$ and $D^*K^*$ are also predicted to be the main
decay channels in the other works~\cite{Godfrey:2015dva,Ferretti:2015rsa}, although the predicted partial
width ratios are very different with each other. There are some differences between the
predictions in this work and those obtained with the SHO wave function
in our previous work~\cite{Xiao:2014ura}, where the decay rate into $DK$ is tiny.
To see the dependence of the decay properties of $D_s(2^3P_2)$ on its mass,
we also plot the main partial widths and the total width as functions of the mass in
Fig.~\ref{DDs2PwaveMirunning}. It is found that the partial and total decay widths increase smoothly with the
mass. With a mass uncertainty of $50$ MeV, the total width of $D_s(2^3P_2)$ varies in the range of $\sim 70-140$ MeV.
It should be mentioned that the typical mass gap between
$D_s(2^3P_2)$ and $D(2^3P_2)$ is around 100 MeV. If $D_J^*(3000)$ corresponds to $D(2^3P_2)$ indeed,
the mass of $D_s(2^3P_2)$ is most likely to be $\sim 3100$ MeV.
Searching for the missing $D_s(2^3P_2)$ may be helpful to understand the nature of $D_J^*(3000)$.
To establish the missing $D_s(2^3P_2)$ state, the
main decay channels, such as $DK^*$ and $DK$, are worth to observing in future experiments.

\subsubsection{$2P_1$ and $2P'_1$  }


The physical states $2P_1$ and $2P'_1$ are mixed states between states $2^1P_1 $ and $2^3P_1 $
via the following mixing scheme:
\begin{equation}
\left(
  \begin{array}{c}
   2P_1\\
   2P'_1\\
  \end{array}\right)=
  \left(
  \begin{array}{cc}
   \cos\theta_{2P} &\sin\theta_{2P}\\
  -\sin\theta_{2P} &\cos\theta_{2P}\\
  \end{array}
\right)
\left(
  \begin{array}{c}
  2^{1}P_{1}\\
  2^{3}P_{1}\\
  \end{array}\right).
\end{equation}
In this work, the $2P_1$ and $2P'_1$ correspond to the low-mass and high-mass mixed states, respectively.

In the $D$-meson family, the masses for the two mixed states
$D(2P_1)$ and $D(2P'_1)$ are determined to be $M=2900$ MeV and $M=2936$ MeV, respectively.
The mass splitting between $D(2P_1)$ and $D(2P'_1)$ is estimated to be $\Delta M\simeq 36$ MeV,
which is close to the predictions in Refs.~\cite{Li:2010vx,Godfrey:2015dva}. The mixing angle
$\theta_{2P}=-23.5^{\circ}$ determined in this work is similar to the determinations
in Refs.~\cite{Lu:2014zua,Godfrey:2015dva}.
Our predicted strong decay properties have been shown in Table~\ref{DDecay2Pwave}.
The low mass state $D(2P_1)$ should be a broad state with a width of about $\Gamma\sim 600$ MeV,
while the high mass state $D(2P_1')$ is a relatively narrow state with a width of
\begin{equation}
\Gamma\simeq 240 \ \ \ \mathrm{MeV }.
\end{equation}
Both $D(2P_1)$ and $D(2P'_1)$ have large decay rates into
the $D^*\pi$, $D^*\eta$, $D_s^*K$, and $D(2^3S_1)\pi$ channels.
To see the dependence of the decay properties of $D(2P_1')$ on its mass,
we also plot the main partial widths and the total width as functions of the mass in
Fig.~\ref{DDs2PwaveMirunning}. It is found that the partial and total decay widths increase smoothly with the
mass. With a mass uncertainty of $50$ MeV, the total width of $D(2P_1')$ varies in the range $\sim 200-300$ MeV.
It should be mentioned that the predicted decay properties strongly depend on the approaches adopted in
the literature~\cite{Xiao:2014ura,Lu:2014zua,Godfrey:2015dva,Song:2015fha}.

It is interesting to find that the $D_J(3000)^0$ resonance observed in the $D^{*+}\pi^-$ channel at LHCb~\cite{LHCb:2013jjb}
might be a good candidate of the high mass mixed state $D(2P'_1)$. With this assignment, both our predicted mass and width
are consistent with the data $M_{exp}=2972\pm 9$ MeV and $\Gamma_{exp}=188 \pm45$ MeV~\cite{LHCb:2013jjb}.
The $D(2P'_1)$ mainly decays into $D^*\pi$ and $D(2^3S_1)\pi$ channels, which can naturedly explain why
$D_J(3000)^0$ is first observed in the $D^{*+}\pi^-$ channel. It should be mentioned that in our previous work
the study with the SHO wave function shows that $D_J(3000)^0$ may favor the low mass mixed state
$D(2P_1)$~\cite{Xiao:2014ura}. However, in the present work, with this assignment both our predicted mass and width are
inconsistent with the observations. To further clarify the nature of
$D_J(3000)^0$, the other decay modes, such as $D^*\eta$, $D_s^*K$, and $D(2^3S_1)\pi$, are worth to
observing in future experiments.

In the $D_s$-meson sector, the masses of the two mixed states $D_s(2P_1)$ and $D_s(2P'_1)$ are predicted to be $M=3002$ MeV and $M=3026$ MeV, respectively. The mass splitting between $D_s(2P_1)$ and $D_s(2P'_1)$ is estimated to be $\Delta M\simeq 24$ MeV, which is consistent with that of Ref.~\cite{Godfrey:2015dva}. The mixing angle $\theta_{2P}=-21.0^{\circ}$ is similar to that for the charmed sector.
Our predicted strong decay properties have been shown in Table~\ref{DsDecay2Pwave}.
It is seen that the low mass state $D_s(2P_1)$ should be a broad state with a width of
\begin{equation}
\Gamma\simeq 400 \ \ \ \mathrm{MeV },
\end{equation}
and dominantly decays into $D^*K$, $DK^*$, and $D^*_s\eta$ channels.
While the high mass state $D_s(1P_1')$ is a relatively narrow state with a width of
\begin{equation}
\Gamma\simeq 140 \ \ \ \mathrm{MeV },
\end{equation}
and dominantly decays into $D^*K$ and $D^*_s\eta$ channels.
Finally, it should mentioned that if the $D_J(3000)^0$ corresponds to the high mass mixed state $D(2P'_1)$,
by combining the typical mass splitting $\Delta M\simeq 100$ MeV between the charmed and charmed-strange mesons, as its flavor partner,
the mass of  $D_s(2P'_1)$ is estimated to be $M=3072$ MeV,
while the low mass state $D_s(2P_1)$ may have a mass around $M=3050$ MeV.
To see the dependence of the decay properties of $D_s(2P_1')$ on its mass,
we also plot the main partial widths and the total width as functions of the mass in
Fig.~\ref{DDs2PwaveMirunning}. It is found that the partial and total decay widths increase smoothly with the
mass. With a mass uncertainty of $50$ MeV, the total width of $D_s(2P_1')$ varies in the range $\sim 120-180$ MeV.

In 2009, the $D_{sJ}(3040)^+$ resonance with mass $M=3044\pm8^{+30}_{-5}$ MeV and width $\Gamma=239\pm35^{+46}_{-42}$ MeV
was observed in the $D^*K$ channel by the BaBar collaboration~\cite{BaBar:2009rro}.
The $D_{sJ}(3040)^+$ may be favor the $2P$ mixed states in the $D_s$-meson family~\cite{Godfrey:2015dva,Sun:2009tg,
Chen:2009zt,Xiao:2014ura,Song:2015nia,Colangelo:2010te,Li:2017zng}. Comparing our predicted mass and
decay properties with the data, we find that the $D_{sJ}(3040)$ seems to more favor the
low mass state $D_s(2P_1)$, however, the assignment of the $D_s(2P_1')$ cannot be
excluded due to the large uncertainties of the data. The partial width ratio
between $D^*K$ and $DK^*$ can be used to test the nature of $D_{sJ}(3040)^+$.
For the $D_s(2P_1)$ assignment, the partial width ratio is predicted to be
\begin{equation}
R=\frac{\Gamma(D^*K)}{\Gamma(DK^*)} \simeq 1.28 ,
\end{equation}
which is different from the value $R\simeq 14.9$ for the $D_s(2P'_1)$ assignment.
It should be emphasized that the $D_{sJ}(3040)^+$ observed in the $D^*K$ channel may be contributed by
both $D_s(2P_1)$ and $D_s(2P'_1)$, since these two states have similar masses and dominantly
decay into $D^*K$ channel with a large branching fraction of $\sim (40-60)\%$.
To establish the $D_s(2P_1)$ and $D_s(2P'_1)$ states and uncover the nature
of $D_{sJ}(3040)^+$, more accurate observations in these dominant channels,
such as $D^*K$, $D^*_s\eta$ and $DK^*$, are needed in future experiments.

\subsection{$1D$-wave states }

\begin{table*}[htp]
\begin{center}
\caption{Partial decay widths and their branching fractions for the $1D$-wave charmed mesons.}
\label{Dmeson1DDecay}
\scalebox{1.00}{
\begin{tabular}{ccccccccccccccccccccccccc}
\hline\hline
&
~~~~~&\multicolumn{2}{c}{$\underline{~~~~~~D(1^3D_1)[2754]~~~~~~}$}      ~~~~~&\multicolumn{2}{c}{$\underline{~~~~~~D(1^3D_3)~as~D_3^*(2750)~~~~~~}$}
~~~~~&\multicolumn{2}{c}{$\underline{~~~~~~D(1D_2)[2755]~~~~~~}$}
~~~~~&\multicolumn{2}{c}{$\underline{~~~~~~D(1D'_2)~as~D_2(2740)~~~~~~}$}
~~~~~\\
&Channel &$\Gamma_i$~(MeV) &Br~(\%) &$\Gamma_i$~(MeV) &Br~(\%) &$\Gamma_i$~(MeV) &Br~(\%) &$\Gamma_i$~(MeV)    &Br~(\%)\\
\hline
&$D\pi$              &138.9    &29.9                 &18.5               &39.3                    &--      &--                 &--     &--   \tabularnewline
& $D_sK$              &33.3     &7.2                  &1.0                &2.1                   &--      &--                  &--     &--   \tabularnewline
& $D\eta$             &18.3     &3.9                  &0.8                &1.7                   &--      &--                  &--     &--   \tabularnewline
& $D^*\pi$            &45.1     &9.7                  &17.5               &37.2                  &91.6     &35.4                 &30.4    &53.0  \tabularnewline
& $D_s^*K$            &6.8      &1.5                  &0.3                &0.6                   &17.5     &6.8                  &0.9     &1.6   \tabularnewline
& $D^*\eta$           &4.7      &1.0                  &0.2                &0.4                   &12.0     &4.6                  &0.8     &1.4   \tabularnewline
& $D_0(2550)\pi$      &0.01      &$2\times10^{-3}$        &$5\times10^{-4}$   &$1\times10^{-3}$        &--      &--                 &--     &--   \tabularnewline
& $D(2^3S_1)(2627)\pi$    &--      &--                   &$3\times10^{-7}$   &$6\times10^{-7}$        &--      &--                  &--     &--   \tabularnewline
& $D_0^*(2300)\pi$    &--      &--                   &--                &--                     &0.02     &$8\times10^{-3}$     &8.1     &14.1   \tabularnewline
& $D_2^*(2460)\pi$    &1.4      &0.3                   &3.3                &7.0                     &117.1    &45.3                &4.4     &7.7   \tabularnewline
& $D_1(2430)\pi$      &26.4     &5.7                   &5.0                &10.6                    &0.5      &0.2                 &0.1     &0.2  \tabularnewline
& $D_1(2420)\pi$      &189.6    &40.6                  &0.03               &0.06                    &0.2      &0.08                 &5.2     &9.1    \tabularnewline
& $D\rho$             &0.2      &0.04                  &0.4                &0.8                     &15.2     &5.9                &5.8     &10.1   \tabularnewline
& $D\omega$           &0.04     &$9\times10^{-3}$      &0.09               &0.2                     &4.3      &1.7                 &1.7     &3.0   \tabularnewline
\hline
& \textbf{Total}      &$\mathbf{464.7}$    &$\mathbf{100}$    &$\mathbf{47.1}$    &$\mathbf{100}$   &$\mathbf{258.4}$    &$\mathbf{100}$   &$\mathbf{57.4}$    &$\mathbf{100}$   \tabularnewline
\hline\hline
\end{tabular}}
\end{center}
\end{table*}

\begin{table*}[htp]
\begin{center}
\caption{Partial decay widths and their branching fractions for the $1D$-wave charmed-strange mesons.}
\label{Dsmeson1DDecay}
\scalebox{1.00}{
\begin{tabular}{ccccccccccccc}
\hline\hline
&
~~~~~&\multicolumn{2}{c}{$\underline{~~~~~~D_s(1^3D_1)~as~D_{s1}^*(2860)~~~~~~}$}      ~~~~~&\multicolumn{2}{c}{$\underline{~~~~~~D_s(1^3D_3)~as~D_{s3}^*(2860)~~~~~~}$}
~~~~~&\multicolumn{2}{c}{$\underline{~~~~~~D_s(1D_2)[2857]~~~~~~}$}
~~~~~&\multicolumn{2}{c}{$\underline{~~~~~~D_s(1D'_2)[2911]~~~~~~}$}
~~~~~\\
&Channel &$\Gamma_i$~(MeV) &Br~(\%) &$\Gamma_i$~(MeV) &Br~(\%) &$\Gamma_i$~(MeV) &Br~(\%) &$\Gamma_i$~(MeV)    &Br~(\%)\\
\hline
&~~ $DK$                   &132.4    &63.7       &11.8    &54.4     &--                 &--                &--     &--   \tabularnewline
&~~ $D_s\eta$              &25.3     &12.2       &0.8     &3.7      &--                 &--                 &--     &--   \tabularnewline
&~~ $D^*K$                 &43.7     &21.0        &8.6     &39.6     &107.1               &81.5                &28.4    &64.5   \tabularnewline
&~~ $D_s^*\eta$            &6.4      &3.1        &0.3     &1.4      &10.5                &8.0                 &1.8     &4.1   \tabularnewline
&~~ $D_0^*(2300)K$         &--      &--        &--     &--      &$6\times 10^{-3}$   &$5\times 10^{-3}$   &3.2     &7.3\tabularnewline
&~~ $DK^*$                 &0.1      &0.05        &0.2     &0.9      &13.8                &10.5                &10.6    &24.1 \tabularnewline
\hline
&~~ \textbf{Total}        &$\mathbf{207.9}$    &$\mathbf{100}$    &$\mathbf{21.7}$    &$\mathbf{100}$   &$\mathbf{131.4}$    &$\mathbf{100}$   &$\mathbf{44.0}$    &$\mathbf{100}$   \tabularnewline
\hline\hline
\end{tabular}}
\end{center}
\end{table*}

\subsubsection{$1^3D_1$  }

In the $D$-meson family, our predicted mass of $D(1^3D_1)$, $M=2754$ MeV, is
consistent with that predicted in Refs.~\cite{Ebert:2009ua,Li:2010vx,Lahde:1999ih,Song:2015fha}.
From Table \ref{Dmeson1DDecay}, it is found that the $D(1^3D_1)$ may be a broad state with a width
of
\begin{equation}
\Gamma\simeq 460 \ \ \ \mathrm{MeV },
\end{equation}
and dominantly decays into the
$D_1(2420)\pi$ and $D \pi$ channels with branching fractions $\sim41\%$ and $\sim30\%$, respectively.
Furthermore, the decay rates into $D_sK$, $D^*\pi$ and $D_1(2430)\pi$ are also notable,
their branching fractions can reach up to $\sim 8\%$. In the literature,
the $D(1^3D_1)$ state is also predicted to be a broad state with a width of
$\Gamma \simeq 300 \sim550$ MeV~\cite{Zhong:2010vq,Lu:2014zua,
Chen:2015lpa,Close:2005se,Song:2015fha}.

In 2015, the LHCb collaboration observed a $J^P=1^-$ resonance $D_1^*(2760)$ in the $D^+\pi^-$ channel
by using the $B^-\to D^+K^-\pi^-$ process~\cite{LHCb:2015eqv}. The resonance mass and width are determined to be
$M=2781\pm31$ MeV and $\Gamma =177\pm 53$ MeV, respectively. From the point of view of the mass,
$J^{P}$ numbers and decay modes, the $D_1^*(2760)$ favors the $D(1^3D_1)$ assignment,
however, our predicted width is about two times larger than the measured one. The $2^3S_1-1^3D_1$
mixing may overcome this problem, which will be further discussed later.

In the $D_s$-meson sector, the mass of $D_s(1^3D_1)$ is predicted to be $M=2843$ MeV,
which is comparable with the predictions in Refs.~\cite{Song:2015nia,Godfrey:2015dva,Zeng:1994vj,Lahde:1999ih}.
Our predicted strong decay properties have been shown in Table~\ref{Dsmeson1DDecay}.
It is found that the $D_s(1^3D_1)$ may have a medium width of
\begin{equation}
\Gamma\simeq 210 \ \ \ \mathrm{MeV },
\end{equation}
and mainly decays into $DK$, $D^*K$ and $D_s\eta$ channels with branching fractions $64\%$, $21\%$ and
$12\%$, respectively. Our predicted strong decay properties are in good agreement with
the predictions with the $^3P_0$ model in Refs.~\cite{Song:2015nia,Song:2014mha,Godfrey:2015dva}.

In 2014, the LHCb collaboration observed a new $J^P=1^-$ resonance $D_{s1}^*(2860)$
in the $\bar{D}^0K^-$ final state by using the $B_s^0\to \bar{D}^0K^- \pi^+$ process~\cite{LHCb:2014ott}.
Its measured mass and width are $M=2859 \pm 12\pm24$~MeV and $\Gamma=159 \pm 23 \pm77$~MeV, respectively.
From the point of view of the mass, $J^{P}$ numbers, decay modes and width,
the $D_{s1}^*(2860)$ favors the $D_s(1^3D_1)$ assignment. It should be mentioned that the possibility of the $D_{s1}^*(2860)$ resonance
as a mixed state between $2^3S_1$ and $1^3D_1$ cannot be excluded, which will be further discussed later.

\subsubsection{$1^3D_3$  }

In the $D$-meson family, the mass of $D(1^3D_3)$ is predicted to be $M=2782$ MeV, which is consistent
with that predicted in Refs.~\cite{Godfrey:2015dva,Zeng:1994vj,Song:2015fha}.
From Table \ref{Dmeson1DDecay}, it is found that the $D(1^3D_3)$ state is
narrow state with a width of $\Gamma \simeq 47$ MeV, and dominantly decays into the
$D\pi$ and $D^*\pi$ final states. The partial width ratio between $D\pi$ and
$D^*\pi$ channels is predicted to be
\begin{equation}
R=\frac{\Gamma(D \pi)}{\Gamma(D^* \pi)} \simeq 1.1.
\end{equation}
Our predicted width and dominant decay modes for $D(1^3D_3)$ are consistent with the predictions
in the literature~\cite{Wang:2016enc,Godfrey:2015dva,Chen:2015lpa,Yu:2016mez,Yu:2014dda,Li:2010vx}, while
our predicted partial ratio $R=\Gamma(D\pi)/\Gamma(D^* \pi) \simeq 1.1$ is
similar to the predictions in Refs.~\cite{Godfrey:2015dva,Chen:2015lpa,Li:2010vx,Lu:2014zua,Song:2015fha}.

The $D_3^*(2750)$ resonance listed in RPP~\cite{Zyla:2020zbs} favors the assignment of $D(1^3D_3)$.
This resonance was first observed in the $D\pi$ and/or $D^*\pi$ channels
by the BaBar collaboration in 2010~\cite{BaBar:2010zpy}, and confirmed by the LHCb collaboration
by using the $pp$ collision processes~\cite{LHCb:2013jjb} and $B$ decay processes
~\cite{Aaij:2015sqa,Aaij:2016fma,Aaij:2019sqk}. The spin-parity numbers are determined to be $J^P=3^-$ by
the LHCb collaboration ~\cite{Aaij:2015sqa}.
The average measured mass and width of $D_3^*(2750)$ are $M_{exp}=2763.1\pm3.2$ MeV and
$\Gamma_{exp}=66\pm5$ MeV~\cite{Zyla:2020zbs}. As the assignment of $D(1^3D_3)$, the mass and width
of $D_3^*(2750)$ are in good agreement with the theoretical predictions.
However, our predicted ratio $R=\Gamma(D\pi)/\Gamma(D^* \pi)\simeq 1.1$ is notably
larger than the measured value $R_{exp}=0.42\pm 0.16$ at BaBar~\cite{BaBar:2010zpy}.
To confirm the nature of $D_3^*(2750)$, the partial width ratio
$R=\Gamma(D\pi)/\Gamma(D^* \pi)$ is expected to be further measured in future experiments.

In the $D_s$-meson sector, the mass of $D_s(1^3D_3)$ is predicted to be $M=2882$ MeV,
while the mass gap between $D_s(1^3D_3)$ and $D(1^3D_3)$ is estimated to be $\Delta M\simeq 100$ MeV.
Our predictions are consistent with those predicted in Refs.~\cite{Wang:2016enc,Song:2015nia,Zeng:1994vj,Lahde:1999ih}.
The $D_s(1^3D_3)$ state may be a very narrow state with a width of $\Gamma \simeq 22$ MeV,
and mainly decays into $DK$ and $D^*K$ final states. The strong decay properties predicted in this work
are consistent with our previous predictions with SHO wave functions~\cite{Zhong:2009sk} and
other predictions in the literature~\cite{Chen:2015lpa,Wang:2016enc,Godfrey:2015dva,Godfrey:2014fga,
Zhang:2006yj,Li:2009qu}. It should be mention that there are obvious model dependencies in
the predictions of the partial width ratio $\Gamma(D^*K)/\Gamma(DK)$ between $DK$ and $D^*K$,
which scatters in the range of $\sim 0.4-0.7$.

In 2006, the BaBar collaboration observed a new charmed-strange meson structure $D_{sJ}(2860)$ in the $DK$ channel
with mass of $M_{exp}=2856.6 \pm6.5$~MeV and a width of $\Gamma_{exp}=47 \pm17$~MeV~\cite{BaBar:2006gme},
which is consistent with the resonance observed in the $D^*K$ channel in 2009~\cite{BaBar:2009rro}.
In 2014, the LHCb collaboration further studied the structure around $2.86$ GeV in the $B_s^0 \to \bar{D}^0 K^-\pi^+$ decay~\cite{LHCb:2014ott,LHCb:2014ioa}.
They found two overlapping spin-1 resonance $D_{s3}^*(2860)$ and spin-3 resonance $D_{s3}^*(2860)$
in the $\bar{D}^0K^-$ final state. The resonance parameters of $D_{s3}^*(2860)$,
$M_{exp}=2860.5 \pm 2.6\pm2.5\pm6.0$~MeV and $\Gamma_{exp}=53 \pm 7 \pm4 \pm 6$~MeV extracted by LHCb~\cite{LHCb:2014ott,LHCb:2014ioa}, are
consistent with those of $D_{sJ}(2860)$ extracted by BaBar. The spin-3 resonance $D_{s3}^*(2860)$ can be assigned
to the charmed-strange state $D_s(1^3D_3)$. As this assignment, both the mass and decay properties
of $D_{s3}^*(2860)$ can be reasonably understood within the quark model. It should be
mentioned that the $DK$ channel is the optimal channel for establishing spin-1 state $D_s(1^3D_1)$ and
spin-3 state $D_s(1^3D_3)$ due to no contributions from the other two $1D$-wave states with $J^P=2^-$.

\subsubsection{$1D_2$ and $1D'_2$  }

The physical states $1D_2$ and $1D'_2$ are mixed states between states $1^1D_2 $ and $1^3D_2$
via the following mixing scheme:
\begin{equation}
\left(
  \begin{array}{c}
   1D_2\\
   1D'_2\\
  \end{array}\right)=
  \left(
  \begin{array}{cc}
   \cos\theta_{1D} &\sin\theta_{1D}\\
  -\sin\theta_{1D} &\cos\theta_{1D}\\
  \end{array}
\right)
\left(
  \begin{array}{c}
  1^{1}D_{2}\\
  1^{3}D_{2}\\
  \end{array}\right).
\end{equation}
In this work, the $1D_2$ and $1D'_2$ correspond to the low-mass and high-mass mixed states, respectively.

In the $D$-meson family, the masses of the two mixed states $D(1D_2)$ and $D(1D'_2)$ are predicted to
be $M=2755$ MeV and $M=2827$ MeV, respectively. The mass splitting between $D(1D_2)$ and $D(1D'_2)$ is estimated to be $\Delta M\simeq 70$ MeV, which is slightly smaller that of $\Delta M\simeq86$ MeV predicted in~\cite{Li:2010vx}, however, is about a factor of 2 larger than $\Delta M\sim 40$ MeV predicted in Refs.~\cite{Ebert:2009ua,Godfrey:2015dva,Zeng:1994vj,Lahde:1999ih,Lu:2014zua}. Our predicted mixing angle between $D(1D_2)$ and $D(1D'_2)$, $\theta_{1D} =-40.2^{\circ}$, is similar to the angle determined within the relativized
quark model~\cite{Ebert:2009ua,Lu:2014zua}.
The predicted strong decay properties of both $D(1D_2)$ and $D(1D'_2)$ are listed in Table~\ref{Dmeson1DDecay}.
It is found that the low mass state $D(1D_2)$ may be a broad state with a width of
\begin{equation}
\Gamma\simeq 260 \ \ \ \mathrm{MeV },
\end{equation}
and dominantly decays into the $D^*\pi$ and $D^*_2(2460)\pi$ channels with
branching fractions $\sim35\%$ and $\sim45\%$, respectively. While the high mass state $D(1D'_2)$ may have a
narrow width of
\begin{equation}
\Gamma\simeq 60 \ \ \ \mathrm{MeV },
\end{equation}
and dominantly decays into the $D^*\pi$ channel.

Some evidence of the mixed states $D(1D_2)$ and $D(1D'_2)$ may have been observed in experiments.
In 2010, the BaBar collaboration observed a new resonance $D(2750)^0$ with a mass of $M_{exp}=2752.4 \pm4.4$~MeV
and a width of $\Gamma_{exp}=71 \pm17$~MeV in the $D^{*+}\pi^-$ channel~\cite{BaBar:2010zpy}.
In 2013, the LHCb collaboration observed an unnatural parity state $D_J(2740)^0$ in the $D^{*+}\pi^-$ channel.
The measured mass and width $M=2737.0\pm3.5 \pm11.2$ MeV and  $\Gamma=73.2 \pm13.4\pm25.0$ MeV at LHCb~\cite{LHCb:2013jjb}
are consistent with the observations of $D(2750)^0$ at BaBar. The spin-parity numbers are identified as $J^P=2^-$.
In 2019, the LHCb collaboration carried out a determination of quantum numbers for several
excited charmed mesons by using the $B^- \to D^{*+}\pi^- \pi^-$ decays~\cite{Aaij:2019sqk}.
In this experiment, the spin-parity numbers of $D_J(2740)^0$ [denoted by $D_2(2740)^0$] was confirmed to be $J^P=2^-$,
while the measured mass $M=2751 \pm3 \pm7$ MeV and width $\Gamma=102\pm6\pm26$ MeV are slightly
different from their previous measurements~\cite{LHCb:2013jjb}.

In our previous work~\cite{Zhong:2010vq,Xiao:2014ura}, we predicted that $D(2750)^0$/$D_2(2740)^0$ is most
likely to be the high-mass mixed state $D(1D'_2)$, which is consistent with the prediction in
Ref.~\cite{Song:2015fha}. Considering $D_2(2740)^0$ as $D(1D'_2)$, our predicted decay width
$\Gamma\simeq 57$ MeV is in agreement with the PDG average data $\Gamma_{exp}\simeq 88\pm 19$ MeV~\cite{Zyla:2020zbs},
however, our predicted mass $M=2827$ MeV is about $70$ MeV larger than the data.
On the other hand, considering $D_2(2740)^0$ as the low mass state $D(1D_2)$, we find that
although the predicted mass $M=2754$ MeV is consistent with the observations, our predicted width
$\Gamma\simeq 250$ MeV is too broad to comparable with the data. It should be mentioned
that two LHCb experiments~\cite{Aaij:2019sqk,LHCb:2013jjb} do not give very stable resonance parameters for $D_2(2740)^0$.
This indicates that the structure around $D_2(2740)^0$ observed in the $D^{*+}\pi^-$
invariant mass spectrum may be contributed by both the broad state $D(1D_2)$ and the relatively narrow
state $D(1D_2')$ at the same time. To distinguish $D(1D_2)$ and $D(1D'_2)$ and establish
them finally, more observations of the $D^*_2(2460)\pi$, $D_s^*K$ and $D^*\eta$ are
suggested to be carried out in future experiments.

In the $D_s$-meson sector, our predicted masses of $D_s(1D_2)$ and $D_s(1D'_2)$
are $M=2857$ MeV and $M=2911$ MeV, respectively, which are close the predictions
in the literature~\cite{Godfrey:2015dva,Zeng:1994vj,Lahde:1999ih}.
Our determined mixing angle $\theta_{1D}=-40.7^{\circ}$ is similar to that for
the $D$-meson sector. The splitting between $D_s(1D_2)$ and $D_s(1D'_2)$, $\Delta M \simeq 54$ MeV,
is similar to that predicted in Ref.~\cite{Li:2010vx}, however, is a factor of $\sim2$ larger than that predicted in Refs.~\cite{Ebert:2009ua,Godfrey:2015dva,Lahde:1999ih}.
The predicted strong decay properties of both $D_s(1D_2)$ and $D_s(1D'_2)$ are listed in Table~\ref{Dsmeson1DDecay}.
It is found that the low mass state $D_s(1D_2)$ has a width of
\begin{equation}
\Gamma\simeq 130 \ \ \ \mathrm{MeV },
\end{equation}
and dominantly decays into the $D^*K$ and $DK^*$ channels with
branching fractions $\sim82\%$ and $\sim11\%$, respectively. While the high mass state $D_s(1D'_2)$ may have a
relatively narrow width of
\begin{equation}
\Gamma\simeq 44 \ \ \ \mathrm{MeV },
\end{equation}
and dominantly decays into the $D^*K$ and $DK^*$ channels with
branching fractions $\sim65\%$ and $\sim24\%$, respectively. The decay properties predicted with genuine wave functions
extracted from potential model in this work are consistent with those predicted with
the SHO wave functions in our previous works~\cite{Zhong:2009sk,Xiao:2014ura}. The $D^*K$ as the main decay channel
of the $D_s(1D_2)$ state has also been predicted within the $^3P_0$
model~\cite{Song:2015nia,Godfrey:2015dva}, however, for the high mass state $D_s(1D'_2)$, their predicted
decay rates into $D^*K$ is tiny.

Our previous studies~\cite{Zhong:2009sk,Xiao:2014ura} indicates that the $1^1D_2$-$1^3D_2$ mixing might
be crucial to uncover the longstanding puzzle about the narrow structure $D_{sJ}(2860)$ in
the charmed-strange meson family, which was first observed in the $DK$
channel, then confirmed in the $D^*K$ channels by the BaBar collaboration
~\cite{BaBar:2006gme,BaBar:2009rro}. Many people believe that
the $D_{sJ}(2860)$ might be the $1^3D_3$ state due to its narrow width. However, considering the $D_{sJ}(2860)$ as
the $1^3D_3$ state only, one cannot well understand the partial width ratio of
$R=\Gamma(DK)/\Gamma(D^*K)= 1.1\pm 0.34$ measured by BaBar~\cite{BaBar:2009rro}.
To overcome the puzzle about the measured radio, in Refs.~\cite{Zhong:2009sk,Xiao:2014ura} we proposed
that the $J^P=2^-$ mixed state $D_s(1D'_2)$ mainly decaying into $D^*K$ might highly overlap with the $D_s(1^3D_3)$ state
around the mass region $2.86$ GeV, which is compatible with the
theoretical analyses in Refs.~\cite{Godfrey:2013aaa,Gandhi:2020vap}. In 2014, the LHCb collaboration observed two overlapping
spin-1 resonance $D_{s1}^*(2860)$ and spin-3 resonance $D_{s3}^*(2860)$ in the $\bar{D}^0K^-$ final state by analyzing
the $B_s^0 \to \bar{D}^0 K^-\pi^+$ decay~\cite{LHCb:2014ott,LHCb:2014ioa}. The measured partial width ratio $R=\Gamma(DK)/\Gamma(D^*K)\simeq 1.1\pm 0.34$ is considered to belong to the $D_{s1}(2860)$ resonance by PDG~\cite{Zyla:2020zbs}. In fact, for $D_{s1}(2860)$
our predicted partial width ratio $\Gamma(DK)/\Gamma(D^*K)\simeq 3$ is still inconsistent
with the measured value at BaBar~\cite{BaBar:2009rro}. Since the structure around $2.86$ GeV in the $DK$
invariant mass spectrum can be contributed by both $D_{s1}^*(2860)$ and $D_{s3}^*(2860)$,
we may expect that the structure around $2.86$ GeV in the $D^*K$ invariant mass spectrum observed at
BaBar~\cite{BaBar:2009rro} can be contributed by all of the $1D$-wave states with $J^P=1^-,2^-,3^-$,
due to their large decay rates. This is also proposed in Refs.~\cite{Segovia:2015dia,Gandhi:2020vap,Zhong:2009sk}.

To uncover the longstanding puzzle about $D_{sJ}(2860)$, searches for the missing
$D_s(1D_2)$ and $D_s(1D'_2)$ are urgently needed to be carried out in experiments.
The $D^*K$ channel may be the optimal channel for future observations.

\subsection{$2D$-wave states }

\begin{table*}[htp]
\begin{center}
\caption{Partial decay widths and their branching fractions for the $2D$-wave charmed mesons.}
\label{Dmeson2DwaveDecay}
\scalebox{1.00}{
\begin{tabular}{ccccccccccccccccccccccccccccccccccc}
\hline\hline
&
~~~~~&\multicolumn{2}{c}{$\underline{~~~~~~D(2^3D_1)[3143]~~~~~~}$}      ~~~~~&\multicolumn{2}{c}{$\underline{~~~~~~D(2^3D_3)[3202]~~~~~~}$}
~~~~~&\multicolumn{2}{c}{$\underline{~~~~~~D(2D_2)[3168]~~~~~~}$}
~~~~~&\multicolumn{2}{c}{$\underline{~~~~~~D(2D'_2)[3221]~~~~~~}$}
~~~~~\\
&Channel &$\Gamma_i$~(MeV) &Br~(\%) &$\Gamma_i$~(MeV) &Br~(\%) &$\Gamma_i$~(MeV) &Br~(\%) &$\Gamma_i$~(MeV)    &Br~(\%)\\
\hline
&~~ $D\pi$                                       &192.1              &23.2                 &34.4               &20.6                   &--      &--                 &--                     &--                           \tabularnewline
&~~ $D_sK$                                       &55.7               &6.7                  &5.0                &3.0                    &--      &--                 &--                     &--                           \tabularnewline
&~~ $D\eta$                                      &24.9               &3.0                  &2.5                &1.5                    &--      &--                 &--                     &--                           \tabularnewline
&~~ $D\eta'$                                     &12.7               &1.5                  &0.2                &0.1                    &--      &--                 &--                     &--                           \tabularnewline
&~~ $D^*\pi$                                     &44.9               &5.4                  &26.6               &15.9                   &152.3    &24.3                &40.4                    &24.8                      \tabularnewline
&~~ $D_s^*K$                                     &10.6               &1.3                  &2.5                &1.5                    &45.2     &7.2                 &4.9                     &3.0                      \tabularnewline
&~~ $D^*\eta$                                    &5.0                &0.6                  &1.3                &0.8                    &20.9     &3.3                 &2.3                     &1.4                      \tabularnewline
&~~ $D^*\eta'$                                   &0.8                &0.1                  &0.01               &$6\times10^{-3}$       &7.5      &1.2                 &0.4                     &0.2                       \tabularnewline
&~~ $D_0(2550)\pi$                               &50.6               &6.1                  &5.5                &3.3                    &--      &--                 &--                     &--                           \tabularnewline
&~~ $D_{s}(2649)(2^1S_0)K$         &$2\times10^{-3}$   &$2\times10^{-4}$     &0.02               &0.01                   &--      &--                 &--                     &--                           \tabularnewline
&~~ $D_0(2550)\eta$                              &1.8                &0.2                  &0.04               &0.02                   &--      &--                 &--                     &--                           \tabularnewline
&~~ $D(2^3S_1)(2627)\pi$                             &20.1               &2.4                  &7.8                &4.7                    &61.2     &9.8                 &22.6                    &13.9                      \tabularnewline
&~~ $D_{s1}^*(2700)K$                            &--                &--                  &$3\times10^{-5}$               &$2\times10^{-7}$                &--      &--                 &0.1                     &0.06                           \tabularnewline
&~~ $D(2^3S_1)(2627)\eta$                            &--                &--                  &$1\times10^{-6}$   &$6\times10^{-7}$       &--      &--                 &0.1                     &0.06                           \tabularnewline
&~~ $D(3^1S_0)(3029)\pi$          &--                &--                  &$6\times10^{-4}$   &$4\times10^{-4}$       &--      &--                 &--                     &--                           \tabularnewline
&~~ $D_0^*(2300)\pi$                             &--                &--                  &--                &--                    &6.7      &1.1                 &7.3                     &4.5                      \tabularnewline
&~~ $D_{s}(1^3P_0)(2409)K$       &--                &--                  &--                &--                    &0.5      &0.08                &0.6                     &0.4                     \tabularnewline
&~~ $D_0^*(2300)\eta$                            &--                &--                  &--                &--                    &0.3      &0.05                &0.3                     &0.2                      \tabularnewline
&~~ $D(2^3P_0)(2849)\pi$      &--                &--                  &--                &--                    &0.1      &0.02                &5.5                     &3.4                     \tabularnewline
&~~ $D_2^*(2460)\pi$                             &0.5                &0.06                 &8.6                &5.1                    &105.3    &16.8                &14.3                    &8.8                      \tabularnewline
&~~ $D_{s2}^*(2573)K$                            &0.1                &0.01                 &0.03               &0.02                   &14.0     &2.2                 &0.6                     &0.4                      \tabularnewline
&~~ $D_2^*(2460)\eta$                            &0.2                &0.02                 &0.03               &0.02                   &9.5      &1.5                 &0.8                     &0.5                     \tabularnewline
&~~ $D(2^3P_2)(2955)\pi$      &0.1                &0.01                 &1.3                &0.8                    &63.6     &10.2                &3.9                     &2.4                      \tabularnewline
&~~ $D_1(2430)\pi$                               &38.2               &4.6                  &4.3                &2.6                    &6.2      &1.0                 &1.4                     &0.9                     \tabularnewline
&~~ $D_s(1P_1)(2528)K$             &3.9                &0.5                  &0.01               &$6\times10^{-3}$       &0.2      &0.03                &0.2                     &0.1                      \tabularnewline
&~~ $D_1(2430)\eta$                              &10.4               &1.3                  &0.02               &0.01                   &0.1      &0.02                &0.2                     &0.1                    \tabularnewline
&~~ $D_1(2420)\pi$                               &173.7              &21.0                 &3.5                &2.1                    &0.5      &0.08                &8.8                     &5.4                      \tabularnewline
&~~ $D_{s}(1P'_1)(2535)K$            &17.3               &2.1                  &0.06               &0.04                   &0.1      &0.02                &$3\times10^{-4}$        &$2\times10^{-4}$    \tabularnewline
&~~ $D_1(2420)\eta$                              &50.6               &6.1                  &0.1                &0.06                   &0.1      &0.02                &$8\times10^{-5}$        &$5\times10^{-5}$      \tabularnewline
&~~ $D(2P_1)(2900)\pi$            &29.8               &3.6                  &4.1                &2.5                    &0.2      &0.03                &0.03                    &0.02                  \tabularnewline
&~~ $D(2P'_1)(2936)\pi$           &44.6               &5.4                  &$6\times10^{-3}$   &$4\times10^{-3}$       &0.01     &0.02                &4.0                     &2.5                    \tabularnewline
&~~ $D(1^3D_1)(2754)\pi$        &3.0                &0.4                  &0.1                &0.06                   &1.7      &0.3                 &4.7                     &2.9                  \tabularnewline
&~~ $D_3^*(2760)\pi$        &0.5                &0.06                 &2.5                &1.5                    &18.6     &3.0                 &1.6                     &1.0                   \tabularnewline
&~~ $D(1D_2)(2755)\pi$              &2.1                &0.3                  &10.2               &6.1                    &4.6      &0.7                 &3.1                     &1.9                  \tabularnewline
&~~ $D(1D'_2)(2827)\pi$             &20.5               &2.5                  &5.3                &3.2                    &0.8      &0.1                 &2.5                     &1.5                  \tabularnewline
&~~ $D\rho$                                      &1.2                &0.1                  &10.8               &6.5                    &48.8     &7.8                 &3.3                     &2.0                  \tabularnewline
&~~ $D\omega$                                    &0.4                &$0.05$               &3.4                &2.0                    &15.6     &2.5                 &1.1                     &0.7                 \tabularnewline
&~~ $D_sK^*$                                     &0.1                &$0.01$               &1.4                &0.8                    &10.1     &1.6                 &2.4                     &1.5                  \tabularnewline
&~~ $D^*\rho$                                    &8.4                &1.0                  &16.5               &9.8                    &22.9     &3.7                 &17.7                    &10.9                 \tabularnewline
&~~ $D^*\omega$                                  &2.5                &$0.3$                &5.3                &3.2                    &6.9      &1.1                 &5.5                     &3.4                 \tabularnewline
&~~ $D_s^*K^*$                                   &0.5                &$0.06$               &3.9                &2.3                    &1.7      &0.3                 &2.4                     &1.5                  \tabularnewline
\hline
&~~ \textbf{Total}                               &$\mathbf{827.8}$    &$\mathbf{100}$      &$\mathbf{167.3}$    &$\mathbf{100}$  &$\mathbf{626.2}$    &$\mathbf{100}$  &$\mathbf{163.0}$    &$\mathbf{100}$         \tabularnewline
\hline\hline
\end{tabular}}
\end{center}
\end{table*}

\begin{table*}[htp]
\begin{center}
\caption{Partial decay widths and their branching fractions for the $2D$-wave charmed-strange mesons.}
\label{Dsmeson2DDecay}
\scalebox{1.00}{
\begin{tabular}{cccccccccccccccccccc}
\hline\hline
&
~~~~~&\multicolumn{2}{c}{$\underline{~~~~~~D_s(2^3D_1)[3233]~~~~~~}$}      ~~~~~&\multicolumn{2}{c}{$\underline{~~~~~~D_s(2^3D_3)[3267]~~~~~~}$}
~~~~~&\multicolumn{2}{c}{$\underline{~~~~~~D_s(2D_2)[3267]~~~~~~}$}
~~~~~&\multicolumn{2}{c}{$\underline{~~~~~~D_s(2D'_2)[3306]~~~~~~}$}
~~~~~\\
&Channel &$\Gamma_i$~(MeV) &Br~(\%) &$\Gamma_i$~(MeV) &Br~(\%) &$\Gamma_i$~(MeV) &Br~(\%) &$\Gamma_i$~(MeV)    &Br~(\%)\\
\hline
&~~ $DK$                                         &157.0    &30.9                &19.0               &26.6               &--      &--                &--     &--         \tabularnewline
&~~ $D_s\eta$                                    &33.2     &6.5                 &2.0                &2.8                &--      &--                &--     &--         \tabularnewline
&~~ $D_s\eta'$                                   &15.6     &3.1                 &0.2                &0.3                &--      &--                &--     &--         \tabularnewline
&~~ $D^*K$                                       &34.7     &6.8                 &11.6               &16.2               &130.7    &31.0               &19.8    &28.8        \tabularnewline
&~~ $D_s^*\eta$                                  &6.3      &1.2                 &4.3                &6.0                &28.7     &6.8                &2.0     &2.9         \tabularnewline
&~~ $D_s^*\eta'$                                 &0.8      &0.2                 &1.7                &2.4                &5.6      &1.3                &0.3     &0.4         \tabularnewline
&~~ $D_0(2550)K$                                 &47.5     &9.4                 &2.2                &3.1                &--      &--                &--     &--         \tabularnewline
&~~ $D_{s}(2^1S_0)(2649)\eta$      &1.6      &0.3                 &0.03               &0.04                &--      &--                &--     &--         \tabularnewline
&~~ $D(2^3S_1)(2627)K$                               &14.7     &2.9                 &1.4                &2.0                &53.7     &12.7               &7.5     &10.9        \tabularnewline
&~~ $D_{s1}^*(2700)\eta$                         &--      &--                 &0.1                &0.1                &0.1      &0.02                &0.1     &0.1              \tabularnewline
&~~ $D_0^*(2300)K$                               &--      &--                 &--                &--                &3.1      &0.7                &4.9     &7.1         \tabularnewline
&~~ $D_{s}(1^3P_0)(2409)\eta$    &--      &--                 &--                &--                 &0.4      &0.1               &0.2     &0.3        \tabularnewline
&~~ $D_2^*(2460)K$                               &2.4      &0.5                 &1.1                &1.5                &99.2     &23.5               &7.3     &10.6      \tabularnewline
&~~ $D_{s2}^*(2573)\eta$                         &0.2      &0.04                &0.02               &0.03               &26.0     &6.2                &1.3     &1.9       \tabularnewline
&~~ $D_1(2430)K$                                 &25.2     &5.0                 &0.2                &0.3                &3.2      &0.8                &1.2     &1.7       \tabularnewline
&~~ $D_{s}(1P_1)(2528)\eta$          &2.4      &0.5                 &0.2                &0.3                 &0.2      &0.02               &0.2     &0.3        \tabularnewline
&~~ $D_1(2420)K$                                 &146.2    &28.8                &1.7                &2.4                &1.1      &0.3                &0.1     &0.1        \tabularnewline
&~~ $D_{s1}(2535)\eta$                           &13.8     &2.7                 &0.1                &0.1                &0.1      &0.02               &0.01    &0.01        \tabularnewline
&~~ $D(1^3D_1)(2754)K$          &--      &--                 &$1\times10^{-4}$   &$1\times10^{-4}$    &0.02     &$5\times10^{-3}$   &2.4     &3.5        \tabularnewline
&~~ $D_3^*(2760)K$          &--      &--                 &0.03               &0.04                &1.6      &0.4               &0.2     &0.3          \tabularnewline
&~~ $D(1D_2)(2755)K$                &--      &--                 &2.8                &3.9                 &0.1      &0.02               &0.3     &0.4        \tabularnewline
&~~ $DK^*$                                       &0.7      &0.1                 &6.4                &9.0                &43.9     &10.4               &5.5     &8.0        \tabularnewline
&~~ $D_s\phi$                                    &0.02     &$4\times10^{-3}$    &0.4                &0.6                &6.4      &1.5                &2.2     &3.2        \tabularnewline
&~~ $D^*K^*$                                     &5.1      &1.0                 &14.3               &20.0               &17.5     &4.1                &12.2    &17.7        \tabularnewline
&~~ $D_s^*\phi$                                  &0.2      &0.04                &1.7                &2.4                &0.7      &0.2                &1.1     &1.6         \tabularnewline
\hline
&~~ \textbf{Total}      &$\mathbf{507.6}$    &$\mathbf{100}$   &$\mathbf{71.4}$    &$\mathbf{100}$  &$\mathbf{422.3}$    &$\mathbf{100}$  &$\mathbf{68.8}$    &$\mathbf{100}$   \tabularnewline
\hline\hline
\end{tabular}}
\end{center}
\end{table*}

\subsubsection{$2^3D_1$  }

In the $D$-meson family, the mass of the $D(2^3D_1)$ state is predicted to be $M=3143$ MeV, which
is comparable with the predictions in Refs.~\cite{Godfrey:2015dva,Zeng:1994vj,Li:2010vx,Song:2015fha}.
Our predicted strong decay properties for these high $2D$-wave states are listed in Table~\ref{Dmeson2DwaveDecay}.
It is found that the $D(2^3D_1)$ state is a very broad state with a width of $\Gamma \simeq 830$ MeV.
This state may be difficult to be observed in experiments due to its broad width. It should be
mentioned that the width of $D(2^3D_1)$ predicted within our chiral
quark model is about a factor of $4-6$ larger than the predictions within the $^3P_0$ models in
Refs.~\cite{Godfrey:2015dva,Song:2015fha}.

In the $D_s$-meson family, the mass of $D_s(2^3D_1)$ is predicted to be $M= 3233$ MeV, which
is comparable with those in Refs.~\cite{Godfrey:2015dva,Zeng:1994vj}.
Our predicted strong decay properties for these high $2D$-wave states are listed in Table~\ref{Dsmeson2DDecay}.
It is found that the $D_s(2^3D_1)$ state is a very broad state with a width
of $\Gamma \simeq 510$ MeV, and may be difficult to be observed in experiments due to its broad width. It should be
mentioned that the width of $D_s(2^3D_1)$ predicted within our chiral
quark model are a factor of $\sim 3$ larger than the predictions within the $^3P_0$ models in
Refs.~\cite{Godfrey:2015dva,Song:2015nia}.

\begin{figure}
\centering \epsfxsize=8.6 cm \epsfbox{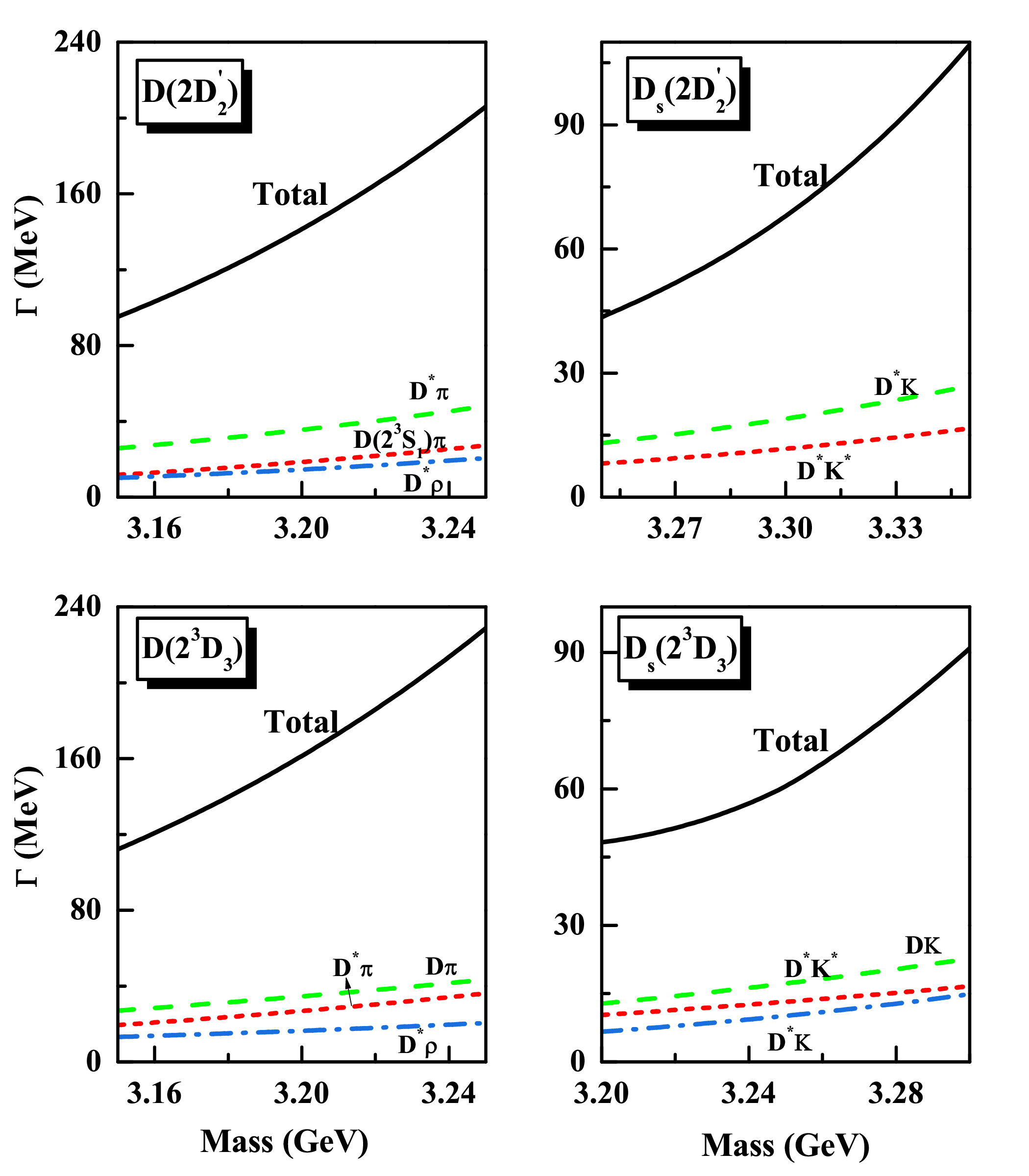} \vspace{-0.2 cm} \caption{Total decay widths and the main partial decay widths for $D(2D'_2)$, $D_s(2D'_2)$, $D(2^3D_3)$ and $D_s(2^3D_3)$ as functions of their masses.}\label{D23D3Mirunning}
\end{figure}


\subsubsection{$2^3D_3$ }

In the $D$-meson family, the mass of the $D(2^3D_3)$ state is predicted to be $M=3202$ MeV, which
is comparable with those predicted in Refs.~\cite{Godfrey:2015dva,Zeng:1994vj,Li:2010vx,Song:2015fha}.
The $D(2^3D_3)$ state is a relatively narrow state with a width of
\begin{equation}
\Gamma\simeq 167 \ \ \ \mathrm{MeV },
\end{equation}
and have relatively large decay rates into $D\pi$ and $D^*\pi$ channels with
branching fractions $\sim 21\%$ and $\sim 16\%$, respectively. More details of the decay properties
can be seen in Table~\ref{Dmeson2DwaveDecay}. To see the dependence of the decay properties of $D(2^3D_3)$ on its mass,
we also plot the main partial widths and the total width as functions of the mass in
Fig.~\ref{D23D3Mirunning}. It is found that the partial and total decay widths increase smoothly with the
mass. With a mass uncertainty of $50$ MeV, the total width of $D(2^3D_3)$ varies in the range $\sim 110-220$ MeV.
A relatively narrow width of $D(2^3D_3)$ is also predicted in Ref.~\cite{Song:2015fha},
although their predicted width $\Gamma \simeq 30$ MeV is about a factor of 5 smaller than ours.
To establish the missing $D(2^3D_3)$ state, the
$D\pi$ and $D^*\pi$ channels are worth to observing in future experiments.
However, it should be pointed out that in Refs.~\cite{Godfrey:2015dva,Song:2015fha},
the $D\pi$ and $D^*\pi$ are not predicted to be the dominant decay modes of $D(2^3D_3)$.

In the $D_s$-meson family, the mass of the $D_s(2^3D_3)$ state is predicted to be $M=3267$ MeV, which
is comparable with those of Refs.~\cite{Godfrey:2015dva,Zeng:1994vj,Li:2010vx,Song:2015fha}.
The $D_s(2^3D_3)$ state has a narrow width of
\begin{equation}
\Gamma\simeq 71 \ \ \ \mathrm{MeV },
\end{equation}
which is in agreement with the prediction of Ref.~\cite{Song:2015nia}.
The $D_s(2^3D_3)$ state dominantly decays into $DK$ , $D^*K^*$ and $D^*K$ channels with
branching fractions $\sim 27\%$ , $\sim 20\%$ and $\sim 16\%$, respectively.
To see the dependence of the decay properties of $D_s(2^3D_3)$ on its mass,
we also plot the main partial widths and the total width as functions of the mass in
Fig.~\ref{D23D3Mirunning}. It is found that the partial and total decay widths increase smoothly with the
mass. With a mass uncertainty of $50$ MeV, the total width of $D_s(2^3D_3)$ varies in the range $\sim 50-90$ MeV.
To establish the missing $D_s(2^3D_3)$ state, the
$DK$ and $D^*K$ channels are worth to observing in future experiments.

\subsubsection{$2D_2$ and $2D'_2$  }

The physical states $2D_2$ and $2D'_2$ are mixed states between states $2^1D_2 $ and $2^3D_2$
via the following mixing scheme:
\begin{equation}\label{mixst}
\left(
  \begin{array}{c}
   2D_2\\
   2D'_2\\
  \end{array}\right)=
  \left(
  \begin{array}{cc}
   \cos\theta_{2D} &\sin\theta_{2D}\\
  -\sin\theta_{2D} &\cos\theta_{2D}\\
  \end{array}
\right)
\left(
  \begin{array}{c}
  2^{1}D_{2}\\
  2^{3}D_{2}\\
  \end{array}\right).
\end{equation}
In this work, the $2D_2$ and $2D'_2$ correspond to the low-mass and high-mass mixed states, respectively.

In the $D$-meson family, the masses of the two mixed states $D(2D_2)$ and $D(2D'_2)$ are predicted to
be $M=3168$ MeV and $M=3221$ MeV, respectively. Our predicted masses are comparable with those
in Refs.~\cite{Li:2010vx,Godfrey:2015dva,Zeng:1994vj}. The mass splitting between $D(2D_2)$ and $D(2D'_2)$ is
estimated to be $\Delta M\simeq 53$ MeV, which is in agreement with that of $\Delta M\simeq52$ MeV predicted
in Ref.~\cite{Ebert:2009ua}. Our predicted mixing angle, $\theta_{2D} =-40.2^{\circ}$,
is similar to that determined within the relativized quark model~\cite{Godfrey:2015dva,Lu:2014zua}.
The predicted strong decay properties of both $D(2D_2)$ and $D(2D'_2)$ are listed in Table~\ref{Dmeson2DwaveDecay}.
It is found that the low mass state $D(2D_2)$ may be a broad state with a width of
$\Gamma\simeq 620$ MeV, and have large decay rates into the $D^*\pi$ and $D^*_2(2460)\pi$ channels with
branching fractions $\sim24\%$ and $\sim17\%$, respectively. The $D(2D_2)$ state may be difficult to
be observed in experiments due to its too broad width. While the high mass state $D(2D'_2)$ may have a
relatively narrow width of
\begin{equation}
\Gamma\simeq 163 \ \ \ \mathrm{MeV },
\end{equation}
and dominantly decays into $D^*\pi$, $D^*_1(2600)\pi$ and $D^*\rho$ channels with
branching fractions $\sim 25\%$, $\sim 14\%$ and $\sim 11\%$, respectively.
To see the dependence of the decay properties of $D(2D'_2)$ on its mass,
we also plot the main partial widths and the total width as functions of the mass in
Fig.~\ref{D23D3Mirunning}. It is found that the partial and total decay widths increase smoothly with the
mass. With a mass uncertainty of $50$ MeV, the total width of $D(2D'_2)$ varies in the range $\sim 90-200$ MeV.
The narrow width nature of $D(2D'_2)$ is also predicted by Song \emph{et al.} in Ref.~\cite{Song:2015fha},
although their predicted width $\Gamma \sim 30$ MeV is a factor of $\sim 5$ smaller than ours.

In the $D_s$-meson family, the masses of the two mixed states $D_s(2D_2)$ and $D_s(2D'_2)$ are predicted to
be $M=3267$ MeV and $M=3306$ MeV, respectively. Our predicted masses are comparable with those
in Refs.~\cite{Li:2010vx,Godfrey:2015dva,Zeng:1994vj}. The mass splitting between $D_s(2D_2)$ and $D_s(2D'_2)$ is
estimated to be $\Delta M\simeq 39$ MeV, which is comparable with that predicted
in~\cite{Li:2010vx,Godfrey:2015dva,Zeng:1994vj}. Our predicted mixing angle, $\theta_{2D} =-41.3^{\circ}$,
is similar to that determined within the relativized
quark model~\cite{Godfrey:2015dva,Lu:2014zua}.
The predicted strong decay properties of both $D_s(2D_2)$ and $D_s(2D'_2)$ are listed in Table~\ref{Dsmeson2DDecay}.
It is found that the low mass state $D_s(2D_2)$ may be a broad state with a width of
$\Gamma\simeq 420$ MeV, and have large decay rates into the $D^*K$ and $D^*_2(2460)K$ channels with
branching fractions $\sim31\%$ and $\sim23\%$, respectively. The $D_s(2D_2)$ state may be difficult to
be observed in experiments due to its broad width. It should be
mentioned that the width of $D_s(2D_2)$ predicted within our chiral
quark model is a factor of $\sim3$ larger than the predictions within the $^3P_0$ models in
Refs.~\cite{Godfrey:2015dva,Song:2015nia}.
While the high mass state $D_s(2D'_2)$ may have a narrow width of
\begin{equation}
\Gamma\simeq 69 \ \ \ \mathrm{MeV },
\end{equation}
which is consistent with the prediction of Ref.~\cite{Song:2015nia}.
The $D_s(2D'_2)$ dominantly decays into $D^*K$, $D^*K^*$ , $D(2^3S_1)K$ and $D^*_2(2460)K$ channels with
branching fractions $\sim 29\%$, $\sim 18\%$ , $\sim 11\%$ and $\sim 11\%$, respectively.
To see the dependence of the decay properties of $D_s(2D'_2)$ on its mass,
we also plot the main partial widths and the total width as functions of the mass in
Fig.~\ref{D23D3Mirunning}. It is found that the partial and total decay widths increase smoothly with the
mass. With a mass uncertainty of $50$ MeV, the total width of $D_s(2D'_2)$ varies in the range $\sim 45-110$ MeV.
The high mass state $D_s(2D'_2)$ may have a large potential to be established in forthcoming experiments.
The $D^*K$ channel may be the optimal channel for future observations.

\begin{table*}[htp]
\begin{center}
\caption{Partial decay widths and their branching fractions for the $1F$-wave charmed mesons.}
\label{Dmeson1FwaveDecay}
\scalebox{1.0}{
\begin{tabular}{cccccccccccccccccccc}
\hline\hline
&
~~~~~&\multicolumn{2}{c}{$\underline{~~~~~~D(1^3F_2)[3096]~~~~~~}$}      ~~~~~&\multicolumn{2}{c}{$\underline{~~~~~~D(1^3F_4)[3034]~~~~~~}$}
~~~~~&\multicolumn{2}{c}{$\underline{~~~~~~D(1F_3)[3022]~~~~~~}$}
~~~~~&\multicolumn{2}{c}{$\underline{~~~~~~D(1F'_3)[3129]~~~~~~}$}
~~~~~\\
&Channel &$\Gamma_i$~(MeV) &Br~(\%) &$\Gamma_i$~(MeV) &Br~(\%) &$\Gamma_i$~(MeV) &Br~(\%) &$\Gamma_i$~(MeV)    &Br~(\%)\\
\hline
&~~ $D\pi$                                       &39.4                  &5.5                 &11.2                  &17.3                 &--                &--                       &--               &--            \tabularnewline
&~~ $D_sK$                                       &11.5                  &1.6                 &1.2                   &1.8                  &--                &--                       &--               &--            \tabularnewline
&~~ $D\eta$                                      &5.7                   &0.8                 &0.8                   &1.2                  &--                &--                       &--               &--            \tabularnewline
&~~ $D\eta'$                                     &3.1                   &0.4                 &0.03                  &0.05                 &--                &--                       &--               &--            \tabularnewline
&~~ $D^*\pi$                                     &25.3                  &3.5                 &13.3                  &20.5                 &42.6               &12.4                     &41.8              &24.9           \tabularnewline
&~~ $D_s^*K$                                     &6.7                   &0.9                 &0.9                   &1.4                  &10.4               &3.0                      &4.3               &2.6            \tabularnewline
&~~ $D^*\eta$                                    &3.5                   &0.5                 &0.6                   &0.9                  &5.8                &1.7                      &0.2               &0.1           \tabularnewline
&~~ $D^*\eta'$                                   &0.7                   &0.1                 &$7\times 10^{-4}$     &$1\times 10^{-3}$    &0.2                &0.06                     &0.1               &0.06            \tabularnewline
&~~ $D_0(2550)\pi$                               &2.2                   &0.3                 &0.4                   &0.6                  &--                &--                       &--               &--              \tabularnewline
&~~ $D(2^3S_1)(2627)\pi$                             &0.4                   &0.06                &0.3                   &0.5                  &0.8                &0.2                      &1.4               &0.8            \tabularnewline
&~~ $D_0^*(2300)\pi$                             &--                   &--                 &--                   &--                  &1.1                &0.3                      &9.5               &5.7           \tabularnewline
&~~ $D_{s}(1^3P_0)(2409)K$       &--                   &--                 &--                   &--                  &$4\times 10^{-3}$  &$1\times 10^{-3}$        &0.7               &0.4          \tabularnewline
&~~ $D_0^*(2300)\eta$                            &--                   &--                 &--                   &--                  &0.01               &$3\times 10^{-3}$        &0.5               &0.3           \tabularnewline
&~~ $D(2^3P_0)(2849)\pi$      &--                   &--                 &--                   &--                   &$2\times 10^{-4}$  &$6\times 10^{-5}$        &$2\times 10^{-3}$ &$1\times 10^{-3}$   \tabularnewline
&~~ $D_2^*(2460)\pi$                             &22.5                  &3.1                 &8.7                   &13.4                 &170.4              &49.7                     &35.7              &21.3          \tabularnewline
&~~ $D_{s2}^*(2573)K$                            &0.6                   &0.1                 &--                   &--                   &--                &--                       &0.2              &0.1                \tabularnewline
&~~ $D_2^*(2460)\eta$                            &1.3                   &0.2                 &$6\times 10^{-4}$        &$9\times 10^{-4}$     &0.01                &$3\times 10^{-3}$          &0.5               &0.3               \tabularnewline
&~~ $D(2^3P_2)(2955)\pi$      &$6\times 10^{-6}$    &$8\times 10^{-7}$    &--                   &--                  &--                &--                       &$5\times 10^{-3}$ &$3\times 10^{-3}$  \tabularnewline
&~~ $D_1(2430)\pi$                               &44.9                 &6.2                  &13.4                  &20.6                 &1.3                &0.4                      &0.1               &0.06          \tabularnewline
&~~ $D_s(1P_1)(2528)K$             &0.6                  &0.1                  &$1\times 10^{-4}$    &$2\times 10^{-4}$   &--                &--                       &0.04              &0.02           \tabularnewline
&~~ $D_1(2430)\eta$                              &3.4                  &0.5                  &0.01                  &0.02                 &--                &--                       &$3\times 10^{-3}$ &$2\times 10^{-3}$   \tabularnewline
&~~ $D_1(2420)\pi$                               &203.1                &28.1                 &0.5                   &0.8                  &0.04               &0.01                     &19.7              &11.7          \tabularnewline
&~~ $D_{s}(1P'_1)(2535)K$            &26.3                 &3.6                  &$5\times 10^{-9}$     &$8\times 10^{-9}$   &--                &--                       &0.1               &0.06      \tabularnewline
&~~ $D_1(2420)\eta$                              &18.5                 &2.6                  &$5\times 10^{-5}$     &$8\times 10^{-5}$    &--                &--                       &0.2               &0.1        \tabularnewline
&~~ $D(2P_1)(2900)\pi$            &0.03                 &0.004                &--                   &--                  &--                &--                       &$5\times 10^{-5}$ &$3\times 10^{-5}$   \tabularnewline
&~~ $D(2P'_1)(2936)\pi$           &$4\times 10^{-3}$    &$5\times 10^{-4}$   &--                   &--                  &--                &--                       &$2\times 10^{-3}$ &$1\times 10^{-3}$ \tabularnewline
&~~ $D(1^3D_1)(2754)\pi$        &0.9                  &0.1                  &$2\times 10^{-4}$     &$3\times 10^{-4}$    &1.2                &0.4                      &18.9              &11.3      \tabularnewline
&~~ $D_3^*(2760)\pi$        &1.9                  &0.3                  &2.4                   &3.7                 &87.9               &25.6                     &6.6               &3.9         \tabularnewline
&~~ $D(1D_2)(2755)\pi$              &91.8                 &12.7                 &2.7                   &4.2                 &0.1                &0.03                     &3.4               &2.0       \tabularnewline
&~~ $D(1D'_2)(2827)\pi$             &88.5                 &12.2                 &0.4                   &0.6                 &0.06               &0.02                     &1.5               &0.9       \tabularnewline
&~~ $D\rho$                                      &2.0                  &0.3                  &1.4                   &2.2                  &10.6               &3.1                      &1.3               &0.8       \tabularnewline
&~~ $D\omega$                                    &0.6                  &0.1                  &0.4                   &0.6                  &3.3                &1.0                      &0.4               &0.2      \tabularnewline
&~~ $D_sK^*$                                     &0.1                  &0.01                 &0.04                  &0.06                 &0.8                &0.2                      &0.6               &0.4       \tabularnewline
&~~ $D^*\rho$                                    &87.5                 &12.1                 &4.7                   &7.2                  &4.8                &1.4                      &15.1              &9.0       \tabularnewline
&~~ $D^*\omega$                                  &26.8                 &3.7                  &1.5                   &2.3                  &1.4                &0.4                      &4.7               &2.8      \tabularnewline
&~~ $D_s^*K^*$                                   &2.7                  &0.4                  &0.04                  &0.06                 &$4\times 10^{-3}$  &$1\times 10^{-3}$        &0.8              &0.5       \tabularnewline
\hline
&~~ \textbf{Total}                               &$\mathbf{722.5}$    &$\mathbf{100}$      &$\mathbf{64.9}$    &$\mathbf{100}$  &$\mathbf{342.8}$    &$\mathbf{100}$  &$\mathbf{168.4}$    &$\mathbf{100}$         \tabularnewline
\hline\hline
\end{tabular}}
\end{center}
\end{table*}

\begin{table*}[htp]
\begin{center}
\caption{Partial decay widths and their branching fractions for the $1F$-wave charmed-strange mesons.}
\label{Dsmeson1FDecay}
\scalebox{1.0}{
\begin{tabular}{cccccccccccccccccccc}
\hline\hline
&
~~~~~&\multicolumn{2}{c}{$\underline{~~~~~~D_s(1^3F_2) [3176]~~~~~~}$}      ~~~~~&\multicolumn{2}{c}{$\underline{~~~~~~D_s(1^3F_4) [3134]~~~~~~}$}
~~~~~&\multicolumn{2}{c}{$\underline{~~~~~~D_s(1F) [3123]~~~~~~}$}
~~~~~&\multicolumn{2}{c}{$\underline{~~~~~~D_s(1F')[3205]~~~~~~}$}
~~~~~\\
&Channel &$\Gamma_i$~(MeV) &Br~(\%) &$\Gamma_i$~(MeV) &Br~(\%) &$\Gamma_i$~(MeV) &Br~(\%) &$\Gamma_i$~(MeV)    &Br~(\%)\\
\hline
&~~ $DK$                                         &33.3    &8.1               &8.1                &26.3               &--               &--                  &--                &--            \tabularnewline
&~~ $D_s\eta$                                    &7.4     &1.8               &0.7                &2.3                &--               &--                  &--                &--            \tabularnewline
&~~ $D_s\eta'$                                   &3.2     &0.8               &0.03               &0.1                &--               &--                  &--                &--            \tabularnewline
&~~ $D^*K$                                       &21.7    &5.3               &8.8                &28.6               &40.3              &19.3                &26.8               &37.4      \tabularnewline
&~~ $D_s^*\eta$                                  &4.4     &1.1               &0.6                &1.9                &7.6               &3.6                 &2.3                &3.2       \tabularnewline
&~~ $D_s^*\eta'$                                 &0.5     &0.1               &$5\times10^{-4}$   &$2\times10^{-3}$   &0.2               &0.1                 &0.1                &0.1       \tabularnewline
&~~ $D_0(2550)K$                                 &1.0     &0.2               &0.01               &0.03               &--               &--                  &--                &--            \tabularnewline
&~~ $D(2^3S_1)(2627)K$                               &0.1     &0.02              &$6\times10^{-6}$   &$2\times10^{-5}$      &$4\times10^{-4}$    &$2\times10^{-4}$     &0.02               &0.03       \tabularnewline
&~~ $D_0^*(2300)K$                               &--     &--                &--                &--                  &0.5               &0.2                 &6.5                &9.1       \tabularnewline
&~~ $D_{s}(1^3P_0)(2409)\eta$    &--     &--                &--                &--                  &$5\times10^{-5}$  &$2\times10^{-5}$    &0.4                &0.6      \tabularnewline
&~~ $D_2^*(2460)K$                               &17.8    &4.3               &2.0                &6.5                &142.4             &68.3                &12.7               &17.7    \tabularnewline
&~~ $D_{s2}^*(2573)\eta$                         &0.9     &0.2               &$5\times10^{-5}$   &$2\times10^{-4}$   &1.8               &0.9                  &0.3                &0.4     \tabularnewline
&~~ $D_1(2430)K$                                 &39.5    &9.6               &3.9                &12.7               &0.3               &0.1                 &0.1                &0.1     \tabularnewline
&~~ $D_{s}(1P_1)(2528)\eta$          &3.5     &0.9               &0.01               &0.03                &$4\times10^{-4}$  &$2\times10^{-4}$    &$5\times10^{-3}$   &$7\times10^{-3}$ \tabularnewline
&~~ $D_1(2420)K$                                 &190.2   &46.3              &0.04               &0.1                &0.02              &0.01                &7.5                &10.5     \tabularnewline
&~~ $D_{s1}(2535)\eta$                           &15.6    &3.8               &$2\times10^{-5}$   &$7\times10^{-5}$   &$5\times10^{-5}$  &$2\times10^{-5}$    &0.1                &0.1      \tabularnewline
&~~ $DK^*$                                       &1.4     &0.3               &1.2                &3.9                &10.5              &5.0                 &2.0                &2.8       \tabularnewline
&~~ $D_s\phi$                                    &0.03    &$7\times10^{-3}$  &0.01               &0.03               &0.5               &0.2                 &0.4                &0.6      \tabularnewline
&~~ $D^*K^*$                                     &69.9    &17.0              &5.4                &17.5               &4.4               &2.1                 &12.0               &16.8      \tabularnewline
&~~ $D_s^*\phi$                                  &0.4     &0.1               &$4\times10^{-5}$   &$1\times10^{-4}$   &--               &--                  &0.4                &0.6       \tabularnewline
\hline
&~~ \textbf{Total}      &$\mathbf{410.8}$    &$\mathbf{100}$   &$\mathbf{30.8}$    &$\mathbf{100}$  &$\mathbf{208.5}$    &$\mathbf{100}$  &$\mathbf{71.6}$    &$\mathbf{100}$   \tabularnewline
\hline\hline
\end{tabular}}
\end{center}
\end{table*}

\subsection{ $1F$-wave states}

\begin{figure}
\centering \epsfxsize=8.6 cm \epsfbox{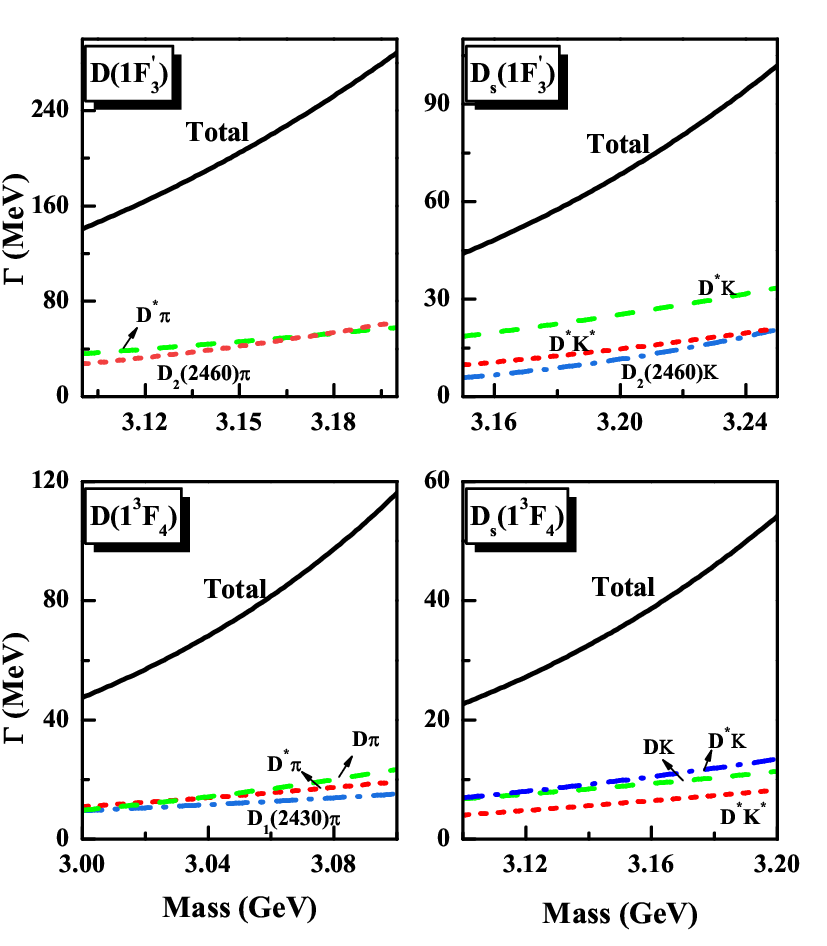} \vspace{-0.2 cm} \caption{Total decay widths and the main partial decay widths for $D(1F'_3)$, $D_s(1F'_3)$, $D(1^3F_4)$ and $D_s(1^3F_4)$ as functions of their masses.}\label{DFwaveMirunning}
\end{figure}

\subsubsection{$1^3F_2$ }

In the $D$-meson family, our predicted mass for $D(1^3F_2)$,
$M=3096$ MeV, is comparable with the predictions in Refs.~\cite{Zeng:1994vj,Song:2015fha}.
From Table~\ref{Dmeson1FwaveDecay}, it is found that the $D(1^3F_2)$ state might be a
very broad state with a width of
\begin{equation}
\Gamma\simeq 720 \ \ \ \mathrm{MeV },
\end{equation}
and dominantly decays into $D_1(2420)\pi$, $D(1D_2)\pi$, $D(1D'_2)\pi$ and $D^*\rho$.
The decay properties predicted in this work roughly agree with those predicted
with the SHO wave functions in our previous work~\cite{Xiao:2014ura},
however, is notably (a factor of $2-4$) broader than those predicted within the
$^3P_0$ models in Refs.~\cite{Godfrey:2015dva,Song:2015fha}. The $D(1^3F_2)$ may
be too broad to be observed in experiments according to our present predictions.

In the $D_s$-meson sector, the predicted masses for $D_s(1^3F_2)$ is
$3176$ MeV, which is comparable with the predictions in Refs.~\cite{Godfrey:2015dva,Zeng:1994vj,Song:2015nia}.
From table Table~\ref{Dsmeson1FDecay}, it is found that the $D_s(1^3F_2)$ state might have a
very broad width of
\begin{equation}
\Gamma\simeq 410 \ \ \ \mathrm{MeV },
\end{equation}
and dominantly decays into $D_1(2420)K$ and $D^*K^*$ with branching fractions $\sim 46\%$ and $\sim 17\%$, respectively.
The $D_s(1^3F_2)$ state is also predicted to be a broad state with a width of $\sim 300-400$ MeV
in Refs.~\cite{Godfrey:2015dva,Song:2015fha,Xiao:2014ura}.
The $D_s(1^3F_2)$ state may be difficult to be established in experiments due to
its too broad width.

\subsubsection{$1^3F_4$ }

In the $D$-meson family, our predicted mass for $D(1^3F_4)$,
$M=3034$ MeV, is comparable with the predictions in Refs.~\cite{Zeng:1994vj,Song:2015fha}.
From Table~\ref{Dmeson1FwaveDecay}, it is found that the $D(1^3F_4)$ is a fairly narrow state with a width of
\begin{equation}
\Gamma\simeq 65 \ \ \ \mathrm{MeV },
\end{equation}
and dominantly decays into the $D\pi$,
$D^*\pi$ and $D_1(2430)\pi$ channels with a comparable branching fraction $\sim 20\%$.
To see the dependence of the decay properties of $D(1^3F_4)$ on its mass,
we also plot the main partial widths and the total width as functions of the mass in
Fig.~\ref{DFwaveMirunning}. It is found that the partial and total decay widths increase smoothly with the
mass. With a mass uncertainty of $50$ MeV, the total width of $D(1^3F_4)$ varies in the range $\sim 50-110$ MeV.
The narrow width of $D(1^3F_4)$ and its relatively large decay rates into $D\pi$ and $D^*\pi$ are
also predicted within the $^3P_0$ models in Refs.~\cite{Godfrey:2015dva,Song:2015fha}.
The $D(1^3F_4)$ might have a large potential to be observed in the $D\pi$ and $D^*\pi$ final
states due to its narrow width.

It is interesting to find that the $D_J^*(3000)$ resonance with a natural parity observed in the $D\pi$ channel by the LHCb collaboration in
2013~\cite{LHCb:2013jjb} might be a good candidate of $D(1^3F_4)$. Our predicted mass $M=3034$ MeV is consistent with the
measured value $M_{exp}=3008.1\pm4.0$ MeV of $D_J^*(3000)$. Taking $D_J^*(3000)$ as the $D(1^3F_4)$ state,
we find that $D_J^*(3000)$ has a large decay rate into the $D\pi$ channel with a branching fraction
of $\sim 17\%$, which is consistent with the observations. The predicted width $\Gamma \simeq 65$ MeV is also comparable with the
data $\Gamma_{exp} =110.5 \pm11.5$ MeV. Considering $D_J^*(3000)$ as the $D(1^3F_4)$ state, it also has a large decay rate
into the $D^*\pi $ channel, the partial width ratio between $D^*\pi $ and $D\pi $ is predicted to be
\begin{equation}
R=\frac{\Gamma(D^* \pi)}{\Gamma(D \pi)} \simeq 1.2,
\end{equation}
which may be useful to test the nature of $D_J^*(3000)$.

In the $D_s$-meson sector, the predicted mass for $D_s(1^3F_4)$,
is $3134$ MeV, which is comparable with the predictions in Refs.~\cite{Godfrey:2015dva,Zeng:1994vj,Song:2015nia}.
Our predicted strong decay properties are listed in Table~\ref{Dsmeson1FDecay}.
It is found that the $D_s(1^3F_4)$ may be a narrow state with a width of
\begin{equation}
\Gamma\simeq 31 \ \ \ \mathrm{MeV },
\end{equation}
and dominantly decays into the $DK$,
$D^*K$ and $D^*K^*$ channels with branching fractions $\sim 26\%$, $\sim 29\%$ and $\sim 18\%$, respectively.
To see the dependence of the decay properties of $D_s(1^3F_4)$ on its mass,
we also plot the main partial widths and the total width as functions of the mass in
Fig.~\ref{DFwaveMirunning}. It is found that the partial and total decay widths increase smoothly with the
mass. With a mass uncertainty of $50$ MeV, the total width of $D_s(1^3F_4)$ varies in the range $\sim 20-50$ MeV.
The $DK$, $D^*K$ and $D^*K^*$ as the main decay channels are also predicted in
Refs.~\cite{Godfrey:2015dva,Song:2015fha,Xiao:2014ura}, however, the predicted width in
these works are much broader than ours. The $D_s(1^3F_4)$ might have large potentials to be observed
in the dominant $DK$ and $D^*K$ channels.

\subsubsection{$1F_3$ and $1F_3'$ }

The physical states $1F_3$ and $1F'_3$ are mixed states between $1^1F_3 $ and $1^3F_3$
via the following mixing scheme:
\begin{equation}
\left(
  \begin{array}{c}
   1F_3\\
   1F'_3\\
  \end{array}\right)=
  \left(
  \begin{array}{cc}
   \cos\theta_{1F} &\sin\theta_{1F}\\
  -\sin\theta_{1F} &\cos\theta_{1F}\\
  \end{array}
\right)
\left(
  \begin{array}{c}
  1^{1}F_{3}\\
  1^{3}F_{3}\\
  \end{array}\right).
\end{equation}
In this work, the $1F_3$ and $1F'_3$ correspond to the low-mass and high-mass mixed states, respectively.

In the $D$-meson family, the predicted masses for the two mixed $F$-wave states
$D(1F_3)$ and $D(1F_3')$ are $M=3022$ and $3129$ MeV, respectively, which are comparable
with the predictions in Refs.~\cite{Zeng:1994vj,Song:2015fha}.
The mixing angle is determined to be $\theta_{1F}=-41^\circ$, which is similar to that determined within the relativized
quark model~\cite{Godfrey:2015dva,Lu:2014zua}.
Our predicted strong decay properties are listed in Table~\ref{Dmeson1FwaveDecay}.
It is found that the low mass mixed state $D(1F_3)$ is a fairly broad state with a width of
\begin{equation}
\Gamma\simeq 340 \ \ \ \mathrm{MeV },
\end{equation}
and dominantly decays into
$D_2^*(2460)\pi$, $D_3^*(2760)\pi$ and $D^*\pi$ with branching fractions $\sim 50\%$, $\sim 26\%$ and $\sim 12\%$, respectively.
The decay properties of $D(1F_3)$ predicted in this work are roughly consistent with those predicted
with the SHO wave functions in our previous work~\cite{Xiao:2014ura},
however, is notably (a factor of $\sim2$) broader than those predictions within the
$^3P_0$ models in Refs.~\cite{Godfrey:2015dva,Song:2015fha}.
While the high mass mixed state $D(1F_3')$ is a relatively narrow state with a width of
\begin{equation}
\Gamma\simeq 170 \ \ \ \mathrm{MeV },
\end{equation}
and dominantly decays into $D^*\pi$, $D_2^*(2460)\pi$, $D_1(2420)\pi$ and $D(1^3D_1)\pi$ with branching fractions
$\sim 25\%$, $\sim 21\%$, $\sim 12\%$ and $\sim 11\%$, respectively.
To see the dependence of the decay properties of $D(1F_3')$ on its mass,
we also plot the main partial widths and the total width as functions of the mass in
Fig.~\ref{DFwaveMirunning}. It is found that the partial and total decay widths increase smoothly with the
mass. With a mass uncertainty of $50$ MeV, the total width of $D(1F_3')$ varies in the range $\sim 140-260$ MeV.
The decay width of $D(1F_3')$ predicted in this work is comparable with our previous prediction with
the SHO wave functions in Ref.~\cite{Xiao:2014ura}, however, is
about a factor of $\sim2$ larger than that predicted in Ref.~\cite{Song:2015fha}. To look for the
missing $D(1F_3)$ and $D(1F_3')$ states, the $D^*\pi$ and $D_2^*(2460)\pi$ final states
are worth to observing in future experiments.

In the $D_s$-meson family, the predicted masses for the two mixed $F$-wave states
$D_s(1F_3)$ and $D_s(1F_3')$ are $M=3123$ and $3205$ MeV, respectively, which are comparable
with the predictions in Refs.~\cite{Zeng:1994vj,Song:2015fha}. The mixing angle is determined
to be $\theta_{1F}=-40.7^\circ$, which is similar to that determined within the relativized
quark model~\cite{Godfrey:2015dva,Lu:2014zua}.
Our predicted strong decay properties are listed in Table~\ref{Dsmeson1FDecay}.
It is found that the low mass mixed state $D_s(1F_3)$ may be a fairly broad state with a width of
\begin{equation}
\Gamma\simeq 210 \ \ \ \mathrm{MeV },
\end{equation}
and dominantly decays into
$D_2^*(2460)K$ and $D^*K$ with branching fractions $\sim 68\%$ and $\sim 19\%$, respectively.
The dominant decay channels of $D_2^*(2460)K$ and $D^*K$ predicted in this work are
consistent with the predictions in Refs.~\cite{Song:2015fha,Xiao:2014ura}, although there are large
uncertainties in the predictions of the total width.
The high mass mixed state $D_s(1F_3')$ is a narrow state with a width of
\begin{equation}
\Gamma\simeq 72 \ \ \ \mathrm{MeV },
\end{equation}
and dominantly decays into
$D^*K$, $D_2^*(2460)K$ and $D^*K^*$ with branching fractions
$\sim 37\%$, $\sim 18\%$ and $\sim 17\%$, respectively.
To see the dependence of the decay properties of $D_s(1F_3')$ on its mass,
we also plot the main partial widths and the total width as functions of the mass in
Fig.~\ref{DFwaveMirunning}. It is found that the partial and total decay widths increase smoothly with the
mass. With a mass uncertainty of $50$ MeV, the total width of $D_s(1F_3')$ varies in the range $\sim 40-100$ MeV. The decay width predicted
in this work is notably narrower than those predicted in Refs.~\cite{Godfrey:2015dva,Song:2015fha,Xiao:2014ura}.
To look for the missing $D_s(1F_3)$ and $D_s(1F_3')$ states, the $D^*K$ and $D_2^*(2460)K$ final states
are worth to observing in future experiments.

\begin{figure}
\centering \epsfxsize=7.6 cm \epsfbox{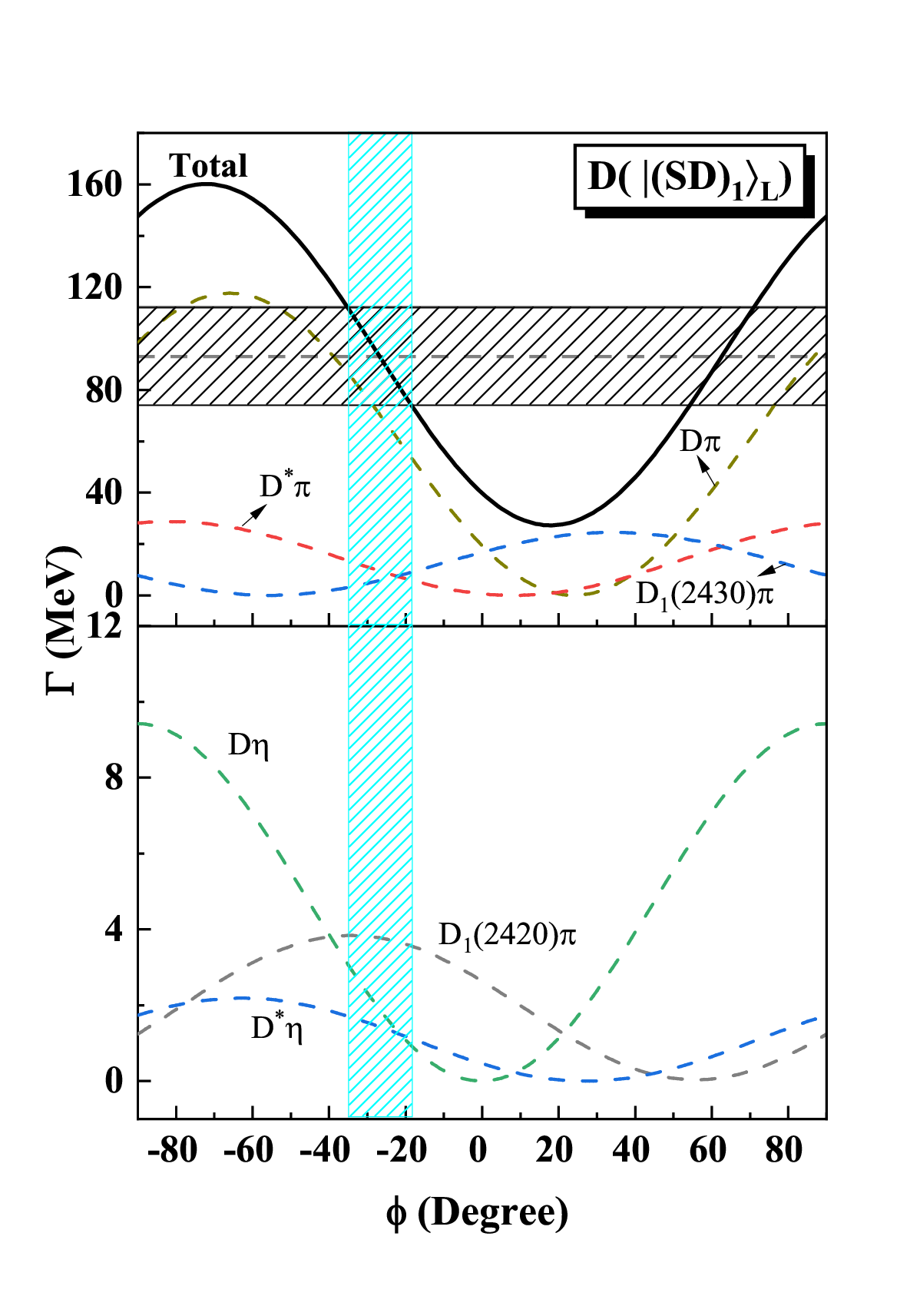} \vspace{-0.6 cm} \caption{Decay properties of $D(|(SD)_1\rangle_L)$ as functions of the mixing angle $\phi$. The horizontally shaded region stands for the measured width $\Gamma_{exp} = 93 \pm6 \pm13$~MeV at BaBar~\cite{BaBar:2010zpy}.
The longitudinally shaded region represents the possible range of the mixing angle $\phi\simeq -(27\pm 8)^\circ$.}\label{DSDLwidthandphi}
\end{figure}

\begin{figure}
\centering \epsfxsize=8.8 cm \epsfbox{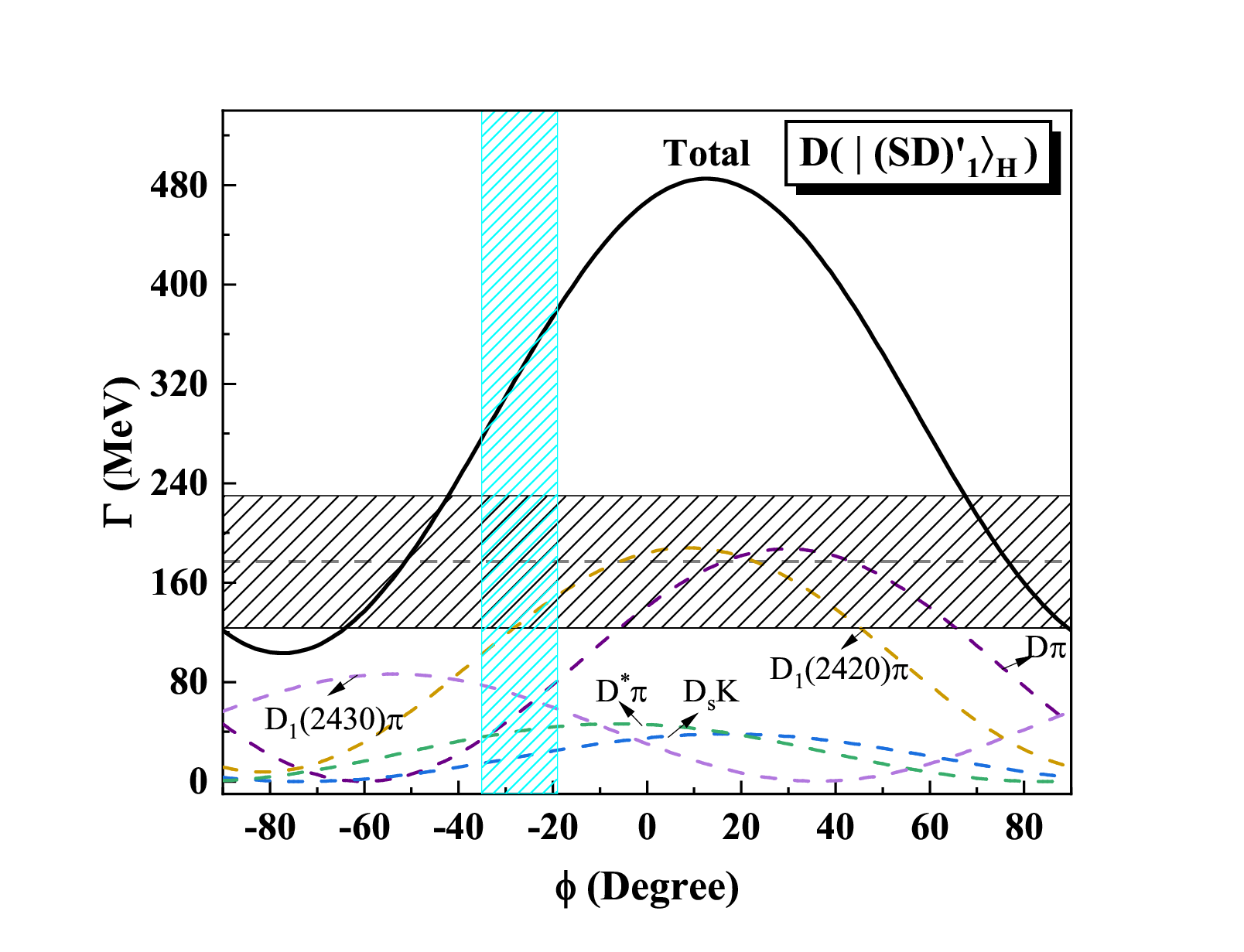} \vspace{-0.0 cm} \caption{Decay properties of $D(|(SD)'_1\rangle_H)$ as functions of the mixing angle $\phi$. The horizontally shaded region stands for the measured width $\Gamma_{exp}\simeq 177\pm53$~MeV of $D_{1}^*(2760)$ at LHCb~\cite{LHCb:2015eqv}. Within the possible mixing angle range $\phi \simeq-(27\pm8)^{\circ}$,  the width is determined to $\Gamma\simeq325\pm53$~MeV, which is shown by longitudinally shaded region. }\label{DSDHwidthandphi}
\end{figure}

\begin{figure}
\centering \epsfxsize=7.6 cm \epsfbox{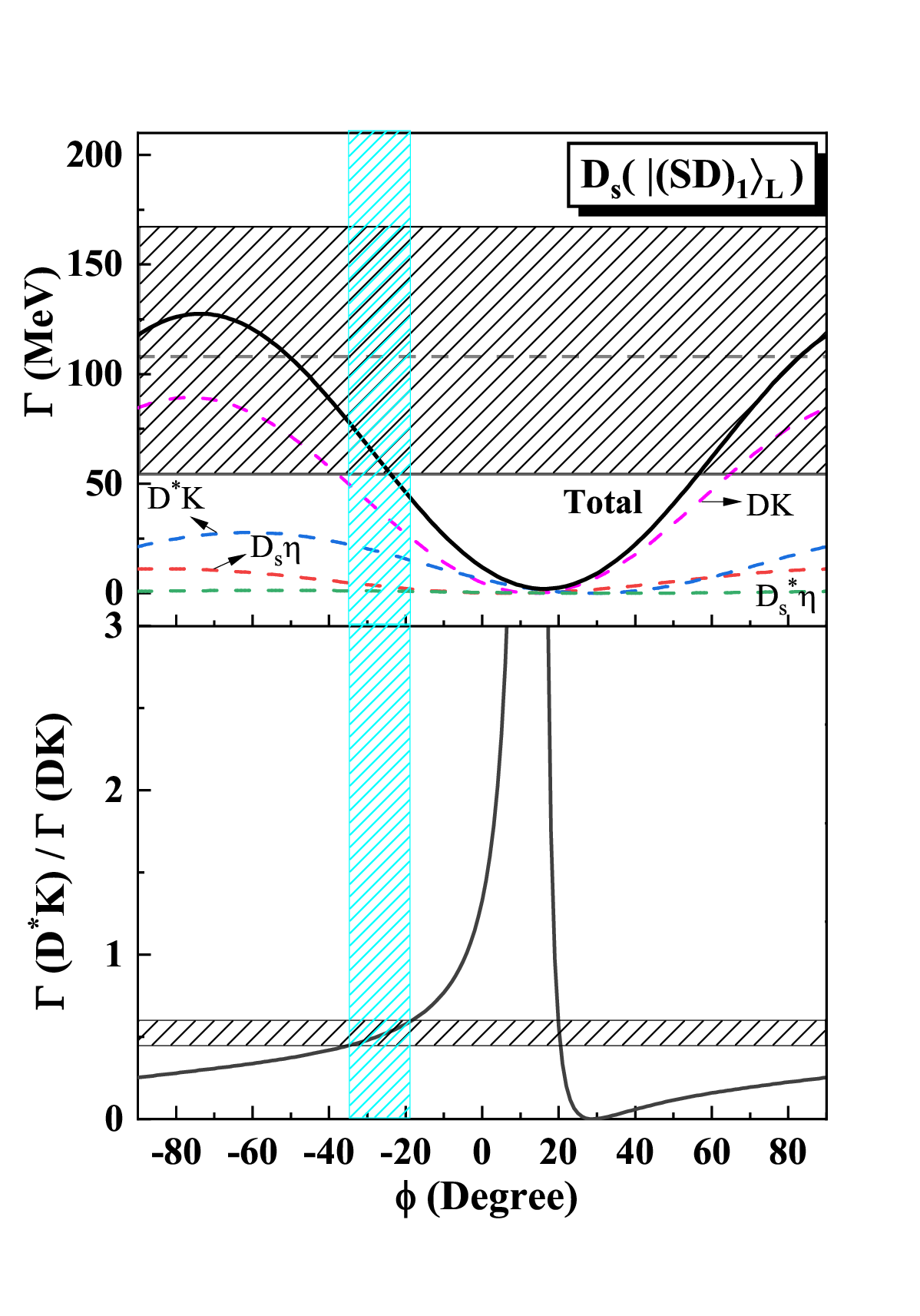} \vspace{-0.6 cm} \caption{Decay properties of $D_s(|(SD)_{1}\rangle_L)$ as functions of the mixing angle $\phi$. The horizontally shaded region stands for the measured width $\Gamma_{exp} = 108 \pm23^{+36}_{-31}$~MeV at Belle~\cite{Belle:2007hht}. Within the possible mixing angle range $\phi \simeq-(27\pm8)^{\circ}$,  the width is determined to $\Gamma\simeq61\pm18$~MeV, which is shown by longitudinally shaded region.}\label{DsSDLwidthandphi}
\end{figure}

\begin{figure}
\centering \epsfxsize=8.8 cm \epsfbox{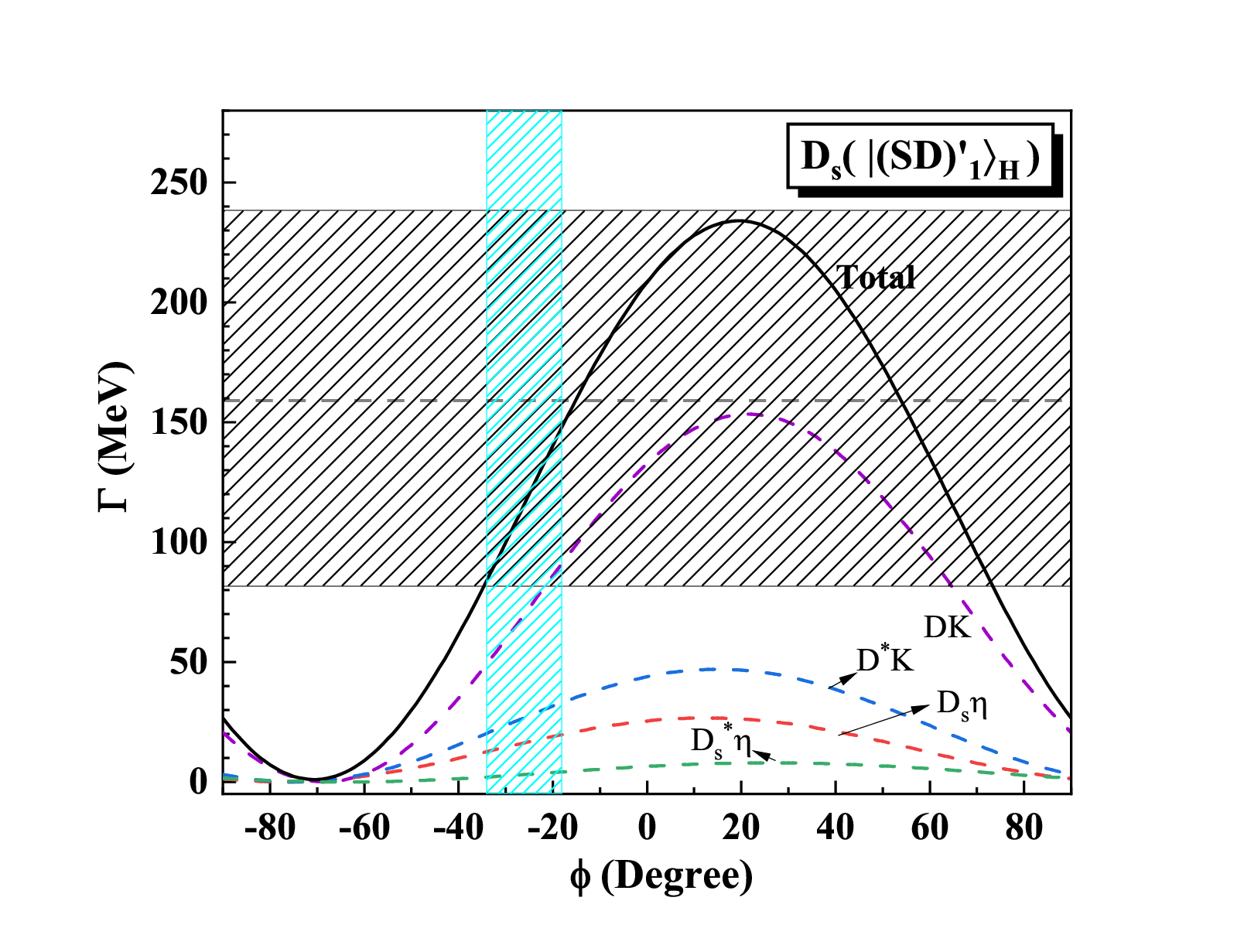} \vspace{-0.6 cm} \caption{Decay properties of $D_s(|(SD)'_{1}\rangle_H)$ as functions of the mixing angle $\phi$. The horizontally shaded region stands for the average measured width $\Gamma_{exp}\simeq 159\pm80$~MeV of $D_{s1}^*(2860)$ from PDG~\cite{Zyla:2020zbs}. Within the possible mixing angle range $\phi \simeq-(27\pm8)^{\circ}$,  the width is determined to $\Gamma\simeq112\pm32$~MeV, which is shown by longitudinally shaded region. }\label{DsSDHwidthandphi}
\end{figure}

\subsection{The $2^3S_1-1^3D_1$ mixing}

In Sec.~\ref{ab}, we have considered the possibility of the $D_1^*(2600)$ and $D_{s1}^*(2700)$ as
the candidates of the pure $2^3S_1$ states in the $D$- and $D_s$-meson families, respectively.
It is found that with these assignments, our predicted widths are too narrow to be
comparable with the data. In our previous works~\cite{Zhong:2010vq,Zhong:2009sk}, we have
studied the strong decay properties of the $D_1^*(2600)$ and
$D_{s1}^*(2700)$ with an SHO approximation. According to our analysis, both $D_1^*(2600)$ and
$D_{s1}^*(2700)$ could be explained as the mixed state
$|(SD)_1\rangle_L$ via the $2^3S_1$-$1^3D_1$ mixing:
\begin{equation}\label{mixsd}
\left(\begin{array}{c}|(SD)_1\rangle_L\cr |(SD)'_1\rangle_H
\end{array}\right)=\left(\begin{array}{cc} \cos\phi &\sin\phi\cr -\sin\phi & \cos\phi
\end{array}\right)
\left(\begin{array}{c}  2^3S_1  \cr  1^3D_1
\end{array}\right),
\end{equation}
where $\mid (SD)_1\rangle_L$ and $\mid (SD)'_1\rangle_H$ are assigned to the
low-mass and high-mass mixed states, respectively. The mixing angle for the charmed sector is $\phi\simeq-(36\pm 6)^\circ$,
while that for charmed-strange sector is $\phi=(-54\pm7)^\circ$. To
explain the strong decay properties of the $D_1^*(2600)$ and/or
$D_{s1}^*(2700)$, configuration mixing between $2^3S_1$ and $1^3D_1$
is also suggested in the literature~\cite{Close:2006gr,Chen:2011rr,Li:2009qu,Yu:2020khh,Sun:2010pg,
Li:2010vx,Chen:2015lpa,Song:2015nia}. In this work we restudy the $D_1^*(2600)$ and $D_{s1}^*(2700)$
resonances as the mixed states via the $2^3S_1$-$1^3D_1$ mixing by using
the genuine wave functions calculated from our potential model.

Considering $D_1^*(2600)$ as the low-mass mixed state $|(SD)_1\rangle_L$, we plot the
strong decay properties as functions of the mixing angle $\phi$ in Figure~\ref{DSDLwidthandphi}.
It is found that if we take a mixing angle $\phi\simeq -(27\pm 8)^\circ$,
the theoretical width can be consistent with the data $\Gamma_{exp}=96\pm6\pm13$ MeV
measured by the BaBar collaboration~\cite{BaBar:2010zpy}. The mixing angle
$\phi\simeq -(27\pm 8)^\circ$ determined in this work is similar to
$\phi\simeq -(36\pm 6)^\circ$ determined  in our previous work~\cite{Zhong:2010vq}.
The $D\pi$ and $D^*\pi$ are the two dominant decay channels of $\mid (SD)_1\rangle_L$,
which can explain why $D_1^*(2600)$ has been first observed in these two channels.
However, the ratio between $D\pi$ and $D^*\pi$
\begin{equation}
R=\frac{\Gamma(D \pi)}{\Gamma(D^* \pi)} \simeq 7.3\pm 1.4,
\end{equation}
is too large to be comparable with the data $\Gamma(D\pi)/\Gamma(D^*\pi ) = 0.32\pm0.11$
measured by the BaBar collaboration~\cite{BaBar:2010zpy}. The ratio predicted with the
genuine wave functions determined from the potential model in this work
is about a factor of $11$ larger than that predicted with the SHO wave
functions in our previous work~\cite{Zhong:2010vq}. The ratio is very sensitive to the details
of the wave function of $2^3S_1$ due to the nodal effects. Thus, the partial width
ratio $\Gamma(D\pi)/\Gamma(D^*\pi )$ is hard to be accurately predicted in theory.


If the $D_1^*(2600)$ is the low-mass mixed state $\mid (SD)_1\rangle_L$ indeed,
the high-mass mixed state $\mid (SD)'_1\rangle_H$ might be observed in experiments as well.
It is interesting to find that the $J^P=1^-$ resonance $D_1^*(2760)$ observed in the $D^+\pi^-$ channel
by the LHCb collaboration~\cite{LHCb:2015eqv} might be a candidate of
the high-mass mixed state $\mid (SD)'_1\rangle_H$ in the $D$-meson family.
Considering the $D_1^*(2760)$ as the $\mid (SD)'_1\rangle_H$ assignment,
the strong decay properties as functions of the mixing angle are plotted in
Figure~\ref{DSDHwidthandphi}. It is found that within the range of the
mixing angle $\phi\simeq -(27\pm 8)^\circ$ determined by the
$D_1^*(2600)$, the width of $D_1^*(2760)$ is predicted to
be $\Gamma\simeq 270-380$ MeV, which is close to the upper limit of the
measured width $\Gamma =177\pm 53$ MeV. As the high-mass mixed state,
$D_1^*(2760)$ should dominantly decay into the $D_1(2420)\pi$,
$D_1(2430)\pi$ and $D\pi$ channels. To confirm the nature of
$D_1^*(2760)$, both $D_1(2420)\pi$ and $D_1(2430)\pi$ channels
are worth observing in future experiments.

In the $D_s$-meson family, considering $D_{s1}^*(2700)$ as the low-mass mixed state $| (SD)_{1}\rangle_L$, we plot the
strong decay properties as functions of the mixing angle $\phi$ in Figure~\ref{DsSDLwidthandphi}.
One sees that if we take the mixing angle with $\phi\simeq -(27\pm 8)^\circ$, the decay width
$\Gamma_{exp}=113_{-37}^{+41}$ MeV and partial width ratio $R_{exp}=\Gamma(D^*K)/ \Gamma(DK) \simeq 0.91 \pm0.25$
of $D_{s1}^*(2700)$ measured by the BaBar collaboration~\cite{Lees:2014abp,BaBar:2009rro}
can be well described within the uncertainties.


If the $D_{s1}^*(2700)$ is the low-mass mixed state $|(SD)_{1}\rangle_L$ indeed,
the high-mass mixed state $|(SD)'_{1}\rangle_H$ might be observed in experiments as well.
The $J^P=1^-$ resonance $D_{s1}^*(2860)$ observed in the $\bar{D}^0K^-$ final state
by the LHCb collaboration~\cite{LHCb:2014ott} might be a candidate of
the high-mass mixed state $|(SD)'_{1}\rangle_H$ in the $D_s$-meson family.
Considering $D_{s1}^*(2860)$ as the $|(SD)'_{1}\rangle_H$ assignment,
the strong decay properties as functions of the mixing angle are plotted in
Figure~\ref{DsSDHwidthandphi}. It is found that if we take mixing angle
$\phi\simeq -(27\pm 8)^\circ$, the predicted width of $D_{s1}^*(2860)$, $\Gamma\simeq 112\pm32$ MeV,
is consistent with the measured width of $\Gamma=159 \pm 23 \pm77$~MeV~\cite{LHCb:2014ott}.
The partial width ratio between $D^*K$ and $DK$ channels is predicted to be
\begin{equation}
R=\frac{\Gamma(D^*K)}{\Gamma(DK)} \simeq 0.38\pm 0.03,
\end{equation}
which can be used to test the nature of $D_{s1}^*(2860)$.

As a whole, our underestimation of the decay widths of $D_1^*(2600)$ and $D_{s1}^*(2700)$
as a pure $2^3S_1$ configuration can be overcome by mixing with some $1^3D_1$-wave components.
Meanwhile, the widths of the $J^P=1^-$ resonances $D_1^*(2760)$ and $D_{s1}^*(2860)$ observed by the
LHCb collaboration can be reasonably explained with the high-mass mixed states $D(|(SD)'_1\rangle_H)$
and $D_s(|(SD)'_1\rangle_H)$, respectively. However, the measured ratio $\Gamma(D\pi)/\Gamma(D^*\pi ) = 0.32\pm0.11$
for $D_1^*(2600)$ is inconsistent with our predictions. To clarify the natures of these
$J^P=1^-$ charmed and charmed-strange meson resonances and test various model predictions, (i) both $D_1^*(2760)$ and $D_{s1}^*(2860)$
are waiting to be confirmed by other experiments; (ii) the partial width ratio
$\Gamma(D\pi)/\Gamma(D^*\pi )$ for $D_1^*(2600)$ and $\Gamma(DK)/\Gamma(D^*K)$ for $D_{s1}^*(2700)$
are waiting to be confirmed by other experiments; (iii) the resonance parameters of
$D_1^*(2600)$ and $D_{s1}^*(2700)$ are waiting to be accurately measured in future experiments.

\section{Summary}\label{Summary}

In this work we systematically calculate the mass spectra of charmed and charmed-strange meson states up to the $2D$ excitations
with a semi-relativistic potential model. Our results are in good agreement with other quark model predictions, although there are some model dependencies in the predicted masses for the higher $2D$- and $1F$-wave states. The strong decay properties are further analyzed with a chiral quark model by using the numerical wave functions obtained from the potential model. To well understand the
$1P$-wave states, we also systematically consider the coupled-channel effects on the masses of the $1P$-wave states by
using the strong decay amplitudes obtained within the chiral quark model. Based on our good descriptions of the mass and decay properties for the low-lying well-established states $D_1(2420)$, $D_1(2430)$, $D_2(2460)$, $D_{s1}(2536)$ and $D_{s2}(2573)$, we give a quark model classification for the high mass resonances observed in recent years. Our main conclusions are summarized as follows.

$\bullet$~There are notable couple-channel corrections to the
bare masses for the $D(1^3P_0)$, $D_s(1^3P_0)$ and $D_s(1P_1)$ states.
The $D_{s0}^*(2317)$ and $D_{s1}(2460)$ can be explained with
the dressed states $D_s(1^3P_0)$ and $D_s(1P_1)$ by the
$DK$ and $D^*K$ loops, respectively. The physical mass for
the dressed $D(1^3P_0)$ state is predicted to be $\sim 2253$ MeV,
which is about 50 MeV lower than the PDG average mass of $D_0^*(2300)$.

$\bullet$~The $D_0(2550)$ resonance can be classified as the $D(2^1S_0)$ state. Considering the newly observed $D_{s0}(2590)$
as the flavor partner of $D_0(2550)$, the physical mass of $D_s(2^1S_0)$, $M_{phy}=2581$ MeV, is close to
the observed mass by including the coupled-channel effects, however, our predicted
width is much smaller than the observed one.

$\bullet$~$D_3^*(2750)$ and $D_2(2740)$ can be classified as the $1D$-wave states with the assignments
$D(1^3D_3)$ and $D(1D'_2)$, respectively. The $D_{s3}^*(2860)$ resonance should be
the flavor partner of $D_3^*(2750)$, and correspond to the $D_s(1^3D_3)$ state. The $D_s(1D'_2)$ state,
as the flavor partner of $D_2(2740)$, is most likely to be observed in the $D^*K$ channel due to its narrow width nature.

$\bullet$~$D_J^*(3000)$ is more favor a candidate of $D(1^3F_4)$ or $D(2^3P_2)$. As the
$D(1^3F_4)$ assignment the predicted width is about a factor of 2 smaller than the observation,
while as the $D(2^3P_2)$ assignment the predicted width is about a factor of 2.5 larger than
the observation.

$\bullet$~$D_J(3000)$ may favor the $2P$-wave high-mass mixed state $D(2P'_1)$.
The $D_{sJ}(3040)$ resonance also favor the $2P$-wave mixed state $D_s(2P_1)$ or $D_s(2P_1')$.
Considering $D_{sJ}(3040)$ as $D_s(2P_1)$, the predicted width is close to the upper limit of the data,
while as the $D_S(2P'_1)$ assignment the predicted width is close to the lower limit of the data.
The $D_{sJ}(3040)$ may be contributed by both $D_s(2P_1)$ and $D_s(2P_1')$.

$\bullet$~$D_{s1}^*(2700)$ and $D_{s1}^*(2860)$ may favor the mixed states
$|(SD)_1\rangle_L$ and $|(SD)'_1\rangle_H$ via the $2^3S_1$-$1^3D_1$ mixing, respectively.

$\bullet$~There still exist puzzles for understanding the natures of $D_1^*(2600)$ and $D_1^*(2760)$.
Considering $D_1^*(2600)$ and $D_1^*(2760)$ as $D(2^3S_1)$ and $D(1^3D_1)$, respectively,
the predicted widths are inconsistent with the data. While considering them as the
mixed states $|(SD)_1\rangle_L$ and $|(SD)'_1\rangle_H$, their widths are reasonably
consistent with the data, however, the ratio $R=(D\pi)/(D^*\pi)$ for $D_1^*(2600)$
is inconsistent with the observation.

$\bullet$~Many missing excited $D$- and $D_s$-meson states, such as $D_s(1D_2)$, $D_s(1D_2')$, $D_s(2^3P_2)$, $D(2^3D_3)$/$D_s(2^3D_3)$, $D(2D_2')$/$D_s(2D_2')$, $D(1^3F_4)$/$D_s(1^3F_4)$ and $D(1F_3')$/$D_s(1F_3')$, have a relatively narrow width,
they are most likely to be observed in their dominant decay channels in future experiments.

\section*{Acknowledgement}

The authors thank Prof. Xiang Liu and Dr. Zhi Yang for very helpful discussions. This work is supported by the National Natural Science Foundation of China (Grants Nos. U1832173, 11775078, 12175065).


\begin{appendix}

%

\section{Coupled-channel model}\label{Coupled-channel quark model}

In this appendix, we give the details of the model including the coupled-channel effects
on the charmed and charmed-strange meson mass spectra.
This simple coupled-channel model has been widely adopted in the literature~\cite{Lu:2017hma,Luo:2019qkm,Luo:2021dvj,Yang:2021tvc,Xie:2021dwe,Ortega:2021fem,Ortega:2016mms,Liu:2011yp,
Lu:2016mbb,Kalashnikova:2005ui,Eichten:1978tg}.

\begin{figure}
\centering \epsfxsize=7.4 cm \epsfbox{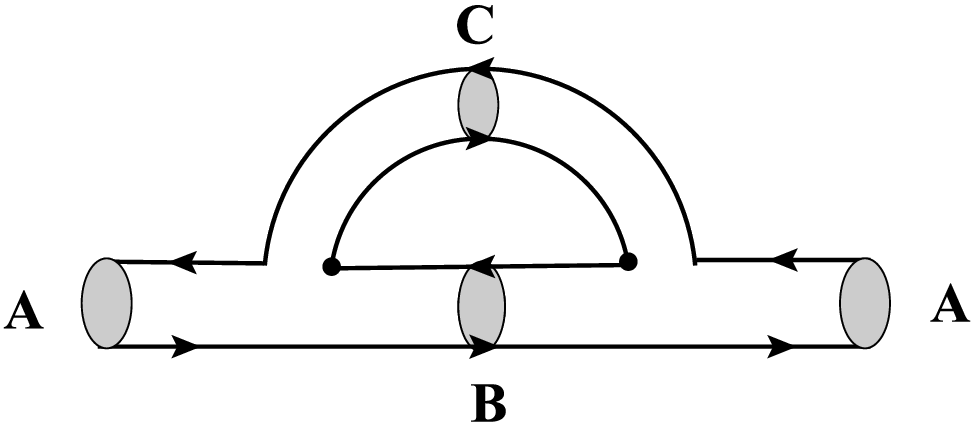} \vspace{-0.0 cm} \caption{The $BC$ hadronic loop coupled to
a bare meson state $|A \rangle$. }\label{CoupledChanneleffectsfig}
\end{figure}

A bare meson state $|A\rangle$ in the quark model can couple to the two-hadron continuum $BC$
by hadronic loops as shown in Fig.~\ref{CoupledChanneleffectsfig}.
The experimentally observed state may be an admixture between the bare state and continuum components, thus,
the wave function of the physical state is given by
\begin{equation}
\begin{aligned}
| \psi \rangle =c_A |A\rangle + \sum_{BC}\int c_{BC}(\mathbf{p})|BC,\mathbf{p}\rangle d^3\mathbf{p},
\end{aligned}
\end{equation}
where $\mathbf{p}=\mathbf{p}_B=-\mathbf{p}_C$ is final two-hadron relative momentum in the initial hadron static system, $c_A$ and $c_{BC}(\mathbf{p})$ denote the probability amplitudes of the bare valence state $|A\rangle$ and $|BC,\mathbf{p}\rangle$ continuum components, respectively.

The full Hamiltonian of this mixed system $| \psi \rangle$ can be written as
\begin{equation}
\begin{aligned}
\hat{H} = \left( \begin{matrix}\hat{H}_0~~~~~~\hat{H}_I
\\  \hat{H}_I~~~~~~\hat{H}_c \end{matrix}\right).
\end{aligned}
\end{equation}
In the above equation, $\hat{H}_0$ is the Hamiltonian of the bare meson state $|A\rangle$ in the
potential model, while $\hat{H}_c$ is the Hamiltonian for the continuum state $|BC,\mathbf{p}\rangle$.
Neglecting the interaction between the hadrons $B$ and $C$, one has
\begin{equation}
\begin{aligned}
\hat{H}_{c}|BC,\mathbf{p}\rangle&=E_{BC}|BC,\mathbf{p}\rangle,
\end{aligned}
\end{equation}
where $E_{BC}=\sqrt{m_B^2+p^2}+\sqrt{m_C^2+p^2}$ represents the energy of $BC$ continuum components.
The mixing between $|A\rangle$ and $|BC,\mathbf{p}\rangle$ is caused by
the Hamiltonian $\hat{H}_I$, which can be borrowed from our chiral quark model.

The Schr\"{o}dinger equation of a mixed system can be written as
\begin{equation}
\begin{aligned}\label{coupled-channel equation}
&\left( \begin{matrix}\hat{H}_0~~~~~~\hat{H}_I
\\  \hat{H}_I~~~~~~\hat{H}_c \end{matrix}\right)
~\left( \begin{matrix} c_A |A\rangle
\\ \sum_{BC}\int c_{BC}(\mathbf{p})|BC,\mathbf{p}\rangle d^3\mathbf{p} \end{matrix}\right) \\
&~~~~~~~~~~~~~~~~~~~~~~~~~~~~~~~=M ~\left( \begin{matrix} c_A |A\rangle
\\  \sum_{BC}\int c_{BC}(\mathbf{p})|BC,\mathbf{p}\rangle d^3\mathbf{p}   \end{matrix}\right).
\end{aligned}
\end{equation}
From Eq.(~\ref{coupled-channel equation}), we have
\begin{equation}\label{coupled-channel equation1}
\begin{aligned}
\langle A | \hat{H} | \psi \rangle =c_A M=c_A M_A+ \sum_{BC}\int c_{BC}(\mathbf{p}) \langle A | \hat{H}_I | BC,\mathbf{p}\rangle d^3\mathbf{p},
\end{aligned}
\end{equation}
\begin{equation}\label{coupled-channel equation2}
\begin{aligned}
\langle BC,\mathbf{p} | \hat{H} | \psi \rangle =c_{BC}(\mathbf{p}) M=c_{BC}(\mathbf{p}) E_{BC}+ c_A\langle BC,\mathbf{p}| \hat{H}_I | A \rangle.
\end{aligned}
\end{equation}
Deriving $c_{BC}(\mathbf{p})$ from Eq.(\ref{coupled-channel equation2}), and substituting it into Eq.(\ref{coupled-channel equation1}), we get a coupled-channel equation
\begin{equation}\label{M=MA+Delta M}
\begin{aligned}
M=M_A+\Delta M(M),
\end{aligned}
\end{equation}
where the mass shift $\Delta M(M)$ is given by
\begin{equation}\label{ReDeltaM}
\begin{aligned}
\Delta M(M) &= \mathrm{Re} \sum_{BC}\int_0^{\infty} \frac{|\langle BC,\mathbf{p} |\hat{H}_I| A \rangle|^2}{(M-E_{BC})} p^2 dp,
\end{aligned}
\end{equation}
and $M_A$ is the bare mass of the meson state $|A\rangle$ obtained from the potential model.
From Eq.(\ref{M=MA+Delta M}) and Eq.(\ref{ReDeltaM}), the physical mass $M$ and the bare state mass shift $\Delta M$ can be determined simultaneously.

\begin{figure}
\centering \epsfxsize=9.8 cm \epsfbox{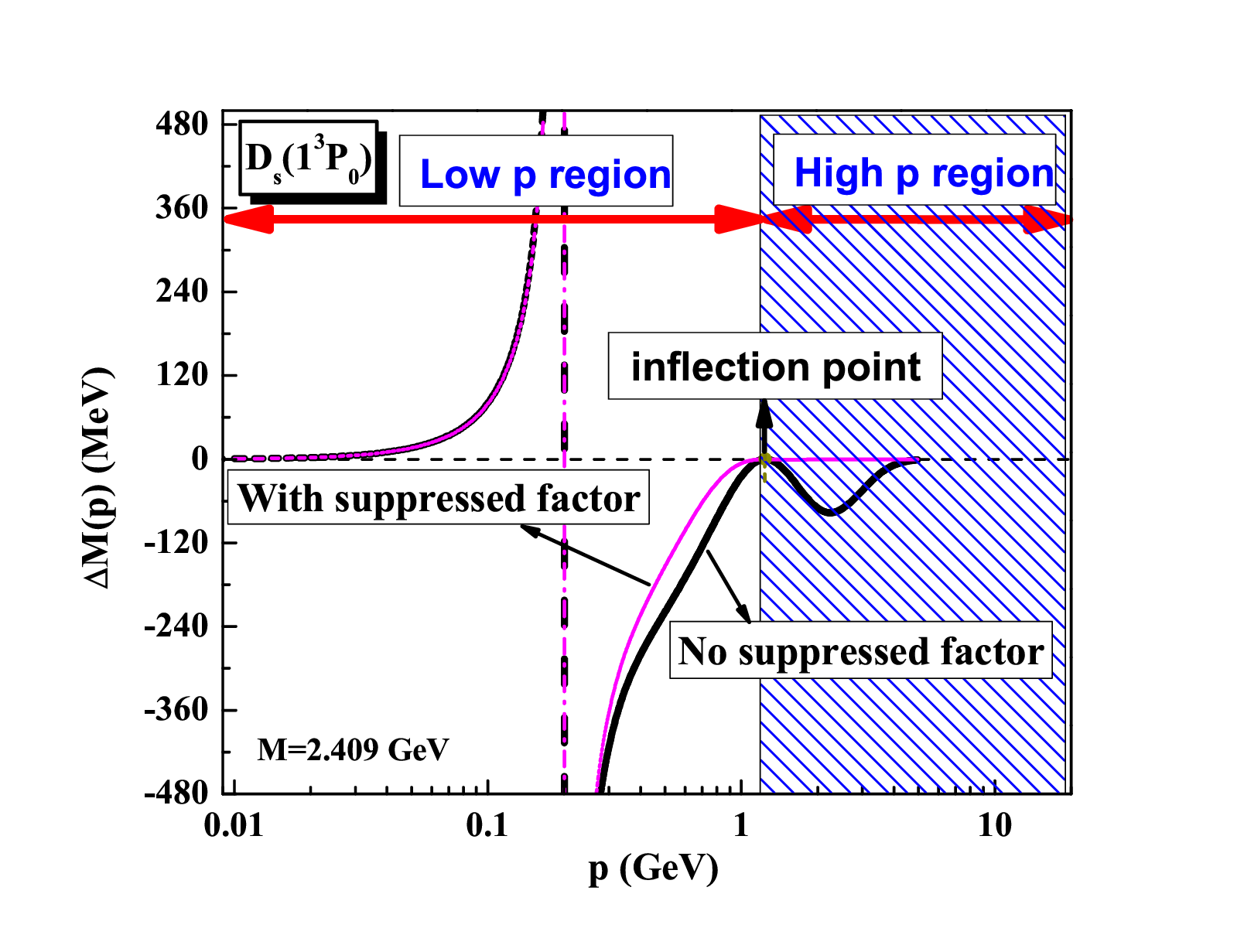} \vspace{-0.6 cm} \caption{The integral function $\Delta M(\mathbf{p})\equiv\frac{|\langle BC,\mathbf{p} |\hat{H}_I| A \rangle|^2}{(M-E_{BC})} p^2$
in Eq.(\ref{ReDeltaM}) for the $D_s(1^3P_0)$ state varies with the momentum $p=|\mathbf{p}|$.
The thin line stands for the results with a suppressed factor $e^{-p^2/(2\Lambda^2)}$ ($\Lambda=0.84$~GeV) as that adopted in Ref.~\cite{Ortega:2016mms},  while the thick line stands for the results without the suppressed factor. }\label{Ds13P0massshiftandp}
\end{figure}

It should be mentioned that when we calculate the mass shift $\Delta M$ by
using the Eq.(\ref{ReDeltaM}), the nonphysical contributions from
higher $\mathbf{p}$ region may be involved. To know the whole momentum region contributions,
as an example, considering the $DK$ loop, in Fig.~\ref{Ds13P0massshiftandp}
we plot the mass shift $\Delta M(\mathbf{p})$ of $D_s(1^3P_0)$ (i.e., the integral function
in Eq.(\ref{ReDeltaM})) as a function of the momentum $p=|\mathbf{p}|$.
It is found that two regions contribute to the mass shift. The main contribution region is
the low $\mathbf{p}$ region dominated by the pole. In the higher $\mathbf{p}$ region
of $p\simeq 1.2-4$ GeV, a small bump structure exists.
This bump contribution may be nonphysical, because the quark pair production rates via the non-perturbative interaction
$\hat{H}_I$ should be strongly suppressed in the high momentum region~\cite{Morel:2002vk,Tan:2021bvl}.
It should be mentioned that in the chiral quark model the
chiral interaction $\hat{H}_I$ is only applicable to the low $\mathbf{p}$ region.

To soften the hard vertices $\hat{H}_I$ in the higher momentum region,
and reasonably describe the mass shifts, an additional factor is suggested to be
introduced into the two-body decay amplitude
$\langle BC,\mathbf{p} |\hat{H}_I| A \rangle$~\cite{Morel:2002vk}.
Adopting suppressed factor $e^{-p^2/(2\Lambda^2)}$
with $\Lambda=0.84$ GeV as that used in Ref.~\cite{Ortega:2016mms},
we also plot the mass shift $\Delta M(\mathbf{p})$ of $D_s(1^3P_0)$ as a function of the momentum $p=|\mathbf{p}|$ in Fig.~\ref{Ds13P0massshiftandp}.
It is found that the factor $e^{-p^2/(2\Lambda^2)}$ indeed eliminates the contributions
from the high momentum region. To eliminate the nonphysical contributions, in our calculations we cut off the momentum $p$ at
the inflection point in $\Delta M(\mathbf{p})$ function as shown in Fig.~\ref{Ds13P0massshiftandp}.
It should be pointed out that the cut-off momentum for each meson states
is different due to the different position of the inflection point.
With this momentum cut-off approach, our predicted mass shifts due to coupled-channel effects for
the $D$ and/or $D_s$ meson states are consistent with the predictions in the
literature~\cite{Yang:2021tvc,Ortega:2016mms,Xie:2021dwe,Ortega:2021fem}.

\end{appendix}

\bibliographystyle{unsrt}

\begin{thebibliography}{120}

\bibitem{BaBar:2010zpy}
P.~del Amo Sanchez \textit{et al.} [BaBar],
Observation of new resonances decaying to $D\pi$ and $D^*\pi$ in inclusive $e^+e^-$ collisions near $\sqrt{s}=$10.58 GeV,
Phys. Rev. D \textbf{82}, 111101 (2010).

\bibitem{LHCb:2013jjb}
R.~Aaij \textit{et al.} [LHCb],
Study of $D_J$ meson decays to $D^+\pi^-$, $D^0 \pi^+$ and $D^{*+}\pi^-$ final states in $pp$ collision,
JHEP \textbf{09}, 145 (2013).
[arXiv:1307.4556 [hep-ex]].


\bibitem{LHCb:2015eqv}
R.~Aaij \textit{et al.} [LHCb],
First observation and amplitude analysis of the $B^{-}\to D^{+}K^{-}\pi^{-}$ decay,
Phys. Rev. D \textbf{91}, 092002 (2015)
[erratum: Phys. Rev. D \textbf{93}, 119901 (2016)].

\bibitem{Aaij:2016fma}
R.~Aaij \textit{et al.} [LHCb],
Amplitude analysis of $B^{-} \to D^{+} \pi^{-} \pi^{-}$ decays,
Phys. Rev. D \textbf{94}, 072001 (2016).

\bibitem{Aaij:2019sqk}
R.~Aaij \textit{et al.} [LHCb],
Determination of quantum numbers for several excited charmed mesons observed in $B^- \to D^{*+} \pi^- \pi^-$ decays,
Phys. Rev. D \textbf{101}, 032005 (2020).

\bibitem{Aaij:2015sqa}
R.~Aaij \textit{et al.} [LHCb],
Dalitz plot analysis of $B^0 \to \overline{D}^0 \pi^+\pi^-$ decays,
Phys. Rev. D \textbf{92}, 032002 (2015).


\bibitem{Zyla:2020zbs}
  P.~A.~Zyla {\it et al.} [Particle Data Group],
  Review of Particle Physics,
  PTEP {\bf 2020}, 083C01 (2020).

\bibitem{BaBar:2006gme}
 B.~Aubert \textit{et al.} [BaBar],
 Observation of a new $D_s$ meson decaying to $D K$ at a mass of 2.86 GeV/c$^2$,
  Phys.\ Rev.\ Lett.\  {\bf 97}, 222001 (2006).

\bibitem{Belle:2007hht}
J.~Brodzicka \textit{et al.} [Belle],
Observation of a new $D_{sJ}$ meson in $B^+\to \bar{D}^0D^0K^+$ decays,
Phys. Rev. Lett. \textbf{100}, 092001 (2008)

\bibitem{BaBar:2009rro}
  B.~Aubert {\it et al.}  [BaBar Collaboration],
  Study of $D_{sJ}$ decays to $D^*K$ in inclusive $e^+ e^-$ interactions,
  Phys.\ Rev.\ D {\bf 80}, 092003 (2009).


\bibitem{LHCb:2012uts}
R.~Aaij \textit{et al.} [LHCb],
Study of $D_{sJ}$ decays to $D^+K^0_S$ and $D^0K^+$ final states in $pp$ collisions,
JHEP \textbf{10}, 151 (2012).


\bibitem{LHCb:2014ott}
R.~Aaij \textit{et al.} [LHCb],
Observation of overlapping spin-1 and spin-3 $\bar{D}^0 K^-$ resonances at mass $2.86 {\rm GeV}/c^2$,
Phys. Rev. Lett. \textbf{113}, 162001 (2014).


\bibitem{LHCb:2014ioa}
R.~Aaij \textit{et al.} [LHCb],
Dalitz plot analysis of $B_s^0 \rightarrow \bar{D}^0 K^- \pi^+$ decays,
Phys. Rev. D \textbf{90}, 072003 (2014).

\bibitem{Aaij:2016utb}
R.~Aaij \textit{et al.} [LHCb],
Study of D$_{sJ}^{(*) +}$ mesons decaying to $D^{*+}K_S^0$ and $D^{*0}K^{+}$ final states,
JHEP \textbf{02}, 133 (2016).

\bibitem{LHCb:2020gnv}
R.~Aaij \textit{et al.} [LHCb],
Observation of a New Excited $D^+_s$ Meson in $B^0 \rightarrow D^- D^+ K^+ \pi^-$ Decays,
Phys. Rev. Lett. \textbf{126}, 122002 (2021).

\bibitem{Lees:2014abp}
J.~P.~Lees \textit{et al.} [BaBar],
Dalitz plot analyses of $B^0 \to D^-D^0K^+$ and $B^+ \to \bar{D}^0D^0K^+$ decays,
Phys. Rev. D \textbf{91}, no.5, 052002 (2015).


\bibitem{Chen:2018nnr}
J.~K.~Chen,
Regge trajectories for the mesons consisting of different quarks,
Eur. Phys. J. C \textbf{78}, 648 (2018).

\bibitem{Allosh:2021biq}
M.~Allosh, Y.~Mustafa, N.~Khalifa Ahmed and A.~Sayed Mustafa,
Ground and Excited State Mass Spectra and Properties of Heavy-Light Mesons,
Few Body Syst. \textbf{62}, 26 (2021).

\bibitem{Patel:2021aas}
V.~Patel, R.~Chaturvedi and A.~K.~Rai,
Spectroscopic properties of $D$-meson using screened potential,
Eur. Phys. J. Plus \textbf{136}, 42 (2021).

\bibitem{Liu:2013maa}
J.~B.~Liu and M.~Z.~Yang,
Spectrum of the charmed and b-flavored mesons in the relativistic potential model,
JHEP \textbf{07}, 106 (2014).

\bibitem{Liu:2015lka}
J.~B.~Liu and M.~Z.~Yang,
Spectrum of Higher excitations of $B$ and $D$ mesons in the relativistic potential model,
Phys. Rev. D \textbf{91}, 094004 (2015).


\bibitem{Liu:2016efm}
J.~B.~Liu and C.~D.~L\"{u},
Spectra of heavy-light mesons in a relativistic model,
Eur. Phys. J. C \textbf{77}, 312 (2017).

\bibitem{Liu:2015uya}
J.~B.~Liu and M.~Z.~Yang,
Heavy-Light Mesons In A Relativistic Model,
Chin. Phys. C \textbf{40}, 073101 (2016).

\bibitem{Badalian:2011tb}
A.~M.~Badalian and B.~L.~G.~Bakker,
Higher excitations of the $D$ and $D_s$ mesons,
Phys. Rev. D \textbf{84}, 034006 (2011).


\bibitem{Ebert:2009ua}
D.~Ebert, R.~N.~Faustov and V.~O.~Galkin,
Heavy-light meson spectroscopy and Regge trajectories in the relativistic quark model,
Eur. Phys. J. C \textbf{66}, 197-206 (2010).

\bibitem{Zhang:2009nu}
A.~Zhang,
Implications to $c\bar{s}$ assignments of $D_{s1}(2700)^{+-}$ and $D_{sJ}(2860)$,
Nucl. Phys. A \textbf{856}, 88-94 (2011).

\bibitem{Zhou:2014ytp}
D.~Zhou, E.~L.~Cui, H.~X.~Chen, L.~S.~Geng, X.~Liu and S.~L.~Zhu,
D-wave heavy-light mesons from QCD sum rules,
Phys. Rev. D \textbf{90}, 114035 (2014).

\bibitem{Chen:2009zt}
B.~Chen, D.~X.~Wang and A.~Zhang,
Interpretation of $D_{sJ}(2632)+$, $D_{s1}(2700)^{+-}$, $D^*_{sJ}(2860)^+$ and $D_{sJ}(3040)^+$,
Phys. Rev. D \textbf{80}, 071502 (2009).


\bibitem{Song:2015fha}
Q.~T.~Song, D.~Y.~Chen, X.~Liu and T.~Matsuki,
Higher radial and orbital excitations in the charmed meson family,
Phys. Rev. D \textbf{92}, 074011 (2015).

\bibitem{Lu:2014zua}
Q.~F.~L\"u and D.~M.~Li,
Understanding the charmed states recently observed by the LHCb and BaBar Collaborations in the quark model,
Phys. Rev. D \textbf{90}, 054024 (2014).

\bibitem{Godfrey:2015dva}
S.~Godfrey and K.~Moats,
Properties of Excited Charm and Charm-Strange Mesons,
Phys. Rev. D \textbf{93}, 034035 (2016).

\bibitem{Kher:2017wsq}
V.~Kher, N.~Devlani and A.~K.~Rai,
Excited state mass spectra, Decay properties and Regge trajectories of charm and charm-strange mesons,
Chin. Phys. C \textbf{41}, 073101 (2017).

\bibitem{Ferretti:2015rsa}
J.~Ferretti and E.~Santopinto,
Open-flavor strong decays of open-charm and open-bottom mesons in the $^3P_0$ model,
Phys. Rev. D \textbf{97}, 114020 (2018).

\bibitem{Segovia:2015dia}
J.~Segovia, D.~R.~Entem and F.~Fernandez,
Charmed-strange Meson Spectrum: Old and New Problems,
Phys. Rev. D \textbf{91}, 094020 (2015).

\bibitem{Gandhi:2019lta}
K.~Gandhi and A.~K.~Rai,
Strong decays analysis of excited nonstrange charmed mesons: Implications for spectroscopy,
Eur. Phys. J. A \textbf{57}, 23 (2021).

\bibitem{Sun:2013qca}
Y.~Sun, X.~Liu and T.~Matsuki,
Newly observed $D_J(3000)^{+,0}$ and $D_J^*(3000)^0$ as $2P$ states in $D$ meson family,
Phys. Rev. D \textbf{88}, 094020 (2013).

\bibitem{Li:2010vx}
D.~M.~Li, P.~F.~Ji and B.~Ma,
The newly observed open-charm states in quark model,
Eur. Phys. J. C \textbf{71}, 1582 (2011).

\bibitem{Yu:2020khh}
H.~Yu, Z.~Zhao and A.~Zhang,
Dynamical mixing between $2^3S_1$ and $1^3D_1$ charmed mesons,
Phys. Rev. D \textbf{102}, 054013 (2020).

\bibitem{Eshraim:2014eka}
W.~I.~Eshraim, F.~Giacosa and D.~H.~Rischke,
Phenomenology of charmed mesons in the extended Linear Sigma Model,
Eur. Phys. J. A \textbf{51}, 112 (2015).

\bibitem{Shah:2014caa}
M.~Shah, B.~Patel and P.~C.~Vinodkumar,
Mass spectra and decay properties of $D_s$ Meson in a relativistic Dirac formalism,
Phys. Rev. D \textbf{90}, 014009 (2014).

\bibitem{Song:2015nia}
Q.~T.~Song, D.~Y.~Chen, X.~Liu and T.~Matsuki,
Charmed-strange mesons revisited: mass spectra and strong decays,
Phys. Rev. D \textbf{91}, 054031 (2015).



\bibitem{Sun:2010pg}
Z.~F.~Sun, J.~S.~Yu, X.~Liu and T.~Matsuki,
Newly observed $D(2550)$, $D(2610)$, and $D(2760)$ as $2S$ and $1D$ charmed mesons,
Phys. Rev. D \textbf{82}, 111501 (2010).

\bibitem{Zhong:2010vq}
  X.~H.~Zhong,
  Strong decays of the newly observed $D(2550)$, $D(2600)$, $D(2750)$, and $D(2760)$,
  Phys.\ Rev.\ D {\bf 82}, 114014 (2010).

\bibitem{Xiao:2014ura}
 L.~Y.~Xiao and X.~H.~Zhong,
 Strong decays of higher excited heavy-light mesons in a chiral quark model,
 Phys. Rev. D \textbf{90}, 074029 (2014).

\bibitem{Yu:2016mez}
G.~L.~Yu, Z.~G.~Wang and Z.~Y.~Li,
Analysis of the charmed mesons $D_{1}^{*}(2680)$, $D_{3}^{*}(2760)$ and $D_{2}^{*}(3000)$,
Phys. Rev. D \textbf{94}, 074024 (2016).

\bibitem{Wang:2010ydc}
Z.~G.~Wang,
Analysis of strong decays of the charmed mesons $D(2550)$, $D(2600)$, $D(2750)$ and $D(2760)$,
Phys. Rev. D \textbf{83}, 014009 (2011).

\bibitem{Wang:2013tka}
Z.~G.~Wang,
Analysis of strong decays of the charmed mesons $D_J(2580), D^*_J(2650), D_J(2740), D^*_J(2760), D_J(3000), D^*_J(3000)$,
Phys. Rev. D \textbf{88}, 114003 (2013).


\bibitem{Yu:2014dda}
G.~L.~Yu, Z.~G.~Wang, Z.~Y.~Li and G.~Q.~Meng,
Systematic analysis of the $D_{J}(2580)$, $D_{J}^{*}(2650)$, $D_{J}(2740)$, $D_{J}^{*}(2760)$, $D_{J}(3000)$ and $D_{J}^{*}(3000)$ in $D$ meson family,
Chin. Phys. C \textbf{39}, 063101 (2015).

\bibitem{Gupta:2018zlg}
P.~Gupta and A.~Upadhyay,
Analysis of strong decays of charmed mesons $D^*_2(2460)$, $D_0(2560)$, $D_2(2740)$, $D_1(3000)$, $D^*_2(3000)$ and their spin partners $D^*_1(2680)$, $D^*_3(2760)$ and $D^*_0(3000)$,
Phys. Rev. D \textbf{97}, 014015 (2018).

\bibitem{Chen:2011rr}
B.~Chen, L.~Yuan and A.~Zhang,
Possible 2S and 1D charmed and charmed-strange mesons,
Phys. Rev. D \textbf{83}, 114025 (2011).

\bibitem{Chen:2015lpa}
B.~Chen, X.~Liu and A.~Zhang,
Combined study of $2S$ and $1D$ open-charm mesons with natural spin-parity,
Phys. Rev. D \textbf{92}, 034005 (2015).


\bibitem{Wang:2016enc}
T.~Wang, Z.~H.~Wang, Y.~Jiang, L.~Jiang and G.~L.~Wang,
Strong decays of $D_{3}^{*}(2760)$ , $D_{s3}^{*}(2860)$ , $B_{3}^{*}$ , and $B_{s3}^{*}$,
Eur. Phys. J. C \textbf{77}, 38 (2017).

\bibitem{Song:2014mha}
Q.~T.~Song, D.~Y.~Chen, X.~Liu and T.~Matsuki,
$D_{s1}^*(2860)$ and $D_{s3}^*(2860)$: candidates for $1D$ charmed-strange mesons,
Eur. Phys. J. C \textbf{75}, 30 (2015).

\bibitem{Wang:2014jua}
Z.~G.~Wang,
$D_{s3}^*(2860)$ and $D_{s1}^*(2860)$ as the 1D $c\bar{s}$ states,
Eur. Phys. J. C \textbf{75}, 25 (2015).

\bibitem{Li:2017sww}
Q.~Li, Y.~Jiang, T.~Wang, H.~Yuan, G.~L.~Wang and C.~H.~Chang,
Study of the excited $1^-$ charm and charm-strange mesons,
Eur. Phys. J. C \textbf{77}, 297 (2017).

\bibitem{Zhang:2016dom}
S.~C.~Zhang, T.~Wang, Y.~Jiang, Q.~Li and G.~L.~Wang,
Strong decays of $2^+$ charm and charm-strange mesons,
Int. J. Mod. Phys. A \textbf{32}, 1750022 (2017).

\bibitem{Tan:2018lao}
X.~Z.~Tan, T.~Wang, Y.~Jiang, S.~C.~Li, Q.~Li, G.~L.~Wang and C.~H.~Chang,
Strong decays of the orbitally excited scalar $D^*_0$ mesons,
Eur. Phys. J. C \textbf{78}, 583 (2018).

\bibitem{Zhong:2008kd}
  X.~H.~Zhong and Q.~Zhao,
  Strong decays of heavy-light mesons in a chiral quark model,
  Phys.\ Rev.\  D {\bf 78}, 014029 (2008).

\bibitem{Zhong:2009sk}
  X.~H.~Zhong and Q.~Zhao,
  Strong decays of newly observed $D_{sJ}$ states in a constituent quark
  model with effective Lagrangians,
  Phys.\ Rev.\  D {\bf 81}, 014031 (2010).

\bibitem{Sun:2009tg}
 Z.~F.~Sun and X.~Liu,
 Newly observed $D_{sJ}(3040)$ and the radial excitations of P-wave charmed-strange mesons,
 Phys. Rev. D \textbf{80}, 074037 (2009).

\bibitem{Wang:2016hkf}
Z.~G.~Wang,
Analysis of the strong decays $D_{s3}^*(2860)\to DK$, $D^{*}K$ with QCD sum rules,
Eur. Phys. J. A \textbf{52}, 303 (2016).

\bibitem{Li:2009qu}
D.~M.~Li and B.~Ma,
Implication of BaBar's new data on the$D_{s1}(2710)$ and $D_{sJ}(2860)$,
Phys. Rev. D \textbf{81}, 014021 (2010).

\bibitem{Godfrey:2013aaa}
S.~Godfrey and I.~T.~Jardine,
Nature of the $D_{s1}^*(2710)$ and $D_{sJ}^*(2860)$ mesons,
Phys. Rev. D \textbf{89}, 074023 (2014).

\bibitem{Godfrey:2014fga}
S.~Godfrey and K.~Moats,
The $D_{sJ}^*(2860)$ Mesons as Excited D-wave $c\bar{s}$ States,
Phys. Rev. D \textbf{90}, 117501 (2014).

\bibitem{Gandhi:2020vap}
K.~Gandhi, A.~K.~Rai and T.~Matsuki,
Identifying $D_{sJ}^*(2860)$ as a four resonance states through strong decay analysis,
[arXiv:2003.00487 [hep-ph]].

\bibitem{Wang:2021orp}
G.~L.~Wang, W.~Li, T.~F.~Feng, Y.~L.~Wang and Y.~B.~Liu,
The newly observed state $D_{s0}(2590)^{+}$,
[arXiv:2107.01751 [hep-ph]].

\bibitem{Tian:2017okw}
Y.~Tian, Z.~Zhao and A.~Zhang,
Study of radially excited $D_s(2^1S_0)$ and $D_s(3P)$,
Chin. Phys. C \textbf{41}, 083107 (2017).

\bibitem{Wang:2018psi}
Z.~H.~Wang, Y.~Zhang, T.~h.~Wang, Y.~Jiang, Q.~Li and G.~L.~Wang,
Strong Decays of $P-$wave Mixing Heavy-Light $1^+$ States,
Chin. Phys. C \textbf{42}, 123101 (2018).

\bibitem{Colangelo:2012xi}
P.~Colangelo, F.~De Fazio, F.~Giannuzzi and S.~Nicotri,
New meson spectroscopy with open charm and beauty,
Phys. Rev. D \textbf{86}, 054024 (2012).

\bibitem{Colangelo:2010te}
P.~Colangelo and F.~De Fazio,
Open charm meson spectroscopy: Where to place the latest piece of the puzzle,
Phys. Rev. D \textbf{81}, 094001 (2010).

\bibitem{Zhao:2016mxc}
Z.~Zhao, Y.~Tian and A.~Zhang,
Hadronic production of $D(2550)$, $D^*(2600)$, $D(2750)$, $D^*_1(2760)$ and $D^*_3(2760)$,
Phys. Rev. D \textbf{94}, 114035 (2016).

\bibitem{Li:2017zng}
S.~C.~Li, T.~Wang, Y.~Jiang, X.~Tan, Q.~Li, G.~L.~Wang and C.~H.~Chang,
Strong decays of $D_J(3000)$ and $D_{sJ}(3040)$,
Phys. Rev. D \textbf{97}, 054002 (2018).


\bibitem{Godfrey:1985xj}
  S.~Godfrey and N.~Isgur,
  Mesons in a relativized quark model with chromodynamics,
  Phys.\ Rev.\ D {\bf 32}, 189 (1985).

\bibitem{Zeng:1994vj}
J.~Zeng, J.~W.~Van Orden and W.~Roberts,
Heavy mesons in a relativistic model,
Phys. Rev. D \textbf{52}, 5229-5241 (1995).

\bibitem{Gupta:1994mw}
S.~N.~Gupta and J.~M.~Johnson,
Quantum chromodynamic potential model for light heavy quarkonia and the heavy quark effective theory,
Phys. Rev. D \textbf{51}, 168-175 (1995).

\bibitem{Lahde:1999ih}
T.~A.~Lahde, C.~J.~Nyfalt and D.~O.~Riska,
Spectra and M1 decay widths of heavy light mesons,
Nucl. Phys. A \textbf{674}, 141-167 (2000).

\bibitem{DiPierro:2001dwf}
M.~Di Pierro and E.~Eichten,
Excited Heavy-Light Systems and Hadronic Transitions,
Phys. Rev. D \textbf{64}, 114004 (2001).

\bibitem{Godfrey:2005ww}
S.~Godfrey, Properties of the charmed P-wave mesons,
Phys. Rev. D \textbf{72}, 054029 (2005).

\bibitem{Close:2005se}
F.~E.~Close and E.~S.~Swanson,
Dynamics and decay of heavy-light hadrons,
Phys. Rev. D \textbf{72}, 094004 (2005)

\bibitem{Vijande:2006hj}
J.~Vijande, F.~Fernandez and A.~Valcarce,
Open-charm meson spectroscopy,
Phys. Rev. D \textbf{73}, 034002 (2006)
[erratum: Phys. Rev. D \textbf{74}, 059903 (2006)].

\bibitem{Close:2006gr}
F.~E.~Close, C.~E.~Thomas, O.~Lakhina and E.~S.~Swanson,
Canonical interpretation of the $D_{sJ}(2860)$ and $D_{sJ}(2690)$,
Phys. Lett. B \textbf{647}, 159-163 (2007).

\bibitem{Li:2007px}
D.~M.~Li, B.~Ma and Y.~H.~Liu,
Understanding masses of $c\bar{s}$ states in Regge phenomenology,
Eur. Phys. J. C \textbf{51}, 359-365 (2007).

\bibitem{Zhang:2006yj}
 B.~Zhang, X.~Liu, W.~Z.~Deng and S.~L.~Zhu,
 $D_ {sJ} (2860)$ and $D_ {sJ} (2715)$,
 Eur. Phys. J. C \textbf{50}, 617-628 (2007).

\bibitem{Wei:2006wa}
W.~Wei, X.~Liu and S.~L.~Zhu,
D wave heavy mesons,
Phys. Rev. D \textbf{75}, 014013 (2007).


\bibitem{Mohler:2011ke}
D.~Mohler and R.~M.~Woloshyn,
$D$ and $D_s$ meson spectroscopy,
Phys. Rev. D \textbf{84}, 054505 (2011).

\bibitem{Moir:2013ub}
G.~Moir, M.~Peardon, S.~M.~Ryan, C.~E.~Thomas and L.~Liu,
Excited spectroscopy of charmed mesons from lattice QCD,
JHEP \textbf{05}, 021 (2013).

\bibitem{Kalinowski:2015bwa}
M.~Kalinowski and M.~Wagner,
Masses of $D$ mesons, $D_s$ mesons and charmonium states from twisted mass lattice QCD,
Phys. Rev. D \textbf{92}, 094508 (2015).

\bibitem{Cichy:2016bci}
K.~Cichy, M.~Kalinowski and M.~Wagner,
Continuum limit of the $D$ meson, $D_s$ meson and charmonium spectrum from $N_f=2+1+1$ twisted mass lattice QCD,
Phys. Rev. D \textbf{94}, 094503 (2016).


\bibitem{Du:2020pui}
M.~L.~Du, F.~K.~Guo, C.~Hanhart, B.~Kubis and U.~G.~Mei\ss{}ner,
Where is the lightest charmed scalar meson?
Phys. Rev. Lett. \textbf{126}, 192001 (2021).

\bibitem{Albaladejo:2016lbb}
M.~Albaladejo, P.~Fernandez-Soler, F.~K.~Guo and J.~Nieves,
Two-pole structure of the $D^\ast_0(2400)$,
Phys. Lett. B \textbf{767}, 465-469 (2017).

\bibitem{Du:2017zvv}
M.~L.~Du, M.~Albaladejo, P.~Fern\'andez-Soler, F.~K.~Guo, C.~Hanhart, U.~G.~Mei\ss{}ner, J.~Nieves and D.~L.~Yao,
Towards a new paradigm for heavy-light meson spectroscopy,
Phys. Rev. D \textbf{98}, 094018 (2018).

\bibitem{Gayer:2021xzv}
L.~Gayer \textit{et al.} [Hadron Spectrum],
Isospin-1/2 $D\pi$ scattering and the lightest $ {D}_0^{\ast } $ resonance from lattice QCD,
JHEP \textbf{07} (2021), 123.


\bibitem{BaBar:2003oey}
B.~Aubert \textit{et al.} [BaBar],
Observation of a narrow meson decaying to $D_s^+ \pi^0$ at a mass of 2.32-GeV/c$^2$,
Phys. Rev. Lett. \textbf{90}, 242001 (2003).

\bibitem{CLEO:2003ggt}
D.~Besson \textit{et al.} [CLEO],
Observation of a narrow resonance of mass 2.46-GeV/c$^2$ decaying to $D^{*+}_s pi0$ and confirmation of the $D^*_{sJ}(2317)$ state,
Phys. Rev. D \textbf{68}, 032002 (2003)
[erratum: Phys. Rev. D \textbf{75}, 119908 (2007)].


\bibitem{Yang:2021tvc}
Z.~Yang, G.~J.~Wang, J.~J.~Wu, M.~Oka and S.~L.~Zhu,
Novel coupled channel framework connecting quark model and lattice QCD: an investigation on near-threshold $D_s$ states,
[arXiv:2107.04860 [hep-ph]].

\bibitem{Ortega:2016mms}
P.~G.~Ortega, J.~Segovia, D.~R.~Entem and F.~Fernandez,
Molecular components in P-wave charmed-strange mesons,
Phys. Rev. D \textbf{94}, 074037 (2016).

\bibitem{Chen:2016spr}
H.~X.~Chen, W.~Chen, X.~Liu, Y.~R.~Liu and S.~L.~Zhu,
A review of the open charm and open bottom systems,
Rept. Prog. Phys. \textbf{80}, 076201 (2017).


\bibitem{Xiao:2013xi}
  L.~Y.~Xiao and X.~H.~Zhong,
  $\Xi$ baryon strong decays in a chiral quark model,
  Phys.\ Rev.\ D {\bf 87}, 094002 (2013).

\bibitem{Liu:2019wdr}
  M.~S.~Liu, K.~L.~Wang, Q.~F.~L\"{u} and X.~H.~Zhong,
  $\Omega$ baryon spectrum and their decays in a constituent quark model,
  Phys.\ Rev.\ D {\bf 101}, 016002 (2020).

\bibitem{Xiao:2018pwe}
L.~Y.~Xiao and X.~H.~Zhong,
Possible interpretation of the newly observed $\Omega(2012)$ state,
Phys. Rev. D \textbf{98}, 034004 (2018).

\bibitem{Zhong:2007gp}
  X.~H.~Zhong and Q.~Zhao,
  Charmed baryon strong decays in a chiral quark model,
  Phys.\ Rev.\  D {\bf 77}, 074008 (2008).

\bibitem{Liu:2012sj}
  L.~H.~Liu, L.~Y.~Xiao and X.~H.~Zhong,
  Charm-strange baryon strong decays in a chiral quark model,
  Phys.\ Rev.\ D {\bf 86}, 034024 (2012).

\bibitem{Yao:2018jmc}
  Y.~X.~Yao, K.~L.~Wang and X.~H.~Zhong,
  Strong and radiative decays of the low-lying $D$-wave singly heavy baryons,
  Phys.\ Rev.\ D {\bf 98}, 076015 (2018).

\bibitem{Wang:2017kfr}
  K.~L.~Wang, Y.~X.~Yao, X.~H.~Zhong and Q.~Zhao,
  Strong and radiative decays of the low-lying $S$- and $P$-wave singly heavy baryons,
  Phys.\ Rev.\ D {\bf 96}, 116016 (2017).

\bibitem{Nagahiro:2016nsx}
  H.~Nagahiro, S.~Yasui, A.~Hosaka, M.~Oka and H.~Noumi,
  Structure of charmed baryons studied by pionic decays,
  Phys.\ Rev.\ D {\bf 95}, 014023 (2017).

\bibitem{Wang:2017hej}
  K.~L.~Wang, L.~Y.~Xiao, X.~H.~Zhong and Q.~Zhao,
  Understanding the newly observed $\Omega_c$ states through their decays,
  Phys.\ Rev.\ D {\bf 95}, 116010 (2017).

\bibitem{Xiao:2020oif}
L.~Y.~Xiao, K.~L.~Wang, M.~S.~Liu and X.~H.~Zhong,
 Possible interpretation of the newly observed $\Omega _b$ states,
Eur. Phys. J. C \textbf{80}, 279 (2020).

\bibitem{Wang:2019uaj}
K.~L.~Wang, Q.~F.~L\"u and X.~H.~Zhong,
Interpretation of the newly observed $\Lambda_b(6146)^{0}$ and $\Lambda_b(6152)^0$ states in a chiral quark model,
Phys. Rev. D \textbf{100}, 114035 (2019).

\bibitem{Wang:2018fjm}
K.~L.~Wang, Q.~F.~L\"u and X.~H.~Zhong,
Interpretation of the newly observed $\Sigma_b(6097)^{\pm}$ and $\Xi_b(6227)^-$ states as the $P$-wave bottom baryons,
Phys. Rev. D \textbf{99}, 014011 (2019).

\bibitem{Xiao:2020gjo}
L.~Y.~Xiao and X.~H.~Zhong,
Toward establishing the low-lying $P$-wave $\Sigma_b$ states,
Phys. Rev. D \textbf{102}, 014009 (2020).

\bibitem{Wang:2020gkn}
K.~L.~Wang, L.~Y.~Xiao and X.~H.~Zhong,
Understanding the newly observed $\Xi_c^0$ states through their decays,
Phys. Rev. D \textbf{102}, 034029 (2020).

\bibitem{Xiao:2017udy}
  L.~Y.~Xiao, K.~L.~Wang, Q.~f.~L\"{u}, X.~H.~Zhong and S.~L.~Zhu,
  Strong and radiative decays of the doubly charmed baryons,
  Phys.\ Rev.\ D {\bf 96}, 094005 (2017).

\bibitem{li:2021hss}
Q.~li, R.~H.~Ni and X.~H.~Zhong,
Towards establishing an abundant $B$ and $B_s$ spectrum up to the second orbital excitations,
Phys. Rev. D \textbf{103}, 116010 (2021).

\bibitem{Eichten:1978tg}
  E.~Eichten, K.~Gottfried, T.~Kinoshita, K.~D.~Lane and T.~M.~Yan,
  Charmonium: The model,
  Phys.\ Rev.\ D {\bf 17}, 3090 (1978);{\bf 21}, 313(E) (1980).

\bibitem{Swanson:2005}
  T.~Barnes, S.~Godfrey, and E.~S.~Swanson,
  Higher charmonia,
  Phys.\ Rev.\ D {\bf 72}, 054026 (2005).

\bibitem{Godfrey:2004ya}
  S.~Godfrey,
  Spectroscopy of $B_c$ mesons in the relativized quark model,
  Phys.\ Rev.\ D {\bf 70}, 054017 (2004).

\bibitem{Hiyama:2003cu}
E.~Hiyama, Y.~Kino and M.~Kamimura,
Gaussian expansion method for few-body systems,
Prog. Part. Nucl. Phys. \textbf{51}, 223-307 (2003).

\bibitem{Deng:2016ktl}
W.~J.~Deng, H.~Liu, L.~C.~Gui and X.~H.~Zhong,
Spectrum and electromagnetic transitions of bottomonium,
Phys. Rev. D \textbf{95}, 074002 (2017).

\bibitem{Deng:2016stx}
W.~J.~Deng, H.~Liu, L.~C.~Gui and X.~H.~Zhong,
Charmonium spectrum and their electromagnetic transitions with higher multipole contributions,
Phys. Rev. D \textbf{95}, 034026 (2017).

\bibitem{Li:2019qsg}
Q.~Li, M.~S.~Liu, Q.~F.~L\"u, L.~C.~Gui and X.~H.~Zhong,
Canonical interpretation of $Y(10750)$ and $\Upsilon(10860)$ in the $\Upsilon$ family,
Eur. Phys. J. C \textbf{80}, 59 (2020).

\bibitem{Li:2019tbn}
Q.~Li, M.~S.~Liu, L.~S.~Lu, Q.~F.~L\"u, L.~C.~Gui and X.~H.~Zhong,
Excited bottom-charmed mesons in a nonrelativistic quark model,
Phys. Rev. D \textbf{99}, 096020 (2019).

\bibitem{Li:2020xzs}
Q.~Li, L.~C.~Gui, M.~S.~Liu, Q.~F.~L\"u and X.~H.~Zhong,
Mass spectrum and strong decays of strangeonium in a constituent quark model,
Chin. Phys. C \textbf{45}, 023116 (2021).



\bibitem{Manohar:1983md}
  A.~Manohar and H.~Georgi,
  Chiral Quarks and the Nonrelativistic Quark Model,
  Nucl.\ Phys.\ B {\bf 234}, 189 (1984).

\bibitem{Li:1994cy}
Z.~P.~Li,
The Threshold pion photoproduction of nucleons in the chiral quark model,
Phys. Rev. D {\bf 50}, 5639 (1994).

\bibitem{Li:1997gda}
Z.~P.~Li, H.~X.~Ye and M.~H.~Lu,
An Unified approach to pseudoscalar meson photoproductions off nucleons in the quark model,
Phys. Rev. C {\bf 56}, 1099 (1997).

\bibitem{Zhao:2002id}
Q.~Zhao, J.~S.~Al-Khalili, Z.~P.~Li and R.~L.~Workman,
Pion photoproduction on the nucleon in the quark model,
Phys. Rev. C {\bf 65}, 065204 (2002).


\bibitem{Zhao:1998fn}
  Q.~Zhao, Z.~P.~Li and C.~Bennhold,
  Vector meson photoproduction with an effective Lagrangian in the quark model,
  Phys.\ Rev.\  C {\bf 58}, 2393 (1998).

\bibitem{Zhao:2000tb}
  Q.~Zhao,
  Nucleonic resonance excitations with linearly polarized photon in $\gamma  p\to \omega p$,
  Phys.\ Rev.\  C {\bf 63}, 025203 (2001).

\bibitem{Zhao:2001jw}
  Q.~Zhao, J.~S.~Al-Khalili and C.~Bennhold,
  Quark model predictions for the $K^*$ photoproduction on the proton,
  Phys.\ Rev.\  C {\bf 64}, 052201 (2001).

\bibitem{Koniuk:1979vy}
  R.~Koniuk and N.~Isgur,
  Baryon Decays in a Quark Model with Chromodynamics,
  Phys.\ Rev.\ D {\bf 21}, 1868 (1980)
  Erratum: [Phys.\ Rev.\ D {\bf 23}, 818 (1981)].

\bibitem{Goity:1998jr}
J.~L.~Goity and W.~Roberts,
A Relativistic chiral quark model for pseudoscalar emission from heavy mesons,
Phys. Rev. D \textbf{60}, 034001 (1999).

\bibitem{Capstick:2000qj}
  S.~Capstick and W.~Roberts,
  Quark models of baryon masses and decays,
  Prog.\ Part.\ Nucl.\ Phys.\  {\bf 45}, S241 (2000).

\bibitem{Wang:2012wk}
Z.~H.~Wang, G.~L.~Wang, J.~M.~Zhang and T.~H.~Wang,
The Productions and Strong Decays of $D_q(2S)$ and $B_q(2S)$,
J. Phys. G \textbf{39}, 085006 (2012).



\bibitem{Xie:2021dwe}
J.~M.~Xie, M.~Z.~Liu and L.~S.~Geng,
$D_{s0}(2590)$ as a dominant cs state with a small $D^*K$ component,
Phys. Rev. D \textbf{104}, no.9, 094051 (2021).

\bibitem{Ortega:2021fem}
P.~G.~Ortega, J.~Segovia, D.~R.~Entem and F.~Fernandez,
The $D_{s0}(2590)^+$ as the dressed $c\bar s(2^1S_0)$ meson in a coupled-channels calculation,
[arXiv:2111.00023 [hep-ph]].


\bibitem{Belle:2003nsh}
K.~Abe \textit{et al.} [Belle],
Study of $B^- \to D^{**0} \pi^- (D^{**0} \to D^{(*)+} \pi^-$ decays,
Phys. Rev. D \textbf{69}, 112002 (2004).

\bibitem{BaBar:2009pnd}
B.~Aubert \textit{et al.} [BaBar],
Dalitz Plot Analysis of $B^- \to D^+ \pi^- \pi^-$,
Phys. Rev. D \textbf{79}, 112004 (2009).

\bibitem{FOCUS:2003gru}
J.~M.~Link \textit{et al.} [FOCUS],
Measurement of masses and widths of excited charm mesons $D_2^*$ and evidence for broad states,
Phys. Lett. B \textbf{586}, 11-20 (2004).

\bibitem{Liu:2012zya}
L.~Liu, K.~Orginos, F.~K.~Guo, C.~Hanhart and U.~G.~Meissner,
Interactions of charmed mesons with light pseudoscalar mesons from lattice QCD and implications on the nature of the $D_{s0}^*(2317)$,
Phys. Rev. D \textbf{87}, 014508 (2013).

\bibitem{Isgur:1998kr}
N.~Isgur,
Spin orbit inversion of excited heavy quark mesons,
Phys. Rev. D \textbf{57}, 4041-4053 (1998).


\bibitem{Morel:2002vk}
D.~Morel and S.~Capstick,
Baryon meson loop effects on the spectrum of nonstrange baryons,
[arXiv:nucl-th/0204014 [nucl-th]].

\bibitem{Tan:2021bvl}
Y.~Tan and J.~Ping,
$D^*_{s0}(2317)$ and $D_{s1}(2460)$ in an unquenched quark model,
[arXiv:2111.04677 [hep-ph]].


\bibitem{Kalashnikova:2005ui}
Y.~S.~Kalashnikova,
Coupled-channel model for charmonium levels and an option for X(3872),
Phys. Rev. D \textbf{72}, 034010 (2005).

\bibitem{Lu:2016mbb}
Y.~Lu, M.~N.~Anwar and B.~S.~Zou,
Coupled-Channel Effects for the Bottomonium with Realistic Wave Functions,
Phys. Rev. D \textbf{94}, 034021 (2016).

\bibitem{Lu:2017hma}
Y.~Lu, M.~N.~Anwar and B.~S.~Zou,
How Large is the Contribution of Excited Mesons in Coupled-Channel Effects?,
Phys. Rev. D \textbf{95}, 034018 (2017).

\bibitem{Liu:2011yp}
J.~F.~Liu and G.~J.~Ding,
Bottomonium Spectrum with Coupled-Channel Effects,
Eur. Phys. J. C \textbf{72}, 1981 (2012).

\bibitem{Luo:2019qkm}
S.~Q.~Luo, B.~Chen, Z.~W.~Liu and X.~Liu,
Resolving the low mass puzzle of $\Lambda_c(2940)^+$,
Eur. Phys. J. C \textbf{80}, 301 (2020).

\bibitem{Luo:2021dvj}
S.~Q.~Luo, B.~Chen, X.~Liu and T.~Matsuki,
Predicting a new resonance as charmed-strange baryonic analog of $D^*_{s0}$(2317),
Phys. Rev. D \textbf{103}, 074027 (2021).




\end{thebibliography}

\end{document}